\documentclass[]{spie}  

\usepackage{booktabs}
\usepackage{float}
\usepackage{multirow}
\usepackage{hyperref}
\usepackage{pgf}

\usepackage{longtable}

\usepackage{tikz}
\usetikzlibrary{shapes,arrows}
\tikzstyle{block} = [rectangle, draw, fill=blue!20, 
    text width=5em, text centered, rounded corners, minimum height=4em]

\usepackage{subfig}
\usepackage{cancel}
\usepackage{amsmath,amsfonts,amssymb}
\usepackage{graphicx}
\usepackage{makecell}

\definecolor{Gray}{gray}{0.9}

\title{Effect of the output activation function on the probabilities and errors in medical image segmentation}

\author[a]{Lars Nieradzik}
\author[a]{Gerik Scheuermann}
\author[b]{Dorothee Saur}
\author[a]{Christina Gillmann}
\affil[a]{Leipzig University, Augustusplatz 10, 04109 Leipzig, Germany}
\affil[b]{Leipzig University Medical Centre, Liebigstraße 20, 04103 Leipzig, Germany}

\authorinfo{Further author information: (Send correspondence to Lars Nieradzik)\\Lars Nieradzik. E-mail: l.nieradzik@gmail.com}

\begin{document}
\maketitle

\begin{abstract}
The sigmoid activation is the standard output activation function in binary classification and segmentation with neural networks. Still, there exist a variety of other potential output activation functions, which may lead to improved results in medical image segmentation. In this work, we consider how the asymptotic behavior of different output activation and loss functions affects the prediction probabilities and the corresponding segmentation errors. For cross entropy, we show that a faster rate of change of the activation function correlates with better predictions, while a slower rate of change can improve the calibration of probabilities. For dice loss, we found that the arctangent activation function is superior to the sigmoid function. Furthermore, we provide a test space for arbitrary output activation functions in the area of medical image segmentation. We tested seven activation functions in combination with three loss functions on four different medical image segmentation tasks to provide a classification of which function is best suited in this application scenario.
\end{abstract}


\section{INTRODUCTION}
\label{sec:intro}  

Image segmentation is an important subtask in medical image analysis \cite{10.4108/eai.12-4-2021.169184}. In recent years, machine learning became a popular tool to complete segmentation tasks as it outperforms analytical methods in many medical disciplines such as tumor detection \cite{Febrianto_2020}, lesion prediction \cite{2021-4} or skin anomaly classification \cite{jimaging7040067}. Here, the output activation function is usually chosen to be the sigmoid function \cite{DBLP:journals/corr/abs-1811-03378} in the case of two potential output classes or the softmax function for more than two potential output classes. Unfortunately, there has not been a testing environment or general test settings that aim to examine the suitability of further potential output activation functions besides the sigmoid function (see Section \ref{sec:rel}). 

A major problem, when considering medical images, is the uncertainty inherent in the medical images themselves and the medical imaging pipeline applied to these images \cite{2021-5}. Gillmann et al. \cite{2021-1} provided a taxonomy of uncertainties that can occur in the medical imaging pipeline and showed how they can affect the decision-making process in clinical daily routine. Therefore, an output activation function that is able to express the uncertainty inherent in a prediction, while increasing the accuracy of the segmentation result, is highly desired. 

In this manuscript, we aim to test and evaluate a variety of output activation functions in combination with loss functions (see Section \ref{sec:met}). Here, we aim to identify a proper combination of output activation (7 in total) and loss functions (3 in total). We aim to test this setting against two different evaluation metrics, namely negative log-likelihood, and dice coefficient while training neural networks on 4 different medical datasets (see Section \ref{sec:eval}).

Therefore, this paper contributes:

\begin{itemize}
    \item A summary of potential output activation functions for medical image segmentation.
    \item A comprehensive evaluation of output activation functions in medical image segmentation.
    \item An analysis of the asymptotic behavior of the output activation functions in relation to the errors and probabilities generated by the neural networks.
    \item An open test space for output activation functions in medical image segmentation.
\end{itemize}

Our results will be discussed in Section \ref{sec:dis}, where advantages, disadvantages, and limitations of the test functions and the provided test space will be shown. One of the conclusions (see Section \ref{sec:con}) that we can draw is that the sigmoid activation function is not the optimal choice of an output activation function regarding medical imaging.

\section{RELATED WORK}
\label{sec:rel}

Medical image segmentation is an important sub-task of the medical image processing pipeline \cite{2019-1}. There exist a variety of network architectures that have been shown to deliver promising segmentation results that outperform classic analytical approaches \cite{lei2020medical}. Depending on the chosen network architecture, multiple parameters need to be fine-tuned during the training process of a neural network. The output activation function is usually not one of these parameters, and is in most cases set to be the sigmoid or softmax function. In this section, we aim to shed light on related work that provides knowledge on other output activation functions that may be used in medical image segmentation.

The standard sigmoid activation is known in statistics as the logistic function and corresponds to the logit model. There are a wide range of papers that compare the results of this logistic function to two other models: the probit and linear model \cite{10.1257/jep.15.4.43, RePEc:aes:jsesro:v:2:y:2013:i:1:p:1-17, https://doi.org/10.1002/smj.582, defaria}. Horowitz et al.\cite{10.1257/jep.15.4.43} reviewed the linear, probit and logit model with examples from biometrics and econometrics. Cakmakyapan et al.\cite{RePEc:aes:jsesro:v:2:y:2013:i:1:p:1-17} showed in a simulation study that the logit model performs better for large sample sizes (500 and 1000) and the probit model is better for small sizes (40, 100, 200). Hoetker et al.\cite{https://doi.org/10.1002/smj.582} reviewed problems when
interpreting the coefficients of logit, probit and linear models.

However, the linear and probit model have never been analyzed in deep learning as output activation functions. Gomes et al.\cite{4626755} compared the complementary log-log and probit model as an alternative to the sigmoid activation, but only in intermediate layers. Another paper by Zeng\cite{ZENG1999} contrasted the logit/probit model with neural networks for social science classification problems. They found that neural network models perform significantly better than logit/probit models. However, they did not try the activation function of the probit model in neural networks.

We use these different statistical models as a starting point for our approach to find alternative output activation functions. In this work, we will use both the activation function of the probit and of the linear model for medical image segmentation and contrast them with the regular sigmoid function.

Apart from the statistical literature, several activation functions were proposed for non-output layers in neural networks. Ramachandran et al.\cite{DBLP:journals/corr/abs-1710-05941} used automated search techniques to find the Swish activation function as a replacement to ReLU. This activation function was then extended and analyzed by Ma et al. \cite{DBLP:journals/corr/abs-2009-04759}. Since neuron inputs tend to follow a normal distribution, Hendrycks et al.\cite{DBLP:journals/corr/HendrycksG16} suggested the use of the standard Gaussian cumulative distribution function (CDF) as part of a new activation function. In statistics, the normal CDF is also used as an activation function in the probit model. This shows that the CDF is a viable option as an output activation function.

A paper by Molina et al.\cite{Molina2020Pade} proposed the Padé Activation Unit, a trainable activation function. By adapting automatically to the data at hand, they obtained improvements relative to the ReLU activation. In this work, we will also use this idea of adaptivity for an output activation function.

Finally, a review by Nwankpa et al.\cite{DBLP:journals/corr/abs-1811-03378} summarized most of the common activation functions used in neural networks. Another paper by Apicella et al.\cite{Apicella_2021} focused more on summarizing trainable activation functions. We also take inspiration from some of these activation functions, which were originally designed for intermediate layers.

Based on this review of the literature, we have seen that finding a better activation function than ReLU is a common topic in the deep learning literature. However, to the best of our knowledge, there are no other publications that focus solely on the activation function in the output layer of neural networks or examine the impact of this choice.

\section{METHODS}
\label{sec:met}

Neural network-based segmentation models consist of an encoder that extracts features and downsamples the image, and a decoder that restores the image to its original size. This process is followed by an output activation function that determines the probability of each class (e.g. foreground and background pixel). During the training process, the loss function determines the correctness of the prediction made and allows readjusting network parameters accordingly.

This work proposes replacing the standard sigmoid activation function that is applied to the decoder outputs with other functions with similar properties. The goal is to clarify if there are more appropriate choices than the sigmoid output activation function for medical image segmentation. Additionally, we combine potential output activation functions with various loss functions when training the network.

One of the results of this work is a test space that allows the user to train segmentation models on various medical segmentation datasets. We can change parameters such as the encoder, the decoder, the output activation, and the loss function, as shown in Figure \ref{fig:structure}. The complete source code is available on GitHub \url{https://github.com/lars76/segmentation_activations}.

\definecolor{celadon}{rgb}{0.67, 0.88, 0.69}
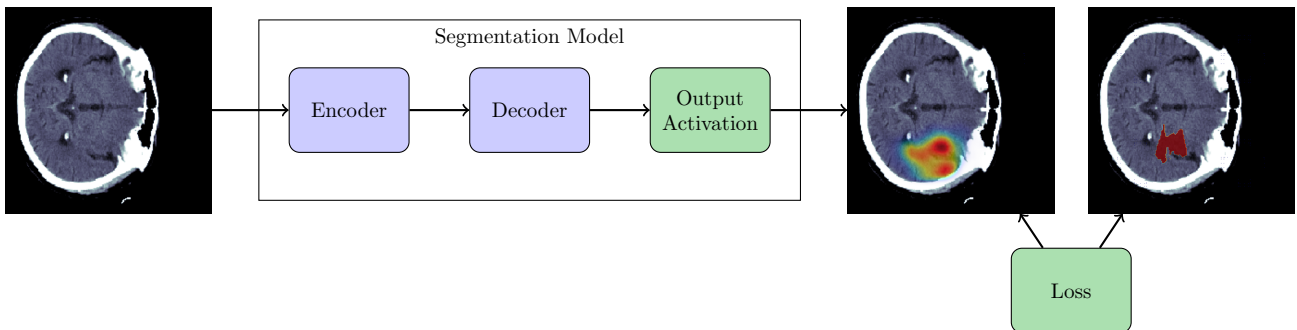
\begin{figure}[H]
\centering
\begin{tikzpicture}[scale=0.8, transform shape]
	\node[inner sep=0pt] (input) at (-1,0)
	    {\includegraphics[scale=0.3]{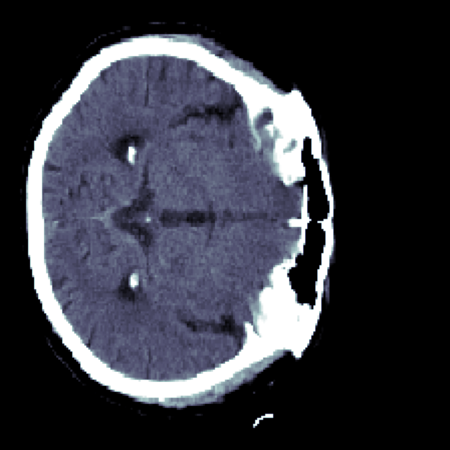}};
	\node at (3,0)  [block] (Encoder) {Encoder};
	\node at (6,0)  [block] (Decoder) {Decoder};
	\node at (9,0)  [block,fill=celadon] (Activation) {Output Activation};
	\node at (15,-3)  [block,fill=celadon] (Loss) {Loss};
	\node[inner sep=0pt] (output) at (13,0)
	    {\includegraphics[scale=0.3]{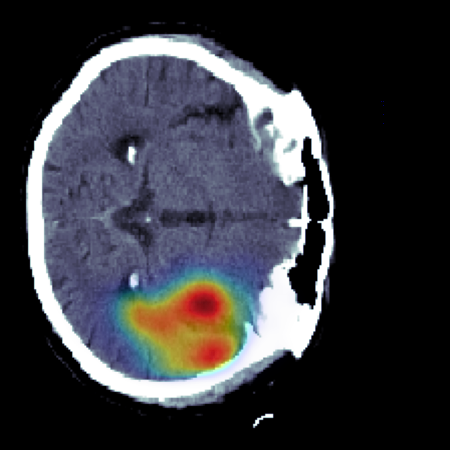}};
	\node[inner sep=0pt] (output2) at (17,0)
	    {\includegraphics[scale=0.3]{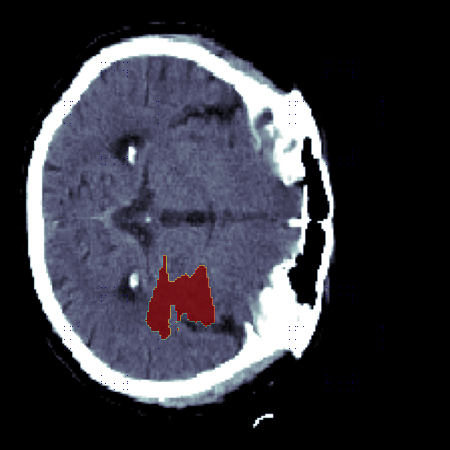}};
	\draw (1.5,-1.5) rectangle (10.5,1.5) node[pos=0.5, anchor=south] [text depth=1cm] {Segmentation Model};

    \draw [->,thick] (Encoder) -- (Decoder);
    \draw [->,thick] (Decoder) -- (Activation);
    \draw [->,thick] (input) -- (Encoder);
    \draw [->,thick] (Activation) -- (output);
    \draw [->,thick] (Loss) -- (output);
    \draw [->,thick] (Loss) -- (output2);
\end{tikzpicture}
    \caption{General workflow of segmentation models. The neural network consists of an encoder, decoder, and activation function. The activation function is applied to the output of the decoder. The loss functions regulate the training process of the neural network. Output activation and loss function are highlighted in green, as they are the main focus of this work.}
  \label{fig:structure}
\end{figure}
  
In the following, we introduce the output activation functions that are evaluated in this paper (see Section \ref{sec:act}), followed by the tested loss functions (see Section \ref{sec:loss}).

\subsection{Output Activation Function}
\label{sec:act}

The sigmoid or logistic function $f(x) = \frac{1}{1+\exp{\left(-x\right)}}$ is the standard output activation function for binary segmentation or classification tasks. It is a simplification of the softmax function that is used for classification or segmentation tasks with more than two classes.

   \begin{figure}[h]
   \begin{center}
   
    \scalebox{0.7}{\input{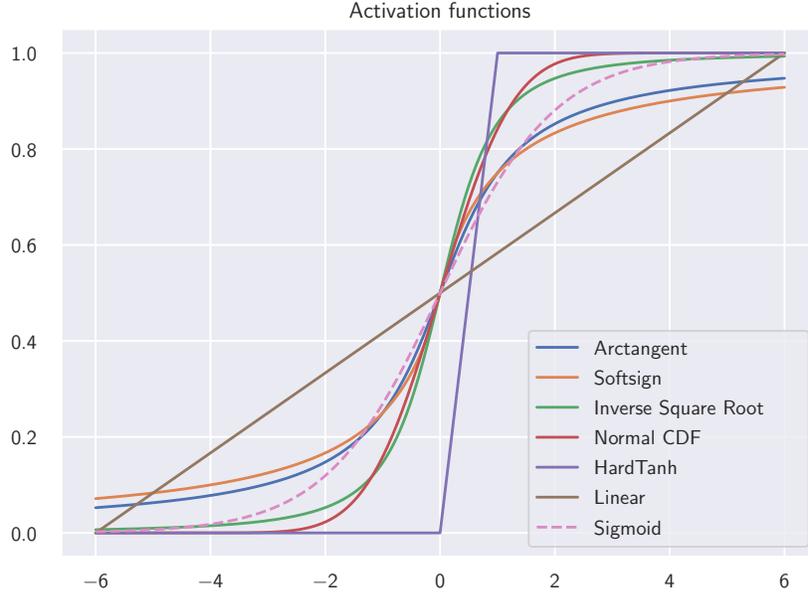}}
   \end{center}
   \caption[example] 
   { \label{fig:activations} Limiting behavior of the activation functions to their asymptotes. The sigmoid function is shown as a dashed line. The remaining functions can mainly be separated into functions with faster or slower rate of change than sigmoid.}
   \end{figure}

The sigmoid function shows an exponential growth around zero that quickly saturates and slows to linear growth (see fig.~\ref{fig:activations}, dashed line). At $x = -6$ and $x = 6$ the sigmoid function is already close to its asymptotes (zero and one). However, this rate of change is not necessarily optimal. We shall consider two cases regarding the asymptotes: a faster rate of change (less freedom of action) and a slower rate of change (more freedom of action).

Both cases will be considered in our analysis using different activation functions on the output layer. The choice of the functions will be explained in the following.

It is possible to use any differentiable function in the output layer of neural networks. There are, however, some requirements. Certain loss functions need activation functions with range $[0, 1]$ (e.g. dice loss). Furthermore, the function should be symmetric such that the probabilities have the same meaning no matter whether the input values are in the intervals $(-\infty, 0)$ or $(0, \infty)$. Many of the activation functions listed by Nwankpa et al. \cite{DBLP:journals/corr/abs-1811-03378} are thus no option. Instead, we choose functions $f : \mathbb{R} \to [0, 1]$ that are for the most part monotonic increasing and symmetric with $f(0) = \frac{1}{2}$.

There is naturally an infinite number of activation functions that satisfy the above requirements. In order to restrict the space of possible functions, we choose functions that are found in statistics (probit model, linear model) and/or have a sigmoidal shape (S-shape). Then we rescale the activations to the range $[0, 1]$.

In order to better understand the limiting behavior of the functions to the horizontal asymptotes at $0$ and $1$, we define the \emph{effective domain} as $X = \{x \mid f(x) \leq 1 - \epsilon, f(x) \geq \epsilon\}$. For most activation functions, we do not need the complete real line $\mathbb{R}$, but can restrict ourselves to a smaller domain $X$.

 \begin{table}[H]
        \centering
        \begin{tabular}{llll}
        \toprule
        \textbf{Function name} & $f(x)$ & $f'(x)$ & Effective domain\\ \hline
        normal CDF & $\Phi(x) = \frac{1}{2}\text{erf}\left(\frac{x}{\sqrt{2}}\right) + \frac{1}{2}$ & $\phi(x) = \frac{\exp{\left(-x^2\right)}}{\sqrt{2\pi}}$ & $[-3, 3]$\\
        sigmoid & $\frac{1}{1 + \exp{\left(-x\right)}}$ & $\frac{1}{1 + \exp{\left(-x\right)}}\left(1 - \frac{1}{1 + \exp{\left(-x\right)}}\right)$ & $[-6, 6]$\\
        inverse square root & $\frac{1}{2}\frac{x}{\sqrt{1 + x^2}} + \frac{1}{2}$ & $\frac{0.5}{\left(x^2 + 1\right)^\frac{3}{2}}$ & $[-10, 10]$ \\
        arctangent & $\frac{\arctan(x)}{\pi} + \frac{1}{2}$ & $\frac{1}{\pi x^2 + \pi}$ & $[-128, 128]$ \\
        softsign & $\frac{1}{2}\frac{x}{1 + |x|} + \frac{1}{2}$ & $\frac{0.5}{(|x| + 1)^2}$ & $[-199, 199]$ \\
        \midrule
        linear & $\frac{x - x_{\text{min}}}{x_{\text{max}} - x_{\text{min}}}$ & $\frac{1}{x_{\text{max}} - x_{\text{min}}}$ & $[x_{\text{min}}, x_{\text{max}}]$\\
        hardtanh & $\begin{cases}1 & \text{if } x > 1\\0 & \text{if } x < 0\\x & \text{otherwise}
        \end{cases}$ & $\begin{cases}1 & \text{if } 0 \leq x \leq 1\\0 & \text{otherwise}
        \end{cases}$ & $[0, 1]$\\
        \bottomrule
        \end{tabular}
        \caption{Output activation functions considered in this work. The table includes the name, the derivative, and the effective domain of each function. The effective domain was rounded by applying the ceiling and floor function. We choose $\epsilon = 0.0025$.}
        \label{tab:activations}
        \end{table}

In Table \ref{tab:activations}, we present the activation functions that are evaluated in this paper. We ordered the functions by their effective domain. Two exceptions are the linear and hardtanh activation.

The linear function depends on the input values and automatically adapts the domain to the image. This function comes originally from the linear probability model in statistics \cite{Chatla2013,10.1257/jep.15.4.43}. As the identity function $f(x) = x$ allows for values outside the interval $[0, 1]$, we perform rescaling by finding the minimum and maximum value in each image or batch of images.

The other exception is the hardtanh function, which has discontinuities and does not output $f(0) = 0$. Thus, it also has a different behavior than the other functions.

\paragraph{Output activation functions with smaller effective domain than sigmoid}

The first case is a faster rate of change than sigmoid. By increasing the rate of change, we can decrease the input values to the activation function $f(x)$. This forces the neural network to decrease the outputs $x$ of the previous layer if probabilities like $f(x) = 0.8$ are needed. A smaller domain than $[-6, 6]$ of the activation function leads to sharper decisions and less uncertainty.

The normal cumulative distribution function (CDF) is an example of a function with a fast rate of change. This activation function can be motivated by considering the probit model from statistics, where the CDF is used as an alternative to the sigmoid/logistic function. Together with the logit model (sigmoid function) it is one of the most frequently used models for binary outputs \cite{Chatla2013,RePEc:aes:jsesro:v:2:y:2013:i:1:p:1-17}.

\paragraph{Output activation functions with larger effective domain than sigmoid}

The second case is a slower rate of change. This brings greater freedom of movement for the output values $f(x)$. It can become easier to output probabilities between $0$ and $1$. A function with a large effective domain allows for more leeway than an exponential function that reaches the asymptotes $0$ or $1$ too quickly. Thus, a slower rate of change can result in better probabilities and counteract problems of probabilities being pushed towards $0$ and $1$.  The function would have effectively a larger domain than sigmoid.

Three examples are the inverse square root function, arctangent, and softsign. While the three functions have a sigmoidal shape, they are not commonly found in statistics. However, the unscaled softsign function is found in the review by Nwankpa et al. \cite{DBLP:journals/corr/abs-1811-03378} for use in non-output layers of neural networks. From the diagram, we can see that the inverse square root function is initially faster than the sigmoid function. For $x = 3$, we have $0.95$ (sigmoid) and $0.97$ (inverse square root). This should also affect the results.

\subsection{Loss function}
\label{sec:loss}

The next step is the definition of the loss function. In the statistical literature, a popular method for estimating the weights is maximum likelihood estimation (MLE). The joint likelihood for binary variables is given by

$${\mathcal {L}}(\beta ;Y,X)=\prod _{i=1}^{n} f(x_{i}^T\beta )^{y_{i}}\left(1-f(x_{i}^T\beta )\right)^{1-y_{i}}\,,$$

where $x_i$ are the inputs and $y_i \in \{0, 1\}$ the outputs. $f(\cdot)$ defines the activation function. By applying the logarithm and changing the sign, we obtain the cross-entropy loss function used in neural networks

$${\mathcal {L}}(\beta ;Y,X)=-\sum_{i=1}^{n} y_{i} \log f(x_{i}^T\beta) + \left(1-y_{i}\right) \log\left(1-f(x_{i}^T\beta )\right)\,.$$

For the binary case, the loss function is also called binary cross-entropy (BCE).

Apart from BCE, a method for finding the coefficients in linear regression is ordinary least squares. The corresponding loss function is the (mean) squared error (MSE) in neural networks

$${\mathcal {L}}(\beta ;Y,X)= \sum_{i=1}^n\left(\hat{y}_i - y_i\right)^2\,,$$

where $\hat{y}_i = f(x_{i}^T\beta)$. However, it is rarely used for segmentation or classification \cite{Goodfellow-et-al-2016}. In the Medical Segmentation Decathlon (MSD) almost all participants used cross entropy \cite{antonelli2021medical}.

A more common loss function than MSE in segmentation is dice loss. The definition is as follows

$${\mathcal {L}}(\beta ;Y,X)= 1 - \frac{2\sum_{i}\hat{y}_{i}y_{i}}{\sum_{i}\hat{y}_{i} + y_{i}}\,.$$

Dice loss optimizes the overlap between the predicted pixels and the ground truth.

We visualize all loss functions for a single prediction with $y_i = 1$ and $\hat{y}_i = f(x)$ in fig.~\ref{fig:loss_func}. As the activation functions differ mainly how fast they approach $0$ or $1$, we refrain from showing all functions. 

\begin{figure}[H]%
    \centering
    \subfloat[\centering The softsign function produces a lower error for bad predictions.]{{\scalebox{0.48}{\input{diagrams/loss_function_Softsign.pgf}} }}
    \qquad
    \subfloat[\centering The sigmoid activation has a much higher error.]{{\scalebox{0.48}{\input{diagrams/loss_function_Sigmoid.pgf}} }}
   \caption[example] 
   { \label{fig:loss_func} Error for a single prediction}
\end{figure}

We use the same domain $[-6, 6]$ for softsign and sigmoid in fig.~\ref{fig:loss_func}. Since softsign has a wider domain, the function takes longer to reach a probability of $0$ or $1$ than sigmoid. This affects also the error. At $-6$ we have a BCE error of approx. $-\log(0.07) \approx 2.6$ for softsign against $\log(1 + e^6) \approx 6$ for sigmoid. 

The mean squared error moves more quickly to an error of $0$ than dice loss. Since real pictures have tens of thousands of pixels, the plots do not show the full behavior of the different loss functions.

We can expect mean squared error and negative log-likelihood to give better probabilistic estimates than dice loss because both functions look at individual pixels. Dice loss, in contrast, jointly optimizes all pixels.

\newpage

\section{EVALUATION}
\label{sec:eval}

We evaluate 21 models that result from the 7 activation functions (see Section \ref{sec:act}) and 3 loss functions (see Section \ref{sec:loss}). In the following subsections, we first describe the evaluation metrics and the segmentation model that were used. Then we present the results on the different medical segmentation datasets.

\subsection{Evaluation Metrics}

All loss functions can also be used as metrics to determine the quality of the trained network. In particular, cross-entropy and mean squared error are known to be proper scoring rules that are suited for model comparisons \cite{Ashukha2020Pitfalls,doi:10.1146/annurev-statistics-062713-085831}. According to tests by Ashukha et al.\cite{Ashukha2020Pitfalls}, mean squared error and cross entropy show high empirical correlation for a wide range of computer vision datasets. Thus, we omit evaluations on the mean squared error, which is also known as Brier score \cite{BRIER1950}.

Cross entropy, often referred to as negative log-likelihood (NLL)\cite{bishop:2006:PRML}, is used for the evaluation of probabilistic errors. However, another approach that is well known in the domain of uncertainty estimation is the reliability diagram\cite{10.5555/3305381.3305518}.

A reliability diagram has on the x-axis the confidence and on the y-axis the fraction of positives. The first step is to put all predicted probabilities into $n$ different intervals or bins. The \emph{evenly spaced} strategy uses the intervals $\left\{\left[0, \frac{1}{n}\right), \left[\frac{1}{n}, \frac{2}{n}\right), \dots, \left[\frac{n-1}{n}, 1\right)\right\}$. The \emph{adaptive} strategy chooses intervals based on quantiles to prevent empty intervals or intervals with few probabilities. In segmentation, probabilities tend to be close to $0$ because there are more background pixels than foreground pixels. For this reason, Nixon et al.\cite{DBLP:journals/corr/abs-1904-01685} proposed thresholding predictions such that predictions below some $\epsilon$ are removed for quantile binning. We choose $n = 15$ bins and $\epsilon = 10^{-2}$ and show both strategies when evaluating the datasets. Additionally, we set another threshold to $1 - 10^{-2}$ to prevent too many bins close to $1$.

The second step is to plot on the $x$-axis the confidence $\frac{1}{|b|}\sum_{(y, \hat{y}) \in b} \hat{y}$ and on the $y$-axis the relative frequency $\frac{1}{|b|}\sum_{(y, \hat{y}) \in b} y$ for all bins or intervals $b$. The result is a reliability diagram.

Apart from probabilistic measures such as NLL or the reliability diagram, we also evaluate the object overlap by considering the dice coefficient as a metric.

The thresholded dice coefficient or F1 score can be defined as follows

$$\text{Dice}_{t} = \frac{2\sum_{i}\mathbf{1}(\hat{y}_{i} > t)y_{i}}{\sum_{i}\mathbf{1}(\hat{y}_{i} > t) + y_{i}}\,,$$

where $t$ is a given threshold. Here, the threshold is chosen that best maximizes the dice coefficient. All thresholds from the set $\{0, 0.05, 0.1, \dots, 0.95, 1.0\}$ were tried out for each cross validation fold. We denote by $\mathbf{1}(\cdot)$ the indicator function that outputs $1$ when the given condition is true.

\subsection{Datasets and Training}
\label{sec:data}

For evaluating the different loss and activation functions, we consider both 2D and 3D medical datasets. As it is computationally infeasible to train neural networks on all possible medical datasets, we choose datasets that are sufficiently different from one another. The four datasets cover different parts of the body: heart (1), brain (2), colon (3), and prostate (4). The objective is to generalize the results of this paper to other medical datasets.

\begin{enumerate}
 \item The \emph{Automated Cardiac Diagnosis Challenge} (ACDC) \cite{8360453} contains cardiac MRI recordings.
 \item \emph{Ischemic Stroke Lesion Segmentation} (ISLES) 2018 \cite{Maier2017,Winzeck2018} deals with CT perfusion data.
 \item \emph{Kvasir-SEG} \cite{jha2020kvasir} is a 2D polyp dataset.
 \item \emph{Medical Segmentation Decathlon}\cite{antonelli2021medical} (MSD) consists of multiple 3D semantic segmentation datasets. The neural networks here are trained on the fifth task, involving T2 and ADC image sequences of prostates.
\end{enumerate}

Since 2D convolutional neural networks (CNN) were used in the winning solutions of competitions such as the \emph{RSNA Intracranial Hemorrhage Detection}\cite{Flanders2020} or ISLES 2018\cite{CLERIGUES2019103487,10.1007/978-3-030-11723-8_25}, we use a standard 2D U-Net architecture for all datasets \cite{DBLP:journals/corr/RonnebergerFB15}. Furthermore, it was shown that performing transfer learning by employing an encoder pretrained on ImageNet, can lead to better results \cite{DBLP:journals/corr/HuhAE16,DBLP:journals/corr/abs-1801-05746, DBLP:journals/corr/abs-1806-00844}. We choose ResNet-34 \cite{DBLP:journals/corr/HeZRS15} as encoder for its simplicity and robustness.

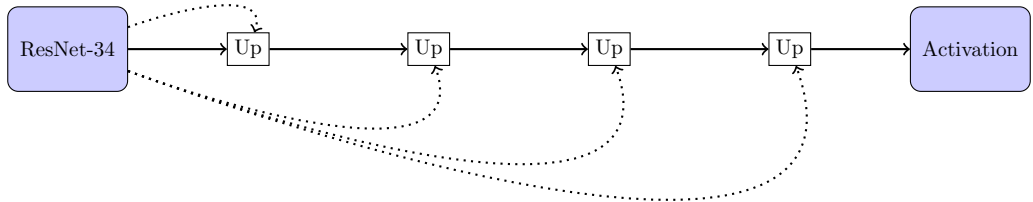
\begin{figure}[H]
\centering
\begin{tikzpicture}[scale=0.8, transform shape]
	\node at (3,0)  [block] (ResNet-34) {ResNet-34};
	\node[minimum width=0.1cm] at (6,0) [draw] (Up) {Up};
	\node[minimum width=0.1cm] at (9,0) [draw] (Up2) {Up};
	\node[minimum width=0.1cm] at (12,0) [draw] (Up3) {Up};
	\node[minimum width=0.1cm] at (15,0) [draw] (Up4) {Up};
	\node at (18,0)  [block] (Activation) {Activation};

    \draw [->,thick] (ResNet-34) -- (Up);
    \draw [->,thick] (Up) -- (Up2);
    \draw [->,thick] (Up2) -- (Up3);
    \draw [->,thick] (Up3) -- (Up4);
    \draw [->,thick] (Up4) -- (Activation);
    
    \draw [->,dotted,thick] (ResNet-34) to[out=-20,in=-70] (Up2);
    \draw [->,dotted,thick] (ResNet-34) to[out=-20,in=-70] (Up3);
    \draw [->,dotted,thick] (ResNet-34) to[out=-20,in=-70] (Up4);
    \draw [->,dotted,thick] (ResNet-34) to[out=20,in=70] (Up);
\end{tikzpicture}
    \caption{ResNet-34 as the encoder and U-Net as the decoder. A dotted line means concatenation with the last layer of the same width and height in the encoder. An up layer consists of a nearest-neighbor interpolation, two convolutions with ReLU activation, and batch normalizations.}
  \label{fig:resnet}
\end{figure}

All models are trained with a batch size of 8 and the adaptive stochastic optimizer Adam with a standard learning rate of $10^{-3}$ \cite{journals/corr/KingmaB14}. We reduce the learning rate when $5$ successive epochs did not result in any improvement, and stop the training after $10$ epochs without improvement. For validating the results, 5-fold cross validation was used. We used PyTorch\cite{paszke2019pytorch} as a deep learning library. For certain activation functions, such as the normal CDF, we had to rescale the outputs to $[10^{-7}, 1-10^{-7}]$ to ensure numerical stability.

\subsection{Evaluation of medical datasets}

\newcolumntype{H}{>{\setbox0=\hbox\bgroup}c<{\egroup}@{}}

To evaluate the selected medical datasets, we provide the mean NLL and dice coefficient of each cross-validated test run and their standard deviation. In addition, we present examples of segmented images that offer high object overlap, as well as examples that offer low overlap according to the selected metric. A deep blue color on the images indicates a low probability, while a deep red means a high probability. At last, we also show a reliability diagram (with the evenly spaced and adaptive strategy) for each tested dataset, comparing the winning activation function(s) to the well-known sigmoid output activation function. With the arrows $\uparrow$ and $\downarrow$, we indicate whether a higher or lower value is better. Except for linear and hardtanh, all activation functions are ordered by their effective domain in ascending order. The appendix \ref{sec:appendix} at the end of the paper contains all the predictions of a test image for each dataset.

\paragraph{Automated Cardiac Diagnosis Challenge (ACDC)}

The first dataset to be tested is the \emph{Automated Cardiac Diagnosis Challenge}. All trained networks and the quality of their outputs are shown in Table \ref{tab:statsACDC}. Since the predictions are quite similar, we only see to a certain extent the presumed influence of the effective domain. If one considers the probabilistic error NLL together with MSE / BCE, normal CDF is better than sigmoid and this function is better than the inverse square root. With a larger effective domain, the error increases. Without inverse square root, this increase in error is even clearer: $0.0542 \to 0.0556 \to 0.0684 \to 0.0698$. The reason that inverse square root does not behave like the other functions is that it is initially faster than sigmoid.

Dice loss shows the inverse effect: a larger effective domain leads to a better NLL error. This can be explained by noting that dice loss does not consider individual pixels like BCE / MSE. A wider domain gives dice loss more flexibility to change individual pixels. We then have more uncertainty and better predictions.

\begin{table}[H]
\centering
\begin{tabular}{llHllH}
\toprule
Activation & Loss & Threshold & NLL $\downarrow$ & Dice $\uparrow$ & $\text{Dice}_{avg}$ $\uparrow$ \\ \hline
\multirow{3}{*}{normal CDF} & BCE & $0.4433 \pm 0.0573$ & $\mathbf{0.0542 \pm 0.0048}$ & $0.9154 \pm 0.0062$ & $0.8219 \pm 0.0047$\\ 
& Dice & $0.38 \pm 0.2632$ & $0.2033 \pm 0.0146$ & $0.9156 \pm 0.007$ & $0.8315 \pm 0.0065$\\ 
& MSE & $0.4433 \pm 0.0442$ & $0.0567 \pm 0.0044$ & $0.9161 \pm 0.0063$ & $0.8231 \pm 0.0051$\\ 
\midrule
\multirow{3}{*}{sigmoid} & BCE & $0.44 \pm 0.0638$ & $0.0556 \pm 0.0051$ & $0.9155 \pm 0.0075$ & $0.8239 \pm 0.0068$\\ 
& Dice & $0.29 \pm 0.2169$ & $0.1409 \pm 0.0185$ & $0.9152 \pm 0.0073$ & $0.831 \pm 0.0067$\\ 
& MSE & $0.4833 \pm 0.0471$ & $0.059 \pm 0.0043$ & $0.9147 \pm 0.0074$ & $0.8212 \pm 0.0052$\\ 
\midrule
\multirow{3}{*}{inverse square root} & BCE & $0.1 \pm 0.0$ & $0.1321 \pm 0.0083$ & $0.9135 \pm 0.007$ & $0.4248 \pm 0.0042$\\ 
& Dice & $0.2067 \pm 0.2272$ & $0.1131 \pm 0.0054$ & $0.9151 \pm 0.0066$ & $0.8312 \pm 0.0062$\\ 
& MSE & $0.4233 \pm 0.0704$ & $0.066 \pm 0.0035$ & $\mathbf{0.9167 \pm 0.0062}$ & $0.8239 \pm 0.005$\\ 
\midrule
\multirow{3}{*}{arctangent} & BCE & $0.3767 \pm 0.1167$ & $0.0684 \pm 0.0071$ & $0.9157 \pm 0.0064$ & $0.8256 \pm 0.0046$\\ 
& Dice & $0.2833 \pm 0.2278$ & $0.0992 \pm 0.0071$ & $0.9159 \pm 0.0052$ & $\mathbf{0.8321 \pm 0.005}$\\ 
 & MSE & $0.4167 \pm 0.0745$ & $0.0887 \pm 0.0033$ & $0.9153 \pm 0.0068$ & $0.82 \pm 0.0061$\\ 
\midrule
\multirow{3}{*}{softsign} & BCE & $0.37 \pm 0.1514$ & $0.0698 \pm 0.0061$ & $\mathbf{0.9167 \pm 0.0065}$ & $0.8265 \pm 0.0066$\\ 
& Dice & $0.3333 \pm 0.2785$ & $0.0965 \pm 0.0073$ & $0.9145 \pm 0.0061$ & $0.8308 \pm 0.0057$\\ 
& MSE & $0.3833 \pm 0.1261$ & $0.088 \pm 0.007$ & $0.915 \pm 0.0061$ & $0.8202 \pm 0.0061$\\ 
\midrule
\multirow{3}{*}{linear} & BCE & $0.46 \pm 0.02$ & $0.0604 \pm 0.0044$ & $0.9086 \pm 0.0061$ & $0.7911 \pm 0.0056$\\
& Dice & $0.4833 \pm 0.0236$ & $0.0701 \pm 0.0063$ & $0.9043 \pm 0.0078$ & $0.7932 \pm 0.0086$\\ 
& MSE & $0.46 \pm 0.02$ & $0.0687 \pm 0.0063$ & $0.9103 \pm 0.0046$ & $0.8005 \pm 0.0043$\\ 
\midrule
\multirow{3}{*}{hardtanh} & BCE & $0.41 \pm 0.0374$ & $0.1178 \pm 0.0178$ & $\underline{0.8191 \pm 0.0159}$ & $0.6239 \pm 0.0417$\\ 
& Dice & $0.2367 \pm 0.2947$ & $\underline{0.2825 \pm 0.0217}$ & $0.9164 \pm 0.0066$ & $\mathbf{0.8321 \pm 0.006}$\\ 
& MSE & $0.44 \pm 0.0374$ & $0.0994 \pm 0.0094$ & $0.9162 \pm 0.0077$ & $0.8237 \pm 0.0071$\\ 
\bottomrule
\end{tabular}
\caption{Results of all tested models trained on the ACDC dataset, where the best (bold) and worst (underline) results were highlighted.}
\label{tab:statsACDC}
\end{table}

The dice coefficient next to the NLL column in the table shows no major differences between the individual activation functions for this dataset. If we then look at the individual predictions of some images, we should not see any big differences either.

\begin{figure}[ht]%
    \centering
    \subfloat[\centering BCE with CDF activation]{\includegraphics[width=35mm]{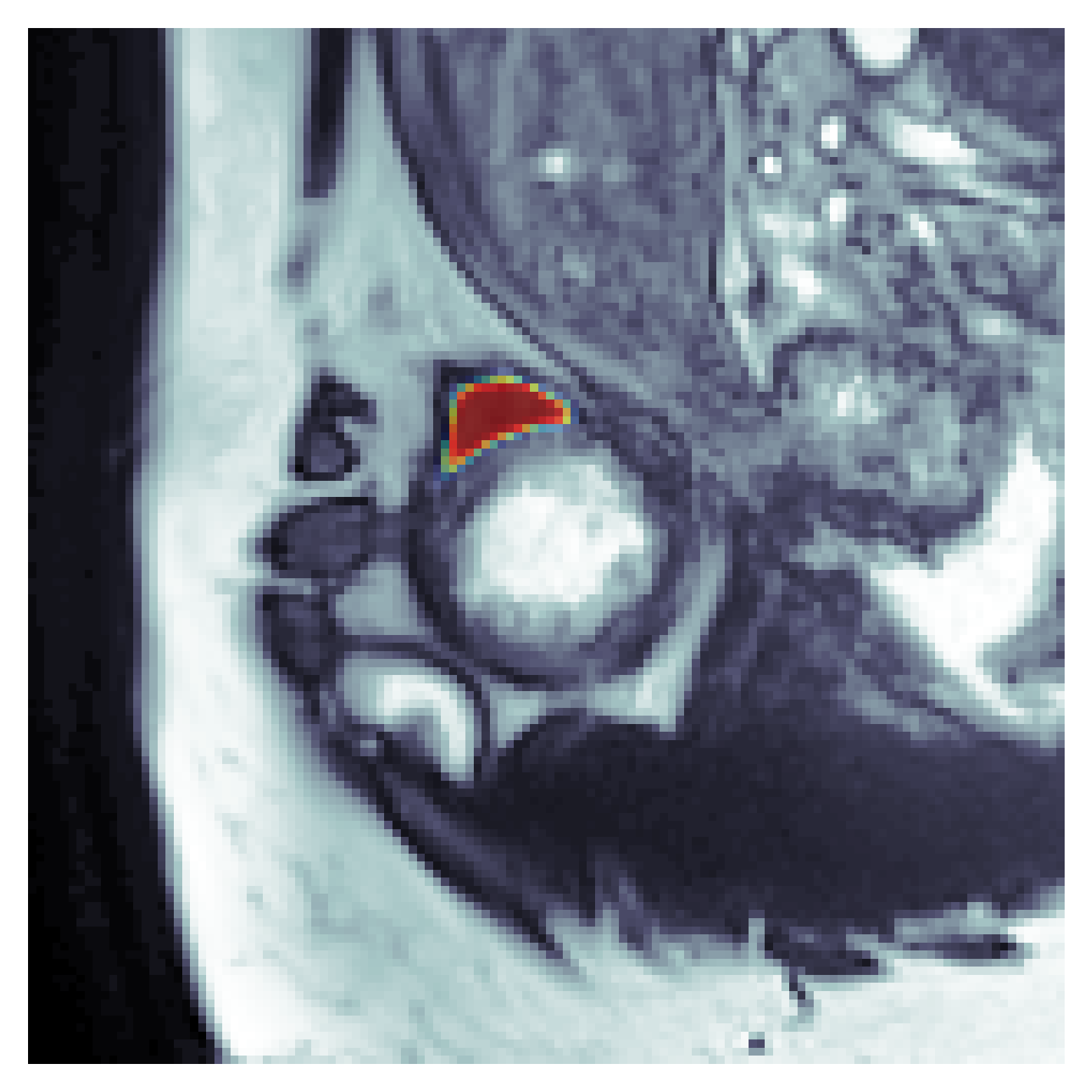}}
    \qquad
    \subfloat[\centering MSE with inverse square root]{\includegraphics[width=35mm]{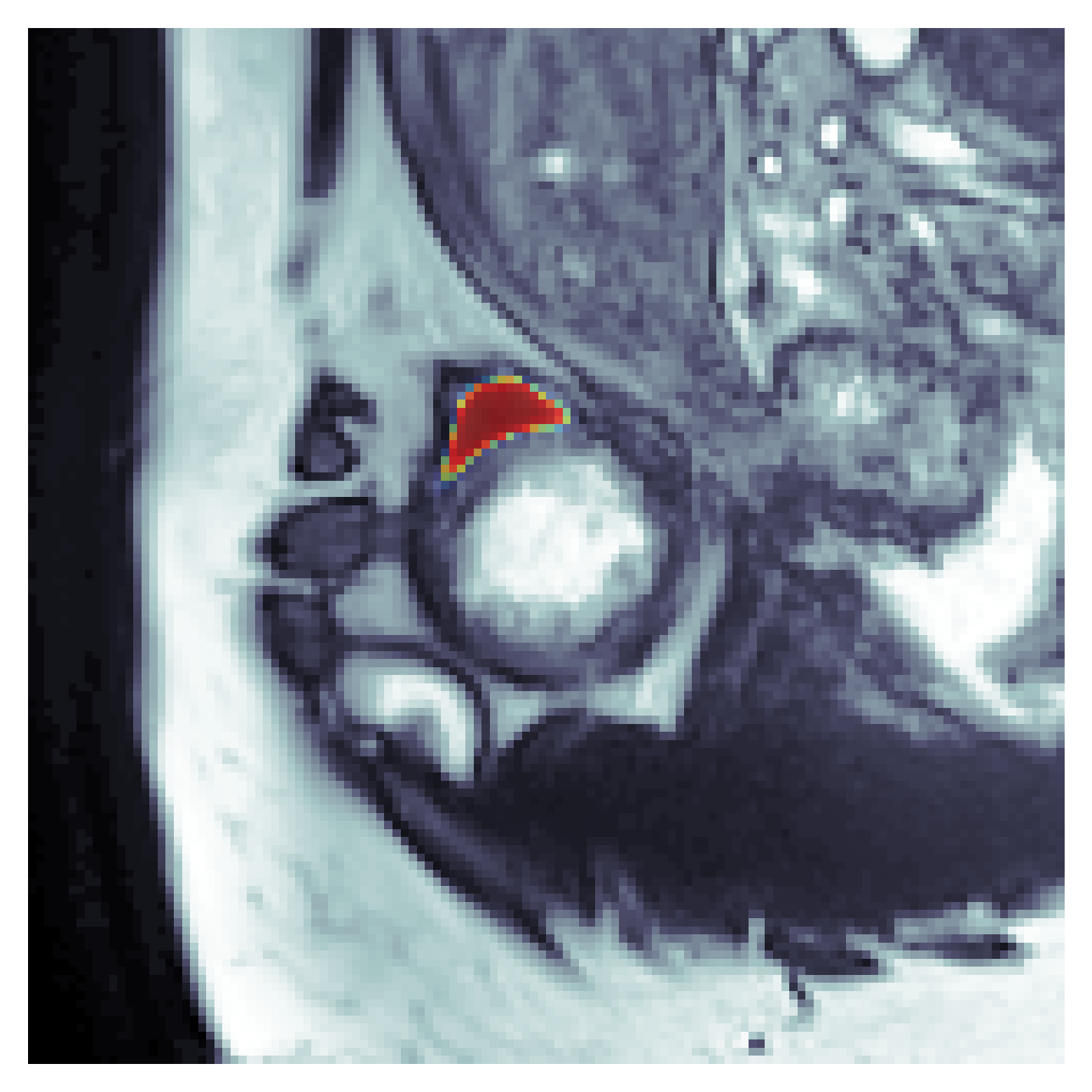} }
    \qquad
    \subfloat[\centering BCE with softsign]{\includegraphics[width=35mm]{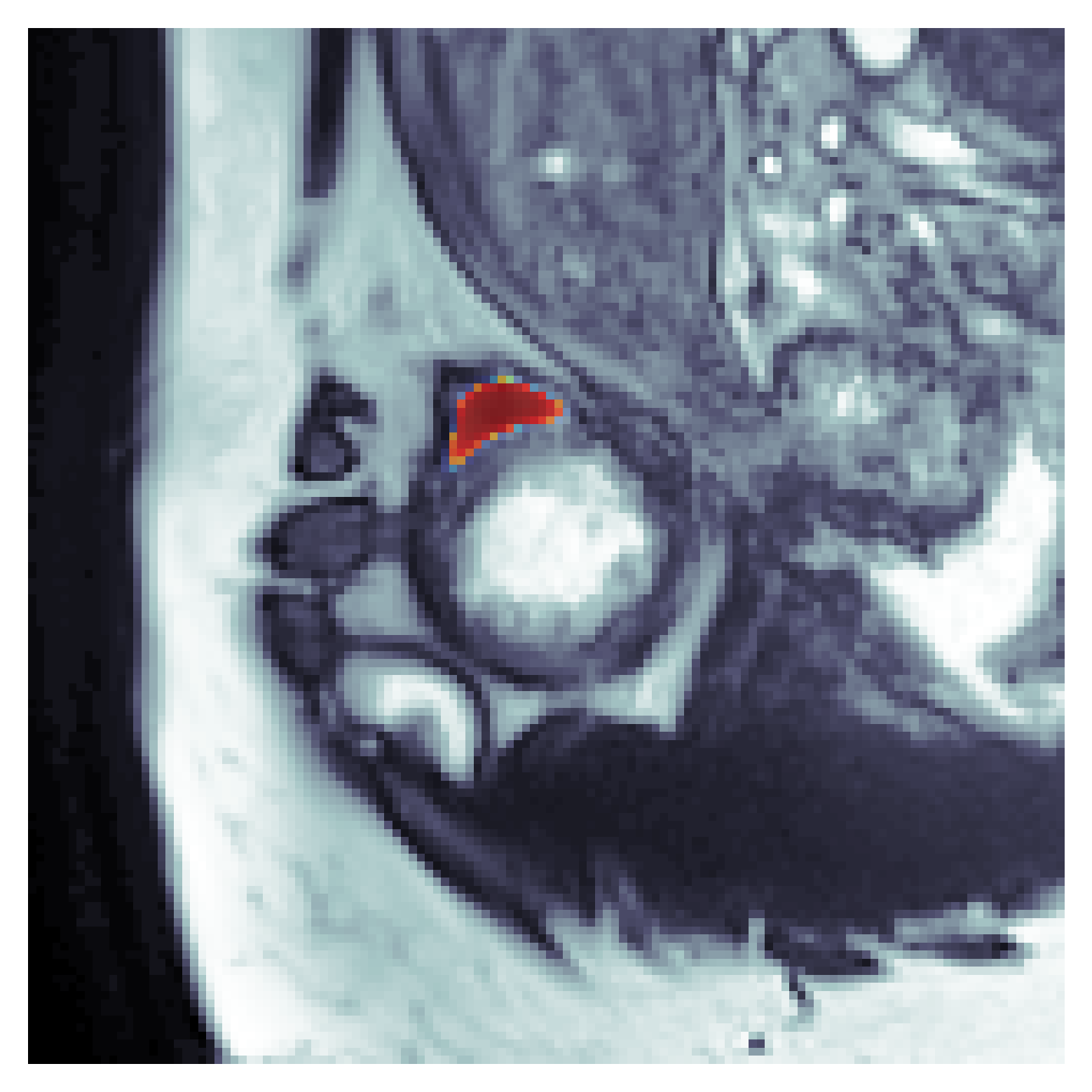} }
    \qquad
    \subfloat[\centering Ground truth]{\includegraphics[width=35mm]{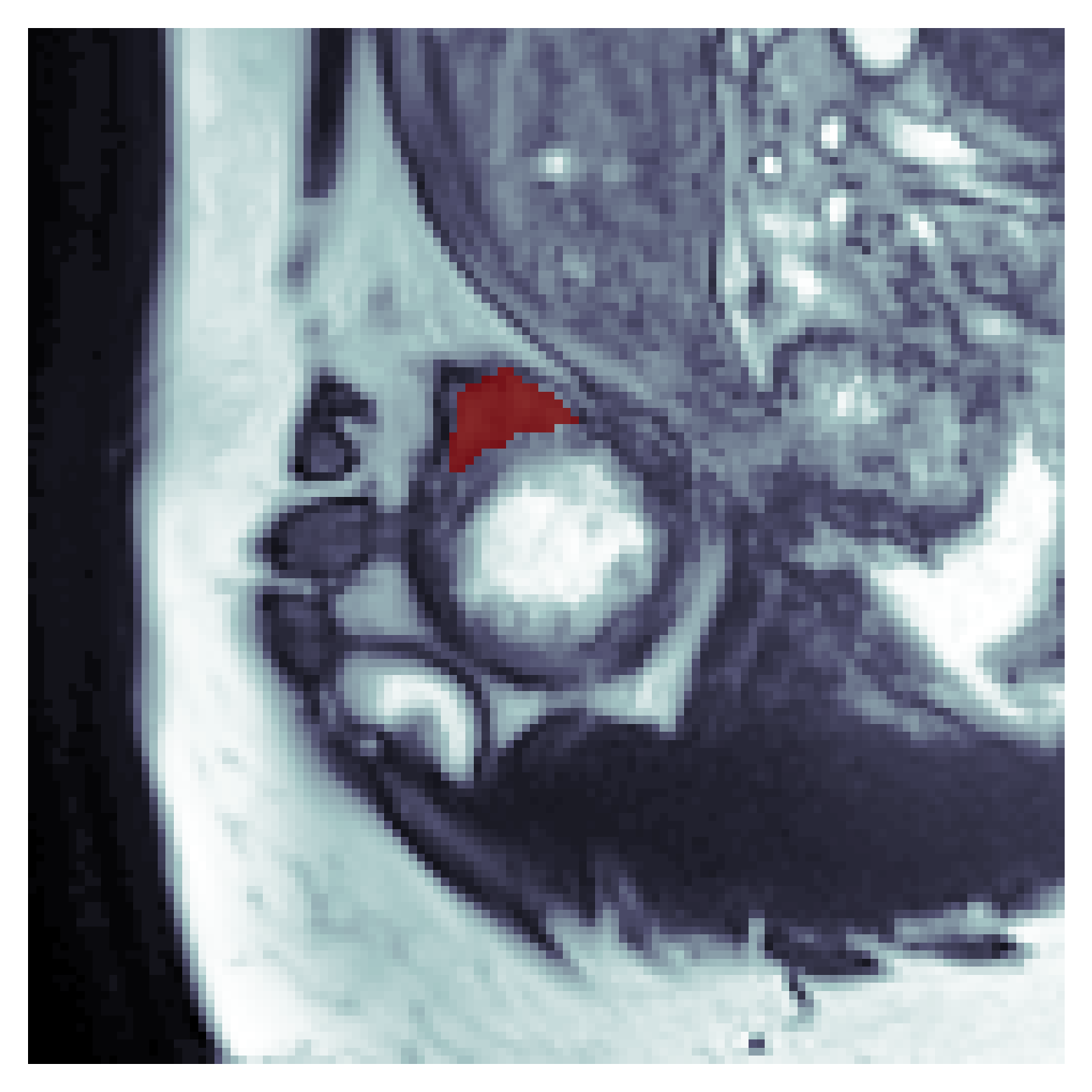} }
   \caption[example] 
   { \label{fig:ACDCbestModels} Best predictions for a single image in ACDC}
\end{figure}

Figure \ref{fig:ACDCbestModels} shows an example of the segmentation result using the best models of our test (highlighted in bold in the table). Figure \ref{fig:ACDCbestModels} (a) comes visually close to the provided ground truth in Figure \ref{fig:ACDCbestModels} (d), while holding a slight uncertainty at the border of the detected area. The mean squared error in combination with inverse square root in Figure \ref{fig:ACDCbestModels} (b) provides similar results, whereas (c) predicted a smaller area relative to the ground truth (d). On the other hand, the model predictions are very certain because the probabilities are highlighted in deep red.

\begin{figure}[ht]%
    \centering
    \subfloat[\centering Dice with hardtanh]{\includegraphics[width=35mm]{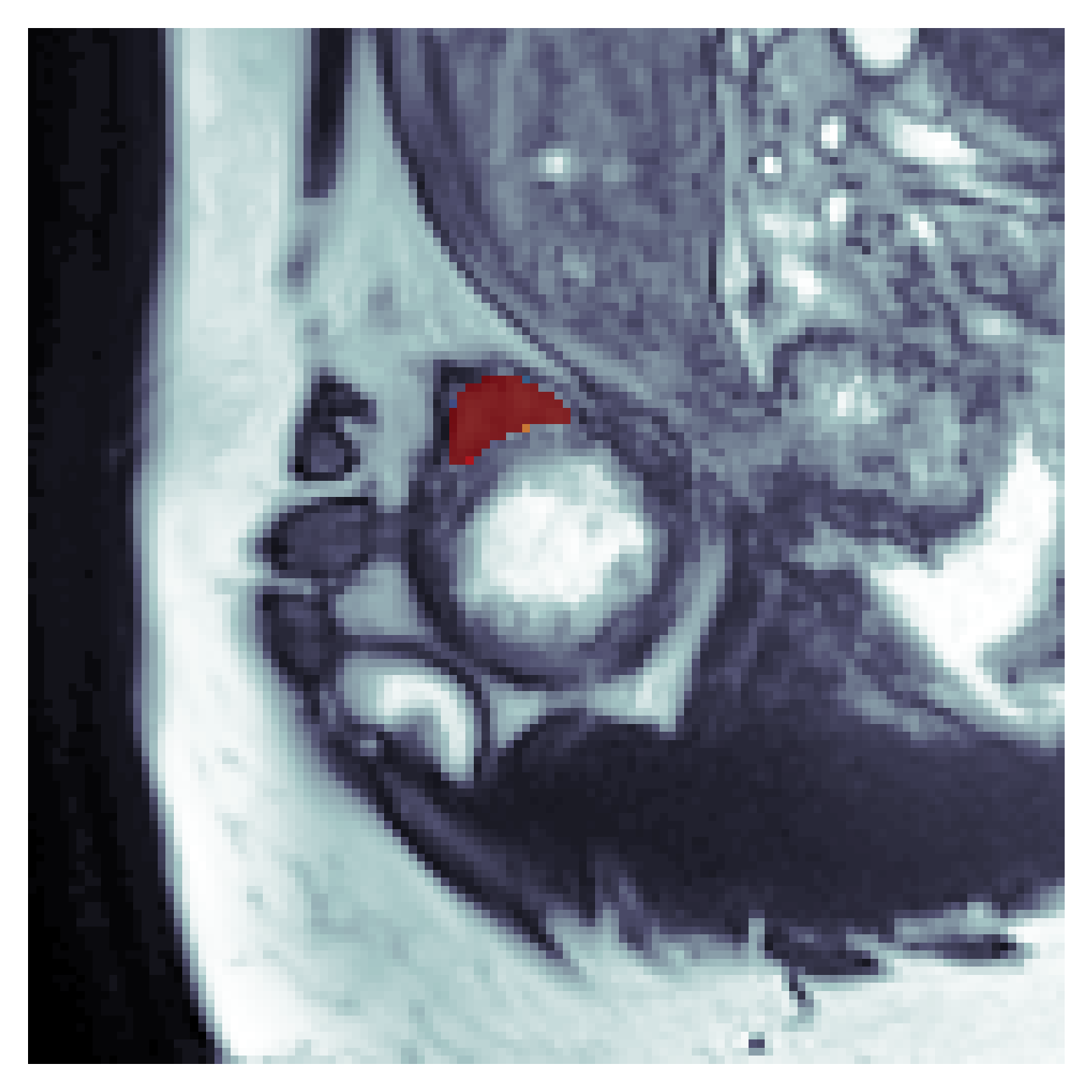}}
    \qquad
    \subfloat[\centering BCE with hardtanh]{\includegraphics[width=35mm]{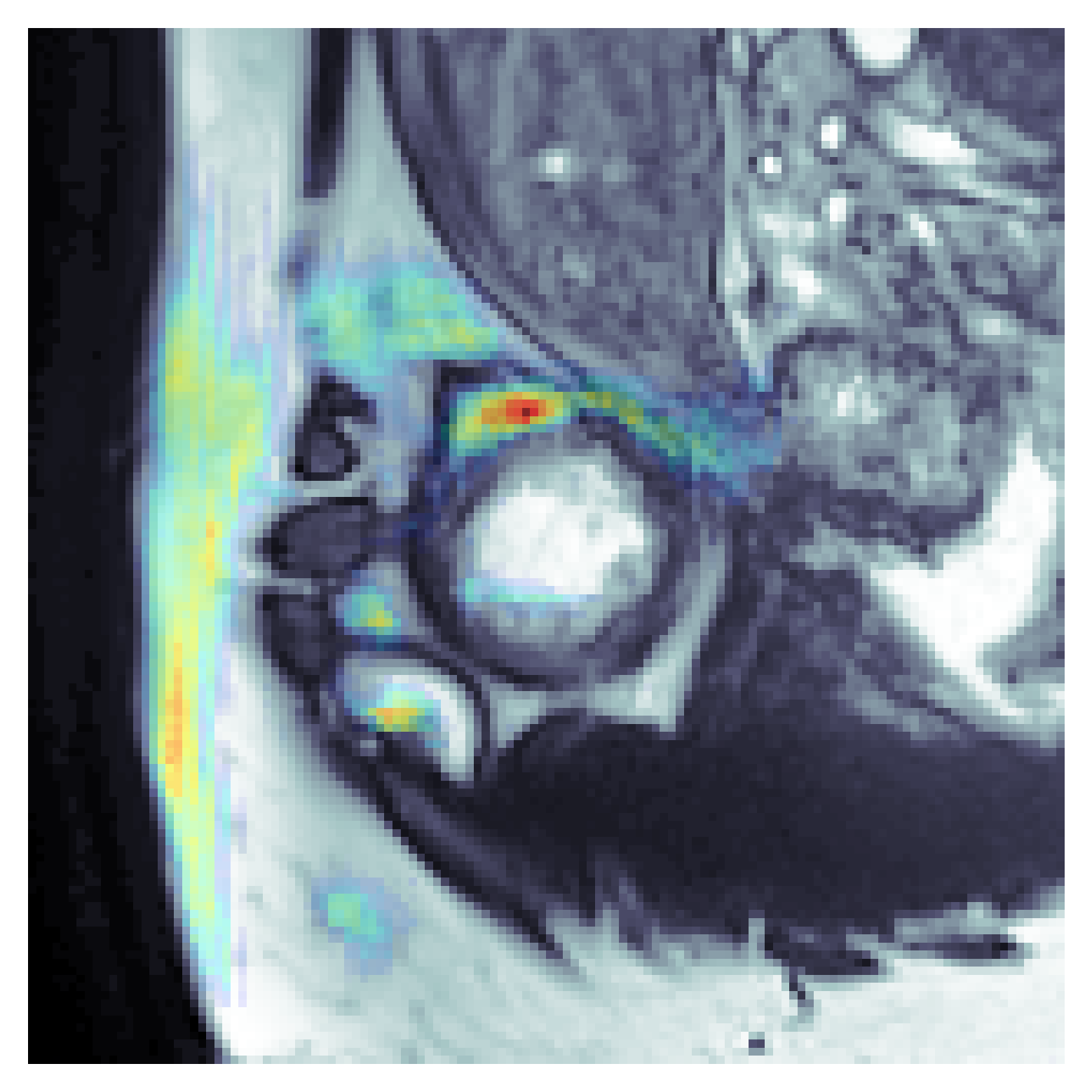} }
   \caption[example] 
   { \label{fig:ACDCworstModels} Worst predictions for a single image in ACDC}
\end{figure}

When reviewing the worst models trained on the ACDC dataset, we can see from Table \ref{tab:statsACDC} that the dice and hardtanh combination has the worst NLL error. The reason is that dice loss optimizes the object overlap and not probabilities. In Figure \ref{fig:ACDCworstModels} (a) we see that there is no uncertainty on the border and the probabilities are close to $1$. In contrast, the combination of hardtanh with BCE optimizes the probabilities and shows more uncertainty. However, Figure \ref{fig:ACDCworstModels} (b) shows massive mispredictions all over the considered image.

\begin{figure}[H]
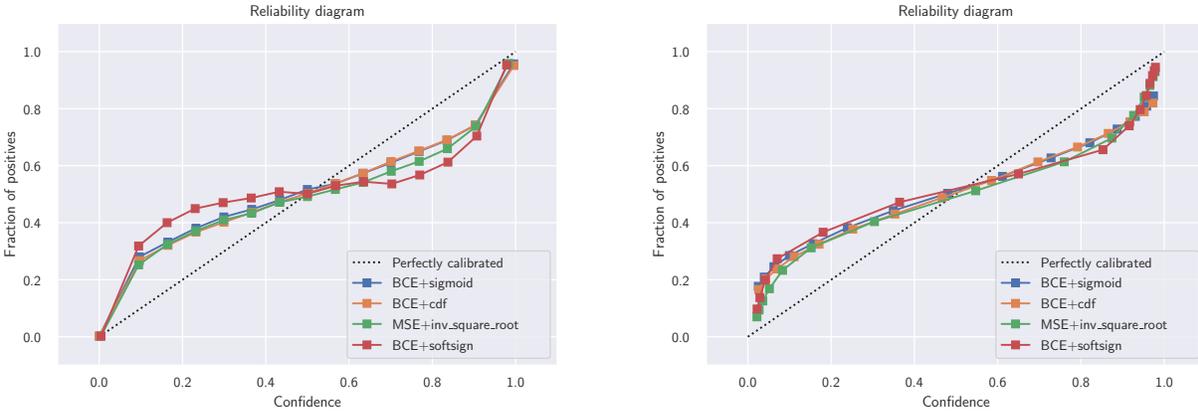
%
    \centering
    \subfloat[\centering Evenly spaced bins]{{\scalebox{0.48}{\input{diagrams/calibration_plot_acac1_grouped.pgf}} }}
    \qquad
    \subfloat[\centering Adaptive bins]{{\scalebox{0.48}{\input{diagrams/calibration_plot_acac1_adaptive_grouped.pgf}} }}
   \caption[example]{Reliability diagrams on the ACDC dataset for the best performing models compared to the sigmoid function as the output activation function.}
  \label{fig:reliabilityACDC}
\end{figure}

Figure \ref{fig:reliabilityACDC} shows the reliability graph of the best activation and loss combinations in comparison to the standard sigmoid function with BCE. We can see that all configurations underestimate the true probabilities for values less than $0.5$ and overestimate the true probabilities for values greater than $0.5$. For BCE+softsign, the difference between confidence and the number of occurrences is at most 20\%. When the prediction states that a pixel is $1$ with a probability of x\%, then this event will not occur exactly on x\% of all occasions. The reason is that the model both underestimates and overestimates the true probabilities. The BCE+softsign combination has also the worst NLL error of the four combinations (BCE+CDF, MSE+inverse square root, BCE+softsign, BCE+sigmoid). Thus, a large effective domain like softsign does not always result in better calibration or probabilistic errors.

When using adaptive binning, the big difference between the combinations vanishes such that the reliability graphs are nearly equal. However, the models are still not perfectly calibrated. Since the dice coefficient did not vary considerably between the activation functions, we do not see a great difference regarding the calibration.

\newpage

\paragraph{Ischemic Stroke Lesion Segmentation Challenge (ISLES)} The second dataset is ISLES. Table \ref{tab:statsISLES} shows that BCE combined with normal CDF achieves the best results for both evaluation metrics. Sigmoid+BCE and linear+BCE have a similar NLL to CDF+BCE, but still a lower dice coefficient.

\begin{table}[H]
\centering
\begin{tabular}{llHllH}
\toprule
Activation & Loss & Threshold & NLL $\downarrow$ & Dice $\uparrow$ & $\text{Dice}_{avg}$ $\uparrow$ \\ \hline
\multirow{3}{*}{normal CDF} & BCE &$0.39 \pm 0.1463$ & $\mathbf{0.0319 \pm 0.0021}$ & $\mathbf{0.661 \pm 0.0309}$ & $0.5188 \pm 0.0668$\\ 
& Dice & $0.47 \pm 0.4069$ & $0.1084 \pm 0.0327$ & $0.6267 \pm 0.0533$ & $0.5664 \pm 0.0482$\\ 
& MSE & $0.37 \pm 0.1327$ & $0.0396 \pm 0.002$ & $0.6217 \pm 0.0419$ & $0.4969 \pm 0.0367$\\ 
\midrule
\multirow{3}{*}{sigmoid} & BCE &$0.45 \pm 0.2569$ & $0.0337 \pm 0.0028$ & $0.6511 \pm 0.0324$ & $0.5319 \pm 0.0379$\\ 
& Dice & $0.24 \pm 0.2354$ & $0.0905 \pm 0.0269$ & $0.6385 \pm 0.0481$ & $0.5781 \pm 0.0429$\\ 
& MSE & $0.31 \pm 0.086$ & $0.0443 \pm 0.0049$ & $0.6375 \pm 0.0548$ & $0.502 \pm 0.0585$\\ 
\midrule
\multirow{3}{*}{inverse square root} & BCE &$0.06 \pm 0.02$ & $0.0518 \pm 0.0042$ & $0.5973 \pm 0.0633$ & $0.2456 \pm 0.0517$\\ 
& Dice & $0.29 \pm 0.3499$ & $0.0712 \pm 0.0098$ & $0.6373 \pm 0.0469$ & $0.5774 \pm 0.0426$\\ 
& MSE & $0.39 \pm 0.1281$ & $0.0548 \pm 0.0031$ & $0.6108 \pm 0.0495$ & $0.4896 \pm 0.0473$\\ 
\midrule
\multirow{3}{*}{arctangent} & BCE & $0.27 \pm 0.1122$ & $0.0556 \pm 0.0083$ & $0.6338 \pm 0.0616$ & $0.4966 \pm 0.0808$\\ 
& Dice & $0.55 \pm 0.4147$ & $0.0608 \pm 0.0086$ & $0.6476 \pm 0.0393$ & $\mathbf{0.5858 \pm 0.0361}$\\ 
& MSE & $0.31 \pm 0.1281$ & $0.0787 \pm 0.0079$ & $0.6348 \pm 0.0469$ & $0.4826 \pm 0.06$\\ 
\midrule
\multirow{3}{*}{softsign} & BCE &$0.42 \pm 0.1749$ & $0.0635 \pm 0.0017$ & $0.6497 \pm 0.0464$ & $0.5129 \pm 0.0355$\\ 
& Dice & $0.42 \pm 0.3736$ & $0.0644 \pm 0.0076$ & $0.6415 \pm 0.0547$ & $0.5809 \pm 0.0508$\\ 
& MSE & $0.39 \pm 0.1463$ & $0.0876 \pm 0.0166$ & $0.6291 \pm 0.0669$ & $0.4825 \pm 0.0645$\\ 
\midrule
\multirow{3}{*}{linear} & BCE &$0.37 \pm 0.0678$ & $0.0333 \pm 0.0029$ & $0.6534 \pm 0.0367$ & $0.4767 \pm 0.0402$\\ 
& Dice & $0.24 \pm 0.201$ & $0.0471 \pm 0.0069$ & $0.6516 \pm 0.0465$ & $0.5826 \pm 0.0436$\\ 
& MSE & $0.39 \pm 0.08$ & $0.0365 \pm 0.0045$ & $0.6537 \pm 0.027$ & $0.5074 \pm 0.0288$\\ 
\midrule
\multirow{3}{*}{hardtanh} & BCE &$0.31 \pm 0.0735$ & $0.0433 \pm 0.0065$ & $\underline{0.5464 \pm 0.0669}$ & $0.3383 \pm 0.0459$\\ 
& Dice & $0.13 \pm 0.16$ & $\underline{0.1878 \pm 0.0333}$ & $0.6244 \pm 0.0562$ & $0.564 \pm 0.051$\\ 
& MSE & $0.31 \pm 0.1241$ & $0.0572 \pm 0.0147$ & $0.602 \pm 0.0391$ & $0.4572 \pm 0.0542$\\
\bottomrule
\end{tabular}
\caption{Results of all tested models trained on the ISLES dataset where the best (bold) and worst (underline) results were highlighted.}
  \label{tab:statsISLES}
\end{table}

For this dataset, we clearly see the influence of the effective domain. As the effective domain increases, the metric NLL gets worse and worse for the loss function BCE. Similar to the last dataset, the loss function dice shows the opposite effect, where a larger effective domain is better for NLL. When considering the dice coefficient next to NLL in the table, there is also a correlation between the domain and the metric, but not as strong as with NLL.

Since the dice coefficient is much lower in this dataset, we expect the predicted images to have a higher uncertainty than the last dataset.

\begin{figure}[H]%
    \centering
    \subfloat[\centering BCE and CDF]{{\includegraphics[width=35mm]{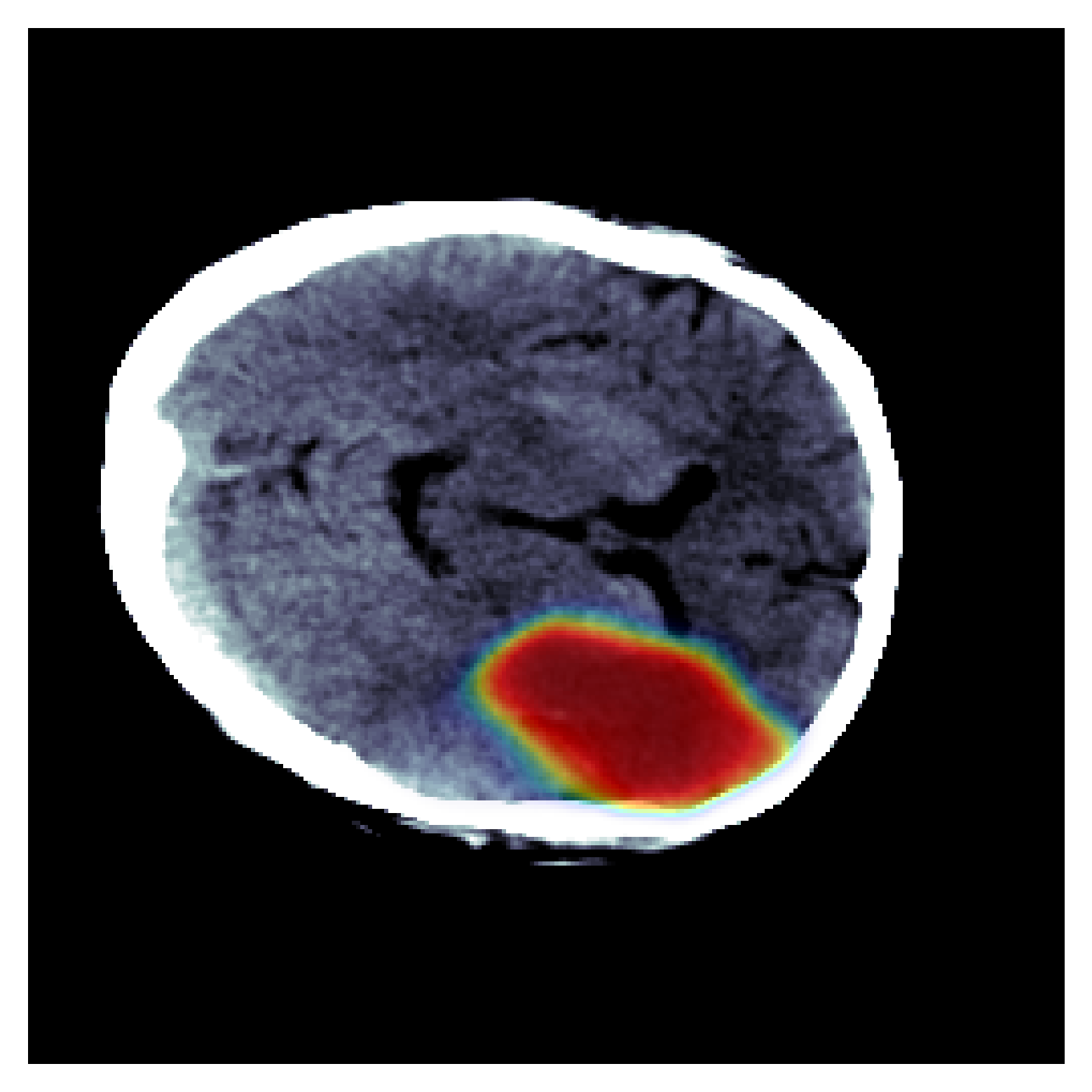} }}
    \qquad
    \subfloat[\centering Ground truth]{\includegraphics[width=35mm]{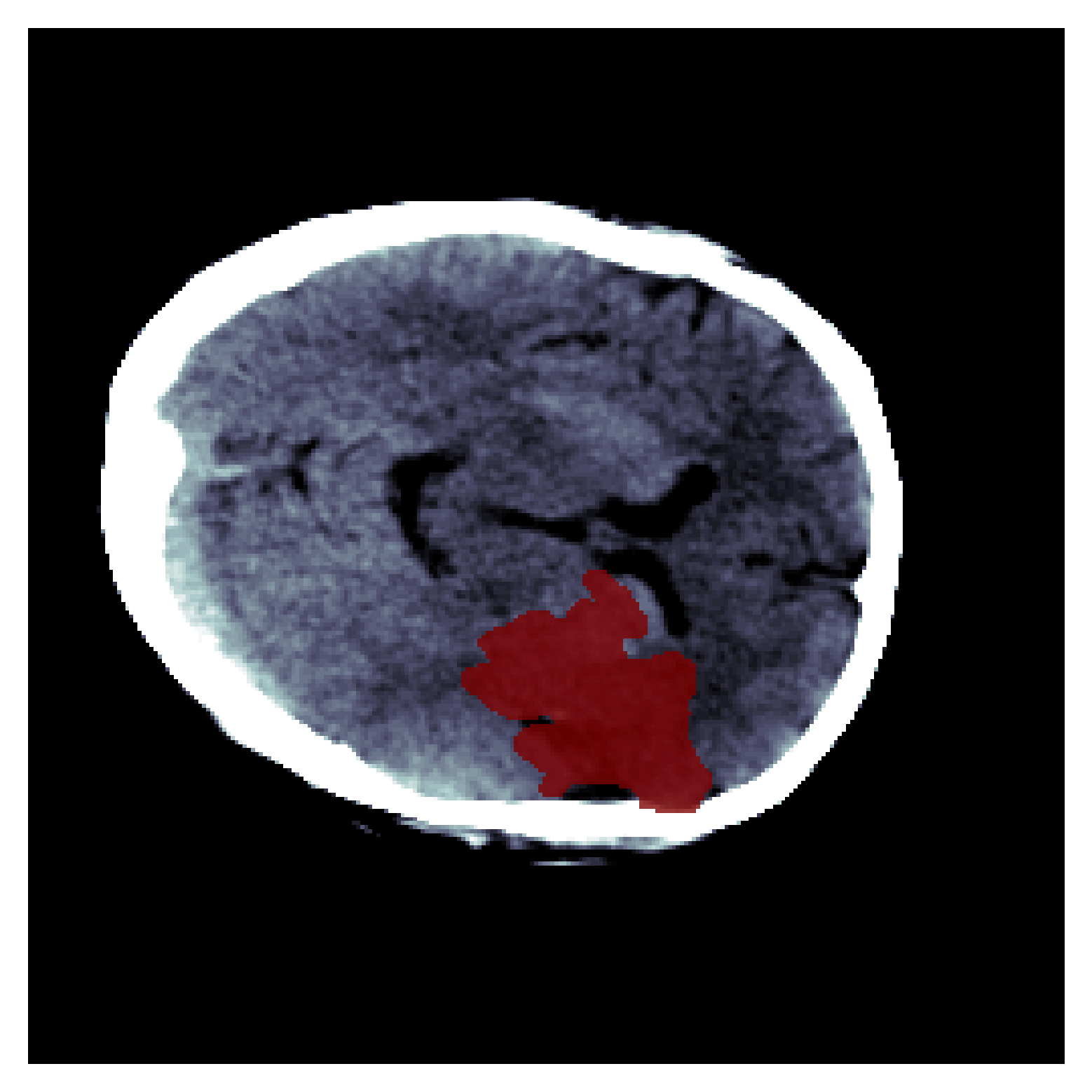}}
   \caption[example]{Best predictions for a single image in ISLES}
  \label{fig:ISLESbestModels}
\end{figure}

Figure \ref{fig:ISLESbestModels} shows a selected slice of a lesion prediction in a patient, with (a) showing the best prediction model and (b) showing the underlying ground truth. The model is able to predict the lesion while maintaining a large interval with a uncertain border around the prediction. Czolbe et al.\cite{DBLP:journals/corr/abs-2103-16265} have shown that uncertainty correlates with segmentation errors. Here, too, we find such a correlation. We can therefore assume that the prediction at the borders was difficult for the network.

\begin{figure}[H]%
    \centering
    \subfloat[\centering Dice and hardtanh]{{\includegraphics[width=35mm]{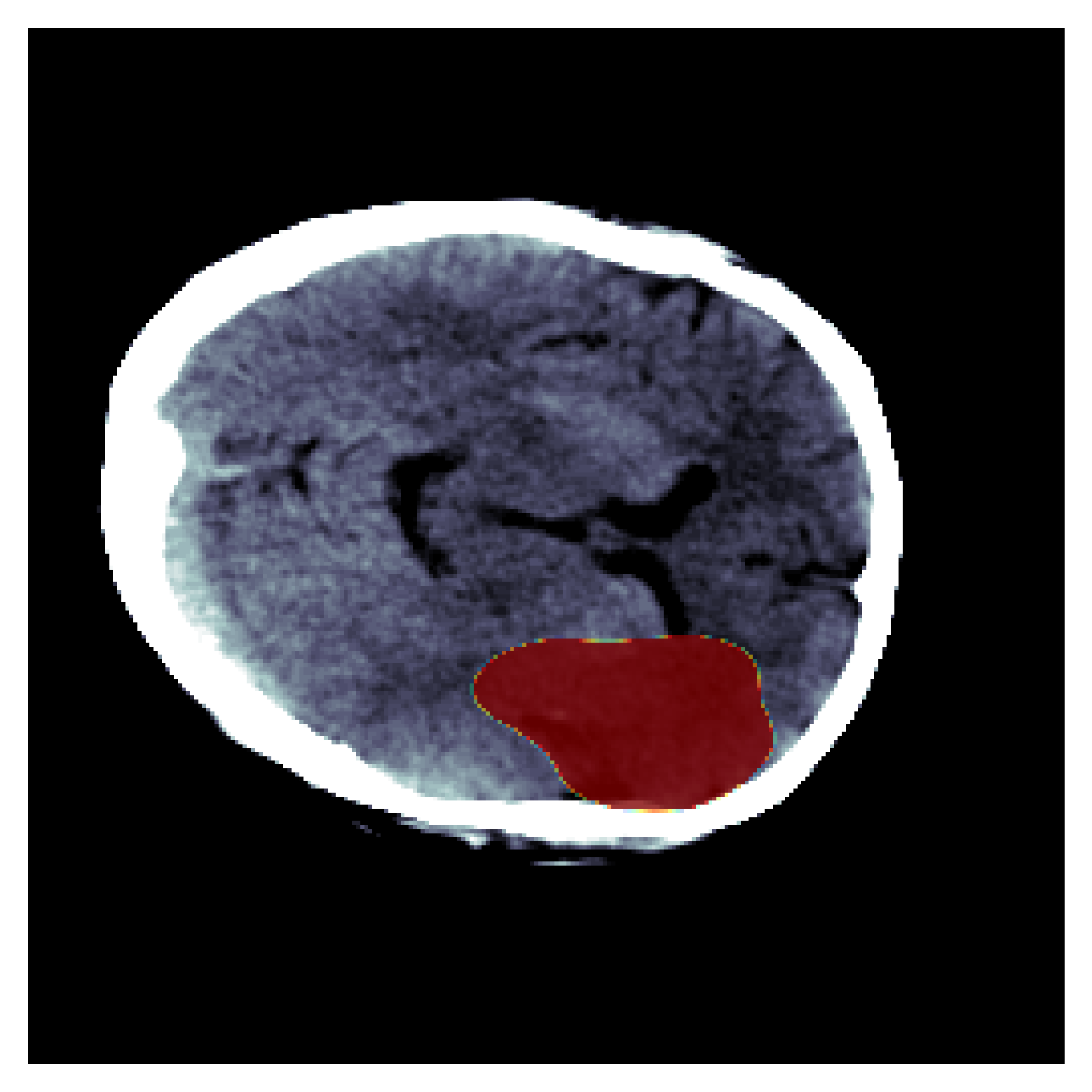} }}
    \qquad
    \subfloat[\centering BCE and hardtanh]{\includegraphics[width=35mm]{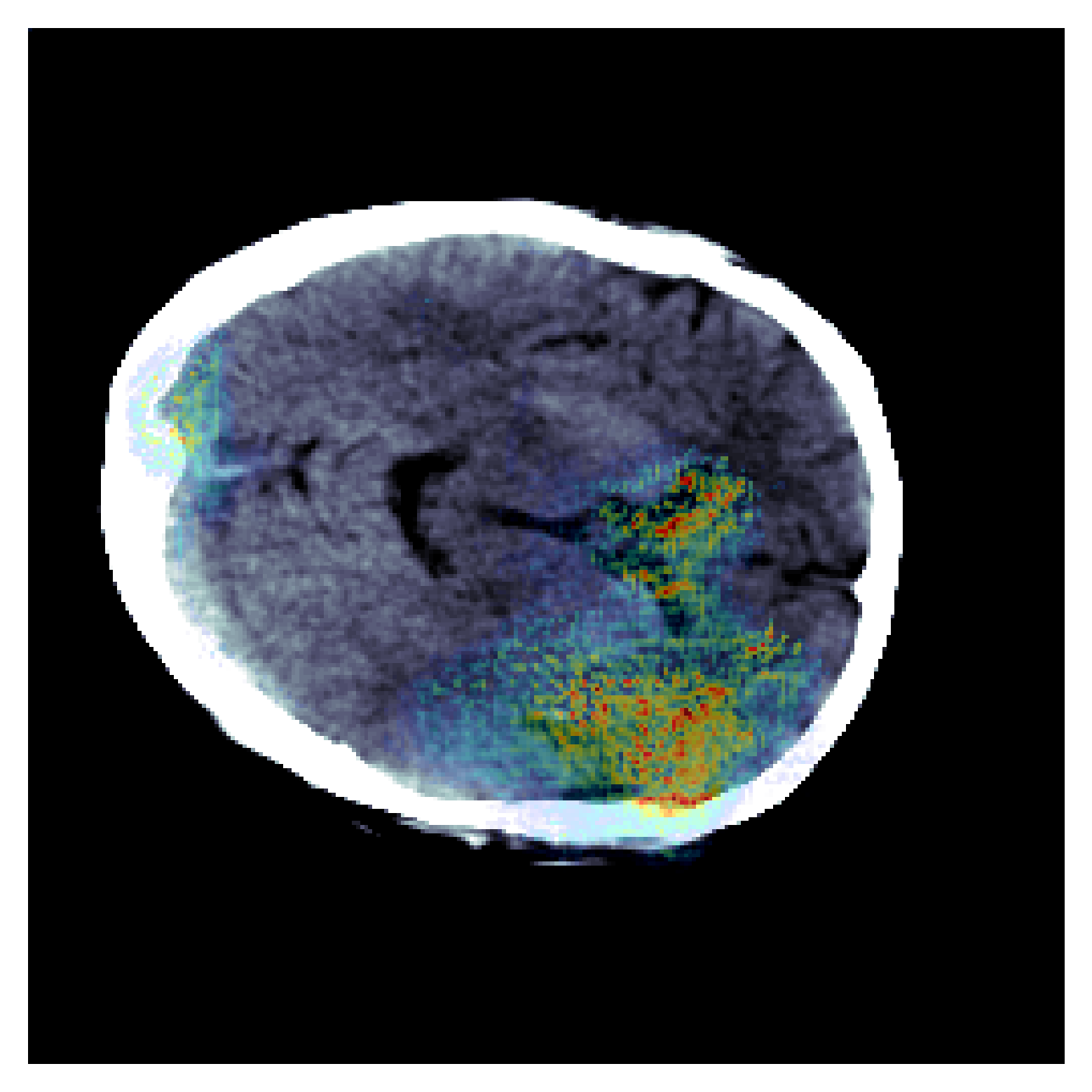}}
   \caption[example]{Worst predictions for a single image in ISLES}
  \label{fig:ISLESworstModels}
\end{figure}

Figure \ref{fig:ISLESworstModels} shows the two models that performed the worst on the ISLES dataset. (a) shows the dice loss in combination with the hardtanh output activation function. We see that this combination has the highest probabilistic error (NLL) because there is no uncertainty. In contrast, the BCE loss in combination with hardtanh (b) provides a very scattered lesion segmentation over large areas in the image. This combination has the highest overlap error and also shows a lot of uncertainty.

\begin{figure}[H]
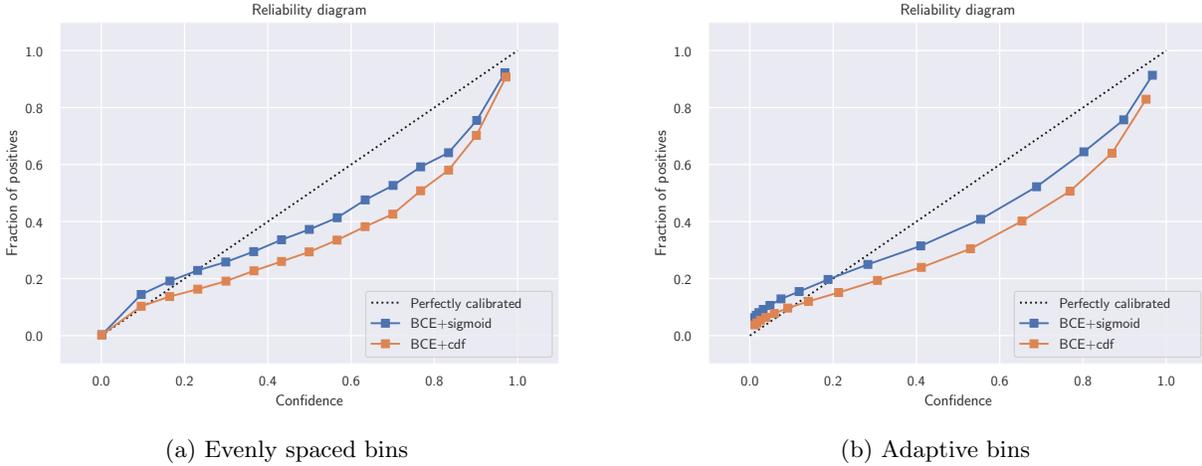
%
    \centering
    \subfloat[\centering Evenly spaced bins]{{\scalebox{0.48}{\input{diagrams/calibration_plot_isles1.pgf}} }}
    \qquad
    \subfloat[\centering Adaptive bins]{{\scalebox{0.48}{\input{diagrams/calibration_plot_isles1_adaptive.pgf}} }}
   \caption[example]{Reliability diagrams on the ISLES dataset for the best performing models compared to the sigmoid function as the output activation function.}
    \label{fig:reliabilityISLES}
\end{figure}

In Table \ref{tab:statsISLES}, BCE in combination with normal CDF has the lowest NLL error. However, in the reliability diagrams in Figure \ref{fig:reliabilityISLES}, we see that sigmoid comes closer to the perfectly calibrated line.

This result is not surprising since CDF has a smaller effective domain than sigmoid. It is more difficult for CDF to return probabilities between $0$ and $1$ than it is for sigmoid. For the last dataset, however, we could not find that big a difference between BCE+CDF and BCE+sigmoid. So it also depends on the dataset.

Even if sigmoid is here better with respect to the calibration error, both combinations overestimate the probabilities for values greater than $0.1$ (BCE+cdf) and for values greater than $0.25$ (BCE+sigmoid) in Figure \ref{fig:reliabilityISLES} (a). Only when the probabilities are close to $0$, the fraction of positives is near the perfectly calibrated line. With the adaptive strategy in (b), there are more bins close to $0$ because most ground truth pixels are $0$.

\paragraph{Kvasir-SEG}

The next dataset is Kvasir-SEG. Hardtanh activation has a higher dice coefficient than both sigmoid and CDF activation. In particular, simple functions such as the hardtanh or the linear function show better results than sigmoid-shaped activations. The winning model using NLL as the metric is MSE with linear output activation function, while the winning model using the dice metric is the combination BCE+hardtanh. 

\begin{longtable}{llHllH}
\toprule
Activation & Loss & Threshold & NLL $\downarrow$ & Dice $\uparrow$ & $\text{Dice}_{avg}$ $\uparrow$ \\ \hline
\multirow{3}{*}{normal CDF} & BCE & $0.31 \pm 0.0917$ & $0.2677 \pm 0.0173$ & $0.6783 \pm 0.0228$ & $0.5981 \pm 0.0199$\\ 
& Dice & $0.16 \pm 0.22$ & $1.3776 \pm 0.0863$ & $0.6475 \pm 0.0312$ & $0.5981 \pm 0.0284$\\ 
& MSE & $0.28 \pm 0.103$ & $0.2674 \pm 0.0124$ & $0.667 \pm 0.0252$ & $0.5891 \pm 0.0261$\\ 
\midrule
\multirow{3}{*}{sigmoid} & BCE & $0.29 \pm 0.0374$ & $0.255 \pm 0.0114$ & $0.6749 \pm 0.0212$ & $0.5977 \pm 0.02$\\ 
& Dice & $0.07 \pm 0.0245$ & $1.1301 \pm 0.0954$ & $0.6486 \pm 0.0134$ & $0.5993 \pm 0.0123$\\ 
& MSE & $0.29 \pm 0.0374$ & $0.2635 \pm 0.0183$ & $0.6686 \pm 0.0252$ & $0.589 \pm 0.0276$\\ 
\midrule
\multirow{3}{*}{inverse square root} & BCE & $0.05 \pm 0.0$ & $0.3314 \pm 0.0169$ & $0.694 \pm 0.0104$ & $0.306 \pm 0.0111$\\ 
& Dice & $0.05 \pm 0.0$ & $0.7643 \pm 0.0344$ & $0.6571 \pm 0.0145$ & $0.6071 \pm 0.0132$\\ 
& MSE & $0.24 \pm 0.049$ & $0.2595 \pm 0.0134$ & $0.6731 \pm 0.0267$ & $0.5868 \pm 0.0229$\\ 
\midrule
\multirow{3}{*}{arctangent} & BCE & $0.35 \pm 0.0837$ & $0.2651 \pm 0.0194$ & $0.6753 \pm 0.0229$ & $0.5974 \pm 0.0251$\\ 
& Dice & $0.41 \pm 0.4409$ & $0.5441 \pm 0.0263$ & $0.6486 \pm 0.0188$ & $0.5994 \pm 0.0166$\\ 
& MSE & $0.28 \pm 0.1249$ & $0.2614 \pm 0.0137$ & $0.6759 \pm 0.0176$ & $0.5853 \pm 0.0208$\\ 
\midrule
\multirow{3}{*}{softsign} & BCE & $0.29 \pm 0.0735$ & $0.2577 \pm 0.0134$ & $0.6797 \pm 0.0165$ & $0.5959 \pm 0.0127$\\ 
& Dice & $0.23 \pm 0.36$ & $0.5072 \pm 0.0094$ & $\underline{0.6427 \pm 0.0171}$ & $0.5942 \pm 0.0157$\\ 
& MSE & $0.3 \pm 0.0837$ & $0.2602 \pm 0.011$ & $0.6774 \pm 0.0155$ & $0.5901 \pm 0.0179$\\ 
\midrule
\multirow{3}{*}{linear} & BCE & $0.29 \pm 0.0735$ & $0.2591 \pm 0.0091$ & $0.6944 \pm 0.01$ & $0.6135 \pm 0.0083$\\ 
& Dice & $0.28 \pm 0.24$ & $0.7848 \pm 0.0579$ & $0.6533 \pm 0.0246$ & $0.6021 \pm 0.0219$\\ 
& MSE & $0.27 \pm 0.0812$ & $\mathbf{0.2493 \pm 0.0145}$ & $0.6956 \pm 0.0186$ & $\mathbf{0.6158 \pm 0.0143}$\\ 
\midrule
\multirow{3}{*}{hardtanh} & BCE & $0.35 \pm 0.0447$ & $0.258 \pm 0.0118$ & $\mathbf{0.7082 \pm 0.0038}$ & $0.6039 \pm 0.0237$\\ 
& Dice & $0.15 \pm 0.2$ & $\underline{1.5785 \pm 0.1143}$ & $0.6516 \pm 0.0242$ & $0.6021 \pm 0.0218$\\ 
& MSE & $0.31 \pm 0.0374$ & $0.3153 \pm 0.0157$ & $0.6837 \pm 0.009$ & $0.6079 \pm 0.0084$\\ 
\bottomrule
\caption{Results of all tested models trained on the Kvasir-SEG dataset where the best (bold) and worst (underline) results were highlighted.}
\end{longtable}

In this table, we again see the opposite effect, that a larger effective domain leads to better probabilities for dice loss + NLL. This time, we cannot find a correlation between BCE + NLL and the effective domain. The order is more random.

\begin{figure}[H]%
    \centering
    \subfloat[\centering BCE+hardtanh]{{\includegraphics[width=35mm]{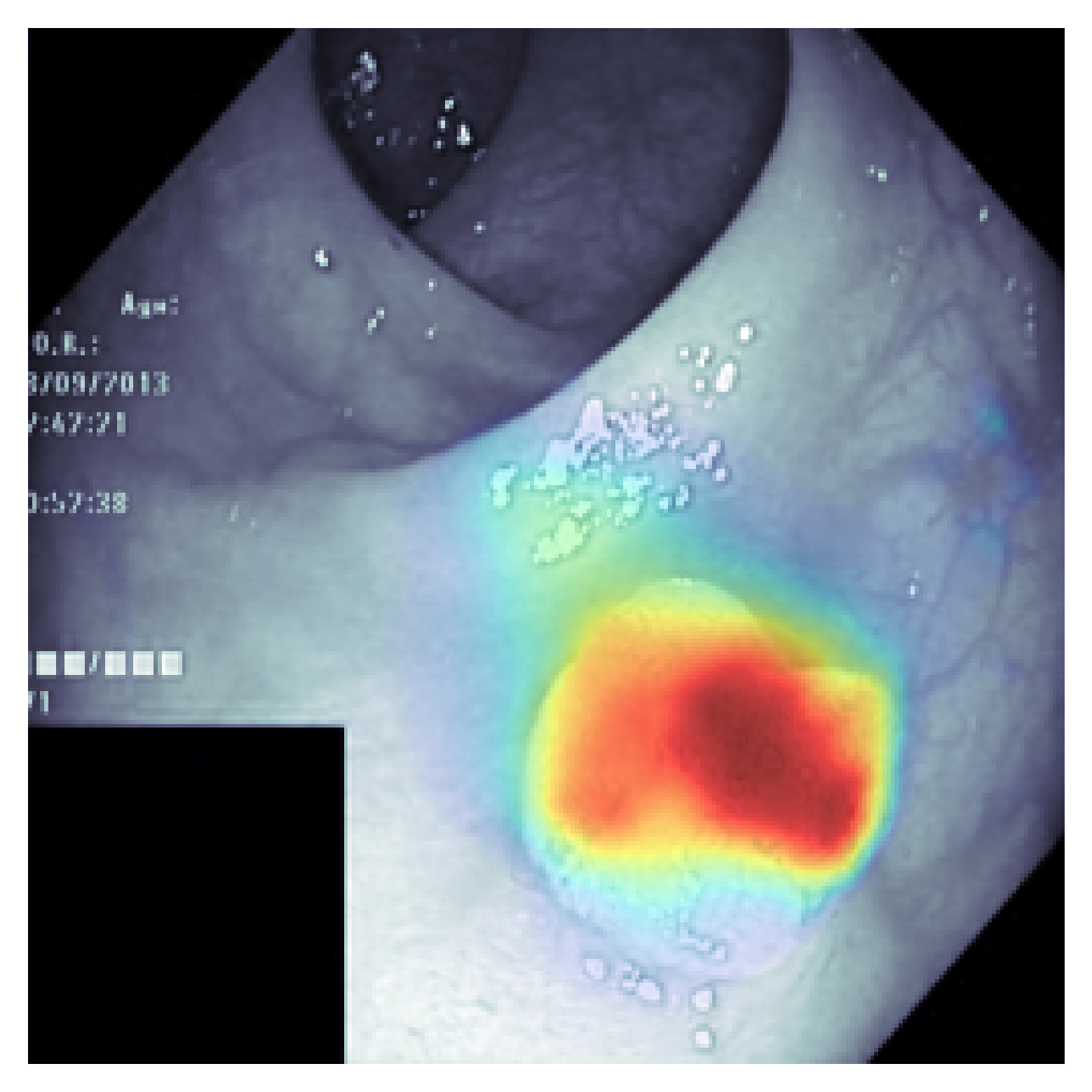} }}
    \qquad
    \subfloat[\centering MSE+linear]{{\includegraphics[width=35mm]{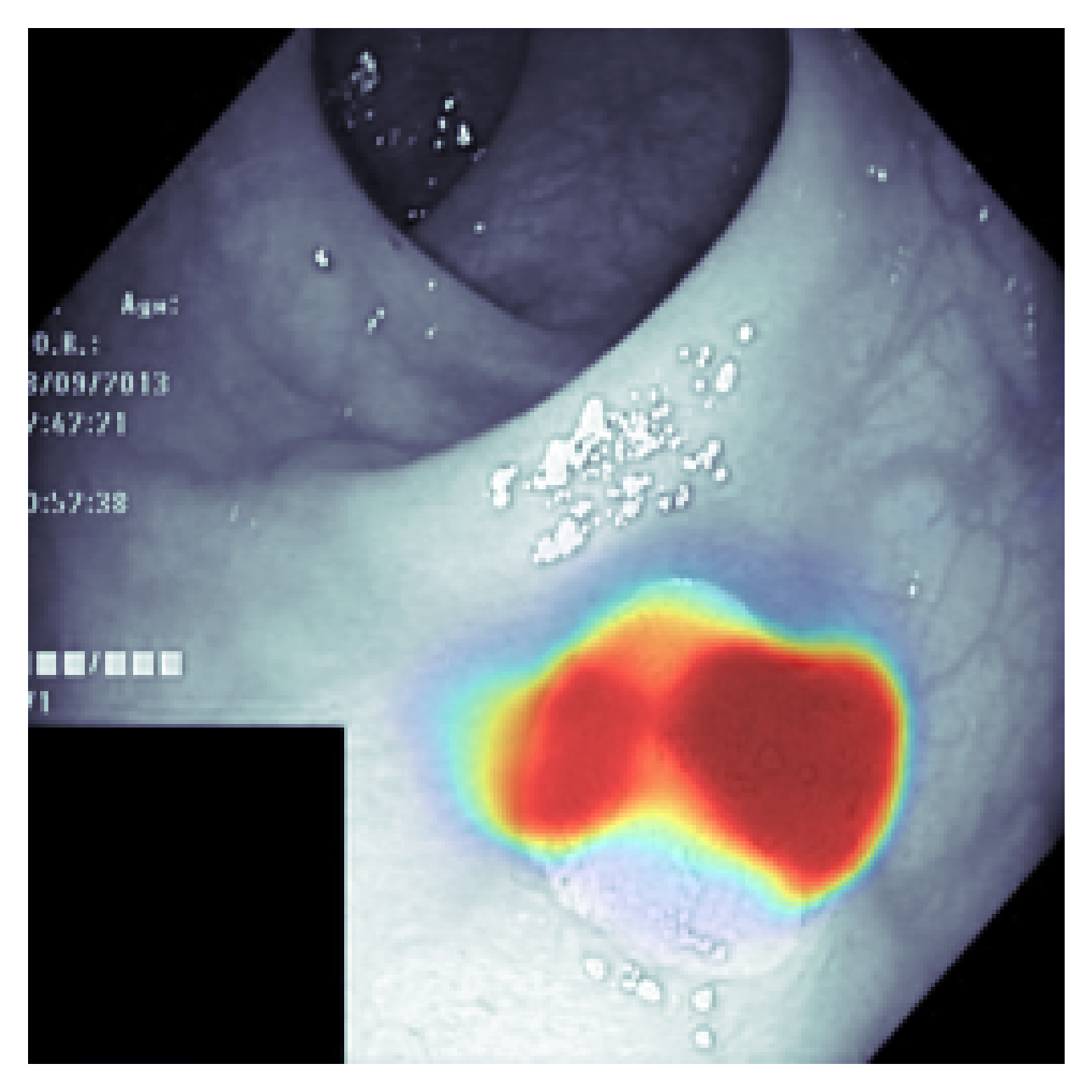} }}
    \qquad
    \subfloat[\centering Ground truth]{{\includegraphics[width=35mm]{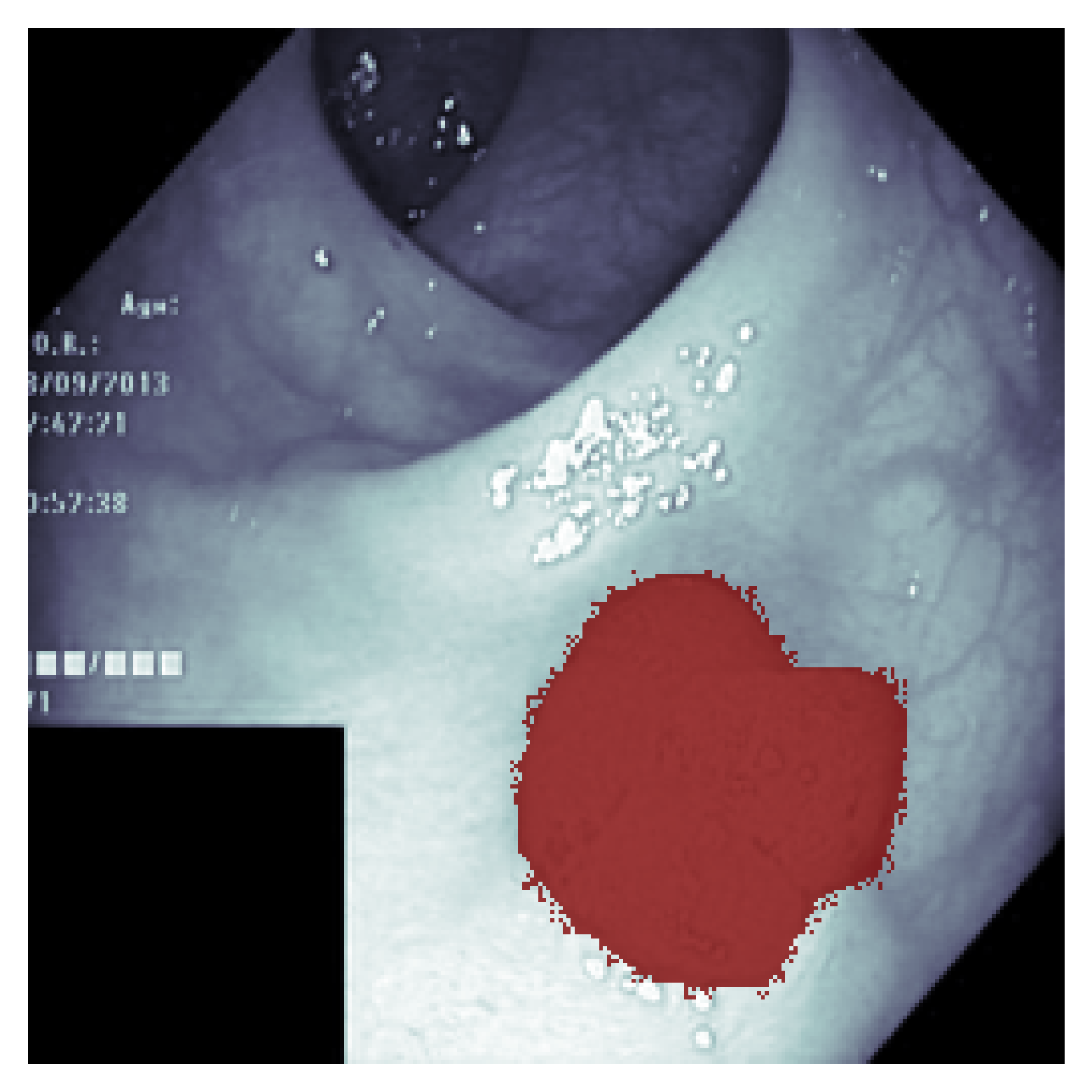}     }}
   \caption[example]{Best predictions for a single image in Kvasir.}
    \label{fig:KvasirbestModels}
\end{figure}

Figure \ref{fig:KvasirbestModels} shows the winning models using the Kvasir dataset with an example patient. (b) shows the prediction of the NLL winning model using MSE and linear activation. Here we can see that the ground truth area is partially predicted while maintaining a large boundary of uncertainty. The model using BCE and hardtanh gives similar results in (a) but has a smaller core of certain predictions compared to the other prediction as shown in (b). 

\begin{figure}[H]%
    \centering
    \subfloat[\centering Dice and hardtanh]{{\includegraphics[width=35mm]{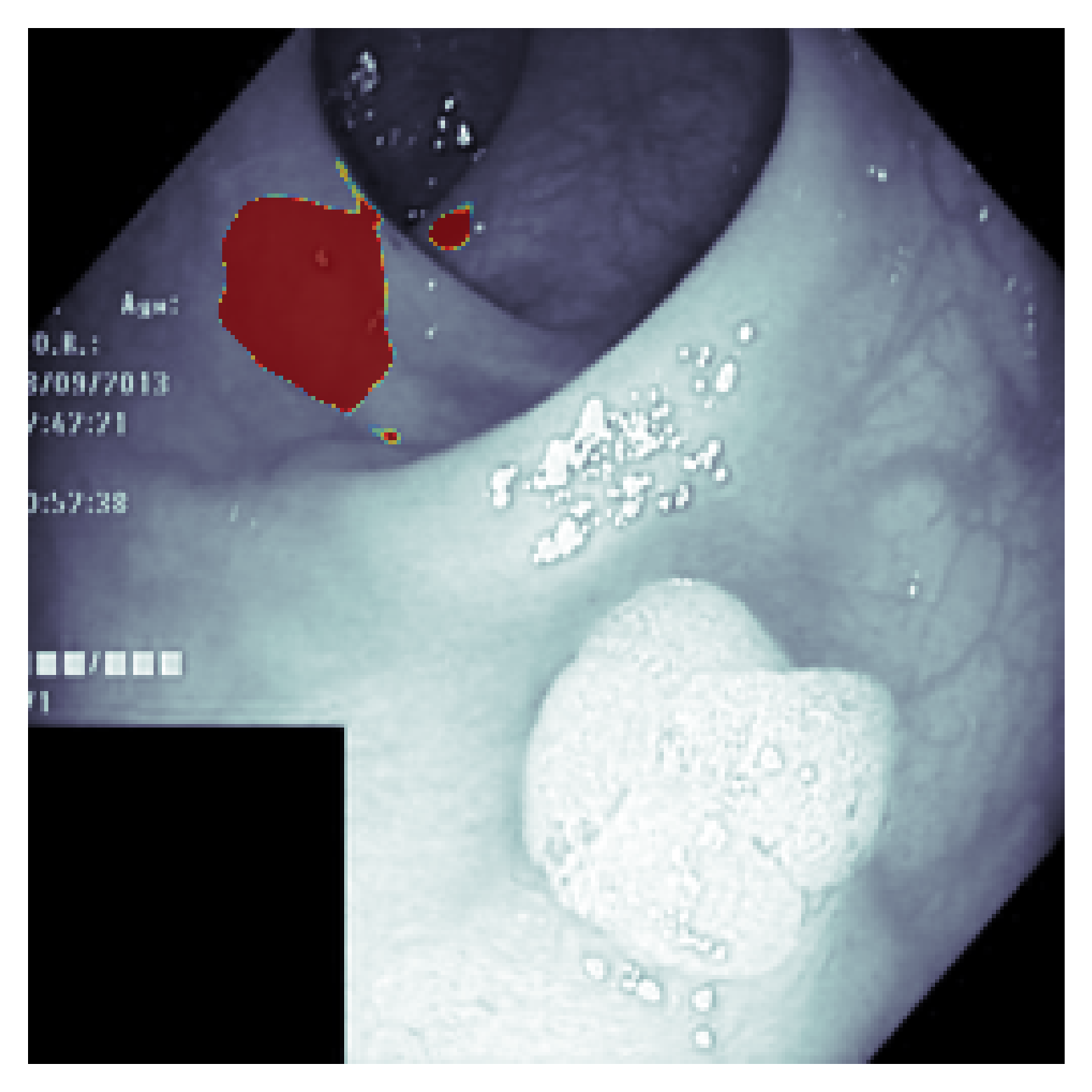} }}
    \qquad
    \subfloat[\centering Dice and softsign]{{\includegraphics[width=35mm]{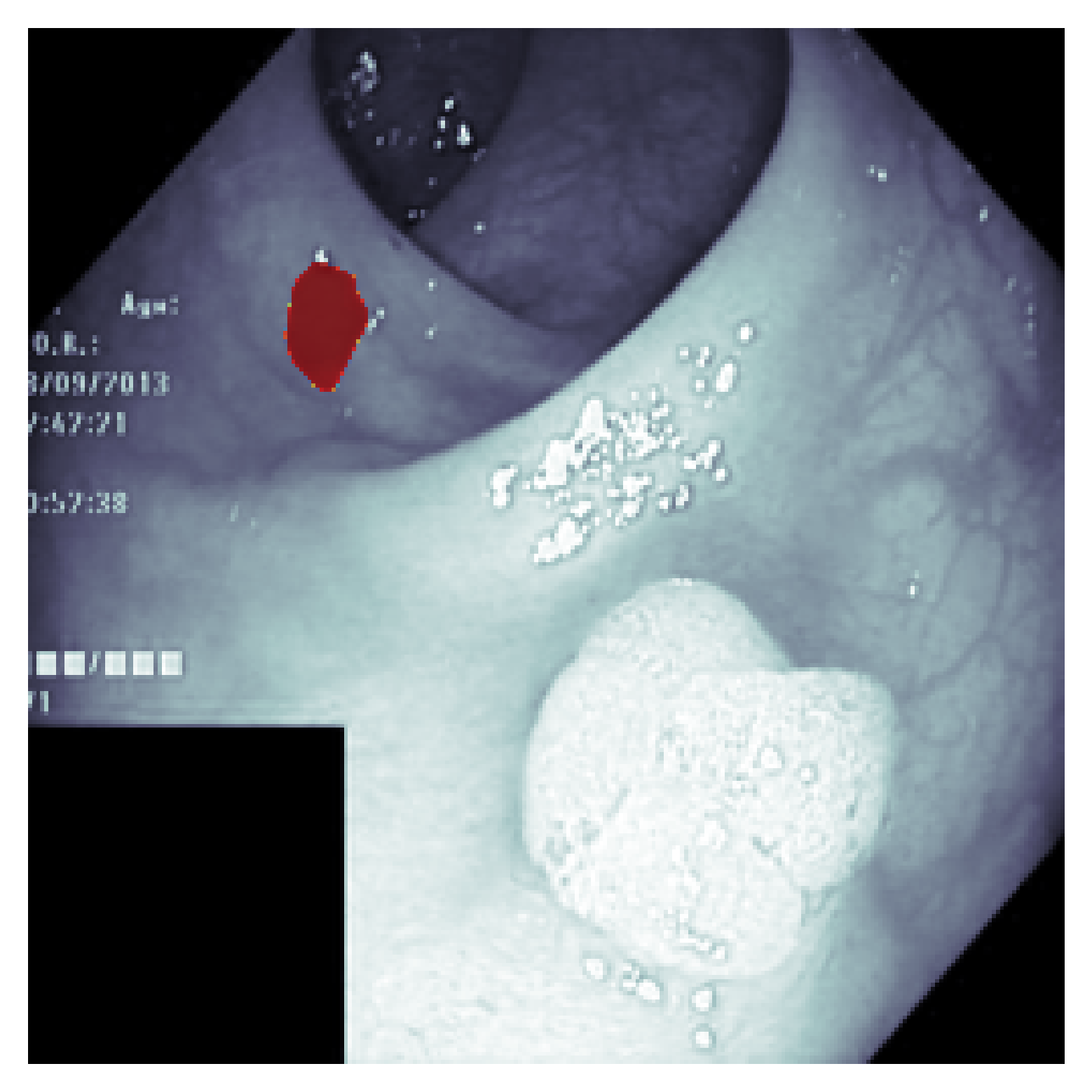} }}
   \caption[example]{Worst predictions for a single image in Kvasir.}
    \label{fig:KvasirWorstModels}
\end{figure}

Figure \ref{fig:KvasirWorstModels} shows the worst two models trained with this dataset. In (a) we can see that the dice and hardtanh combination predicts the segmentation area at a wrong position without any uncertainties. In (b) we see the same situation for the dice and softsign combination. This again illustrates the problem with dice loss that it does not output any uncertainties.

\begin{figure}[H]
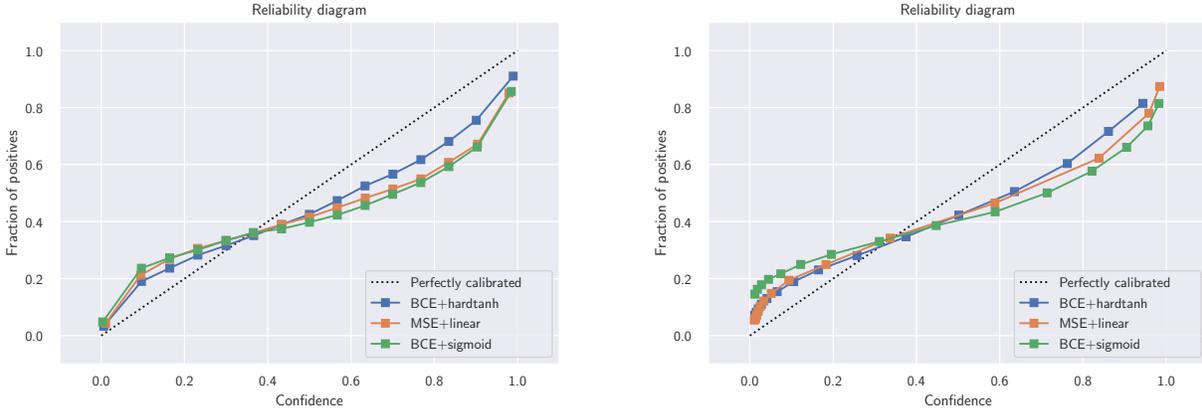
%
    \centering
    \subfloat[\centering Evenly spaced bins]{{\scalebox{0.48}{\input{diagrams/calibration_plot_kvasir1.pgf}} }}
    \qquad
    \subfloat[\centering Adaptive bins]{{\scalebox{0.48}{\input{diagrams/calibration_plot_kvasir1_adaptive.pgf}} }}
   \caption[example]{Reliability diagrams on the Kvasir dataset for the best performing models, compared to the sigmoid function as the output activation function.}
   \label{reliabilityKvasir}
\end{figure}

The reliability diagrams in Figure \ref{reliabilityKvasir} are similar to those for ACDC. The probabilities for values less than 0.35 are underestimated and overestimated for values greater than 0.35. BCE+hardtanh is closer to the perfectly calibrated line than MSE+linear, although the latter function has better NLL. Therefore, a model's calibration level does not always correlate with NLL. The use of quantiles does not have a strong effect on the reliability diagram, as can be seen in (b). Most bins are close to $0$.

\paragraph{Medical Segmentation Decathlon (MSD)}

The final dataset is the \emph{Medical Segmentation Decathlon}. As shown in Table \ref{tab:statsMSD}, BCE with arctangent as the output activation function provides the best prediction using Dice as the evaluation metric. In addition, using NLL as the evaluation metric, the winning model is BCE with sigmoid.

\begin{longtable}{llHllH}
\toprule
Activation & Loss & Threshold & NLL $\downarrow$ & Dice $\uparrow$ & $\text{Dice}_{avg}$ $\uparrow$ \\ \hline
\multirow{3}{*}{normal CDF} & BCE & $0.28 \pm 0.064$ & $0.0411 \pm 0.0125$ & $0.76 \pm 0.0565$ & $0.6737 \pm 0.0445$\\ 
& Dice & $0.44 \pm 0.4085$ & $0.1727 \pm 0.0407$ & $0.6694 \pm 0.0716$ & $0.6034 \pm 0.0628$\\ 
& MSE & $0.305 \pm 0.1293$ & $0.0469 \pm 0.017$ & $0.7642 \pm 0.0546$ & $0.6768 \pm 0.0453$\\
\midrule
\multirow{3}{*}{sigmoid} & BCE & $0.3 \pm 0.0949$ & $\mathbf{0.0383 \pm 0.0107}$ & $0.7682 \pm 0.056$ & $0.6763 \pm 0.039$\\ 
& Dice & $0.265 \pm 0.2656$ & $0.1123 \pm 0.0286$ & $0.704 \pm 0.067$ & $0.6365 \pm 0.0601$\\ 
& MSE & $0.375 \pm 0.068$ & $0.0519 \pm 0.0173$ & $0.7603 \pm 0.0478$ & $0.6649 \pm 0.0426$\\ 
\midrule
\multirow{3}{*}{inverse square root} & BCE & $0.08 \pm 0.0245$ & $0.0593 \pm 0.0118$ & $0.7404 \pm 0.0494$ & $0.33 \pm 0.0166$\\ 
& Dice & $0.37 \pm 0.3964$ & $0.077 \pm 0.0211$ & $0.7293 \pm 0.0812$ & $0.6595 \pm 0.0734$\\ 
& MSE & $0.25 \pm 0.0447$ & $0.0581 \pm 0.0122$ & $0.7605 \pm 0.0531$ & $0.6617 \pm 0.0357$\\ 
\midrule
\multirow{3}{*}{arctangent} & BCE & $0.23 \pm 0.1005$ & $0.0608 \pm 0.0113$ & $\mathbf{0.7686 \pm 0.0564}$ & $0.6748 \pm 0.0403$\\ 
& Dice & $0.345 \pm 0.3889$ & $0.0655 \pm 0.0164$ & $0.7559 \pm 0.0519$ & $\mathbf{0.6835 \pm 0.046}$\\ 
& MSE & $0.235 \pm 0.032$ & $0.0897 \pm 0.0216$ & $0.7568 \pm 0.0451$ & $0.6439 \pm 0.0385$\\ 
\midrule
\multirow{3}{*}{softsign} & BCE & $0.23 \pm 0.1077$ & $0.0672 \pm 0.0192$ & $0.764 \pm 0.0519$ & $0.6617 \pm 0.0459$\\ 
& Dice & $0.36 \pm 0.372$ & $0.0708 \pm 0.0164$ & $0.7485 \pm 0.055$ & $0.6762 \pm 0.048$\\ 
& MSE & $0.325 \pm 0.1188$ & $0.0963 \pm 0.0255$ & $0.7622 \pm 0.0533$ & $0.6445 \pm 0.0306$\\ 
\midrule
\multirow{3}{*}{linear} & BCE & $0.29 \pm 0.0624$ & $0.04 \pm 0.0083$ & $0.7335 \pm 0.0501$ & $0.6104 \pm 0.0436$\\
& Dice & $0.38 \pm 0.1382$ & $0.059 \pm 0.0116$ & $0.6834 \pm 0.0793$ & $0.5955 \pm 0.0472$\\ 
& MSE & $0.285 \pm 0.0808$ & $0.0457 \pm 0.0098$ & $0.7359 \pm 0.0471$ & $0.62 \pm 0.0403$\\ 
\midrule
\multirow{3}{*}{hardtanh} & BCE & $0.395 \pm 0.085$ & $0.0512 \pm 0.0098$ & $0.6517 \pm 0.0664$ & $0.4776 \pm 0.0396$\\ 
& Dice & $0.24 \pm 0.3562$ & $\underline{0.2201 \pm 0.0373}$ & $0.6914 \pm 0.0707$ & $0.6246 \pm 0.0644$\\ 
& MSE & $0.29 \pm 0.1546$ & $0.2076 \pm 0.2268$ & $\underline{0.605 \pm 0.2861}$ & $0.5347 \pm 0.2692$\\ 
\bottomrule
\caption{Results of all tested models trained on the MSD dataset where the best (bold) and worst (underline) results were highlighted.}
\label{tab:statsMSD}
\end{longtable}

The worst results are obtained with the hardtanh activation function. hardtanh+MSE failed to converge on the last cross-validation fold and stopped with a dice coefficient close to $0$. This explains the large standard deviation of the MSE+hardtanh combination.

In this table, the effective domain has a partial influence on NLL+BCE. As we move from sigmoid to softsign, the NLL increases. However, dice loss in combination with NLL shows again the opposite effect.

\begin{figure}[H]%
    \centering
    \subfloat[\centering BCE and arctan]{{\includegraphics[width=35mm]{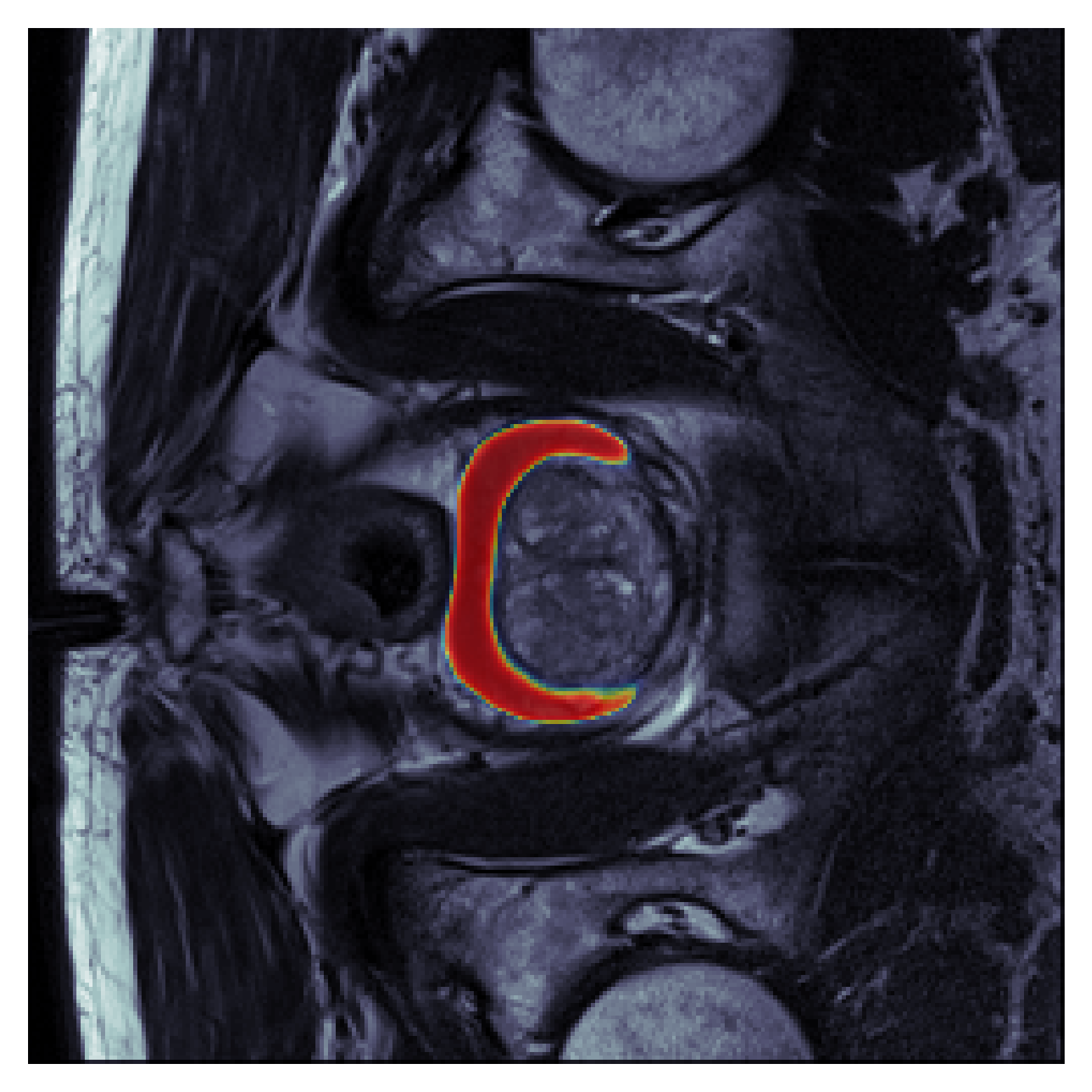} }}
    \qquad
    \subfloat[\centering BCE and sigmoid]{\includegraphics[width=35mm]{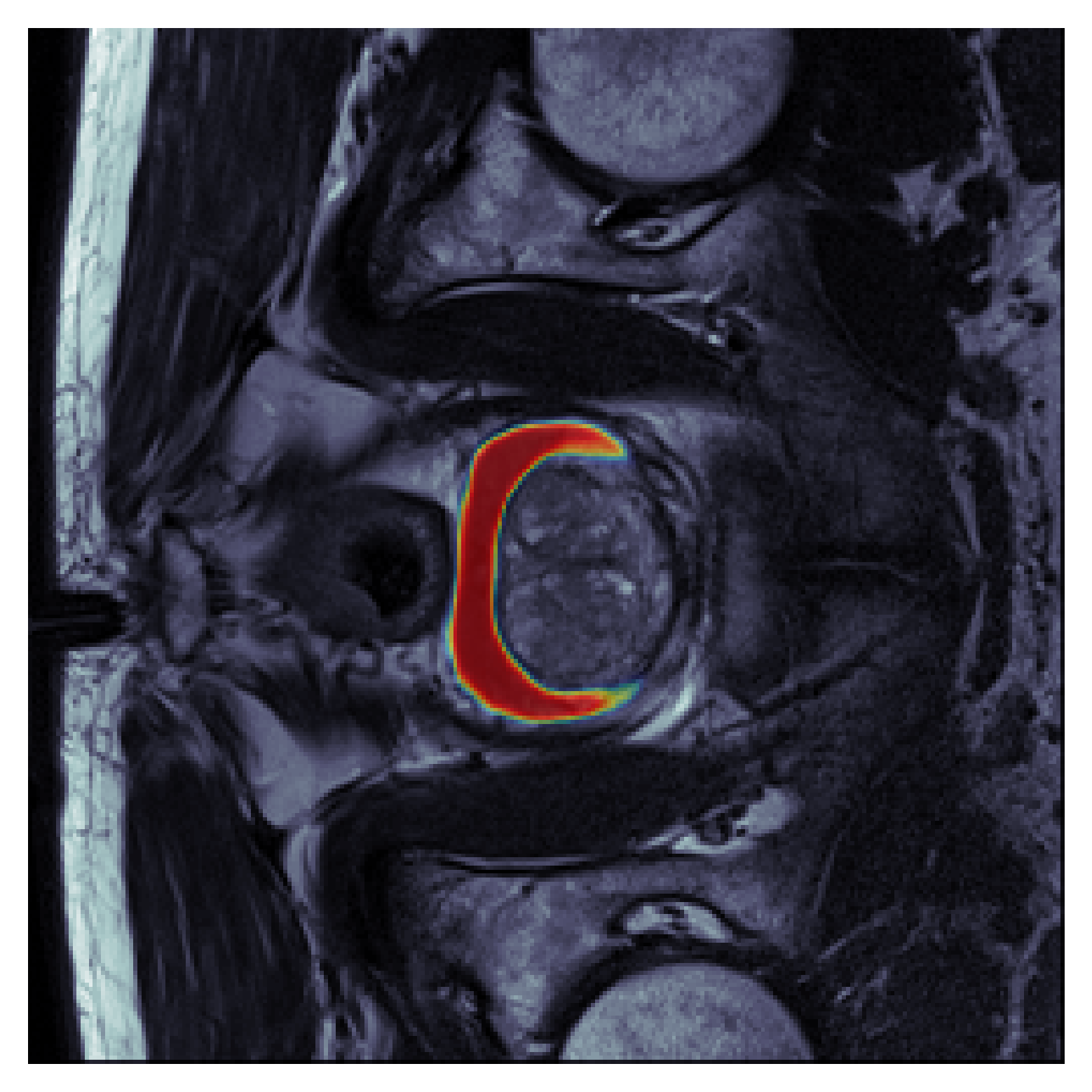}}
    \qquad
    \subfloat[\centering Ground truth]{\includegraphics[width=35mm]{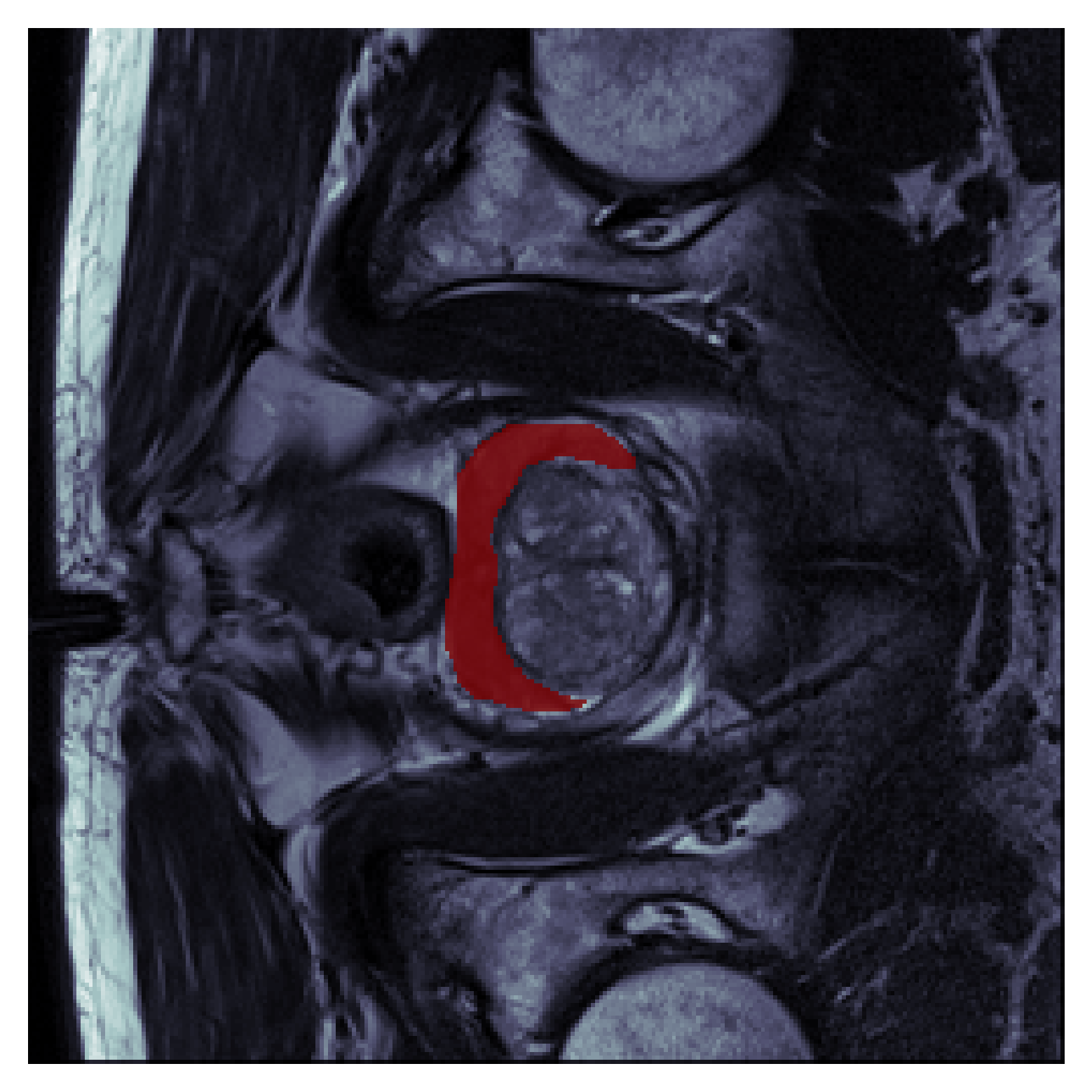}}
   \caption[example]{Best predictions for a single image in MSD.}
  \label{fig:MSDBestModels}
\end{figure}

If we look further at the winning models, as shown in Figure \ref{fig:MSDBestModels}, we can see that the ground truth area is well determined by the neural network. In (b) there is higher uncertainty at the tips of the C-shaped prediction. In contrast, (a) has more certainty than (b) for the top tip. For the non-border area, the probability is close to one.

\begin{figure}[H]%
    \centering
    \subfloat[\centering BCE and hardtanh]{{\includegraphics[width=35mm]{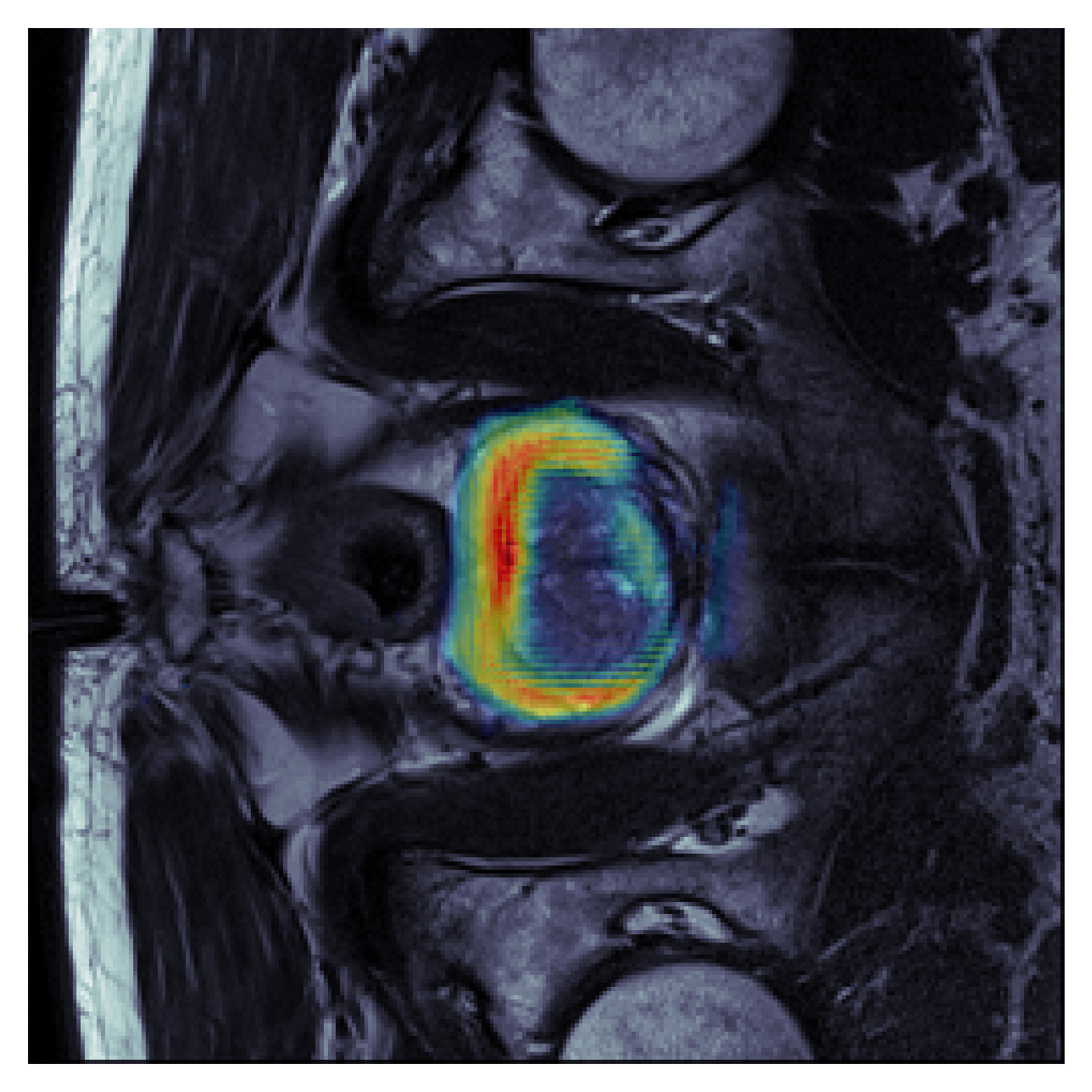} }}
    \qquad
    \subfloat[\centering Dice and hardtanh]{\includegraphics[width=35mm]{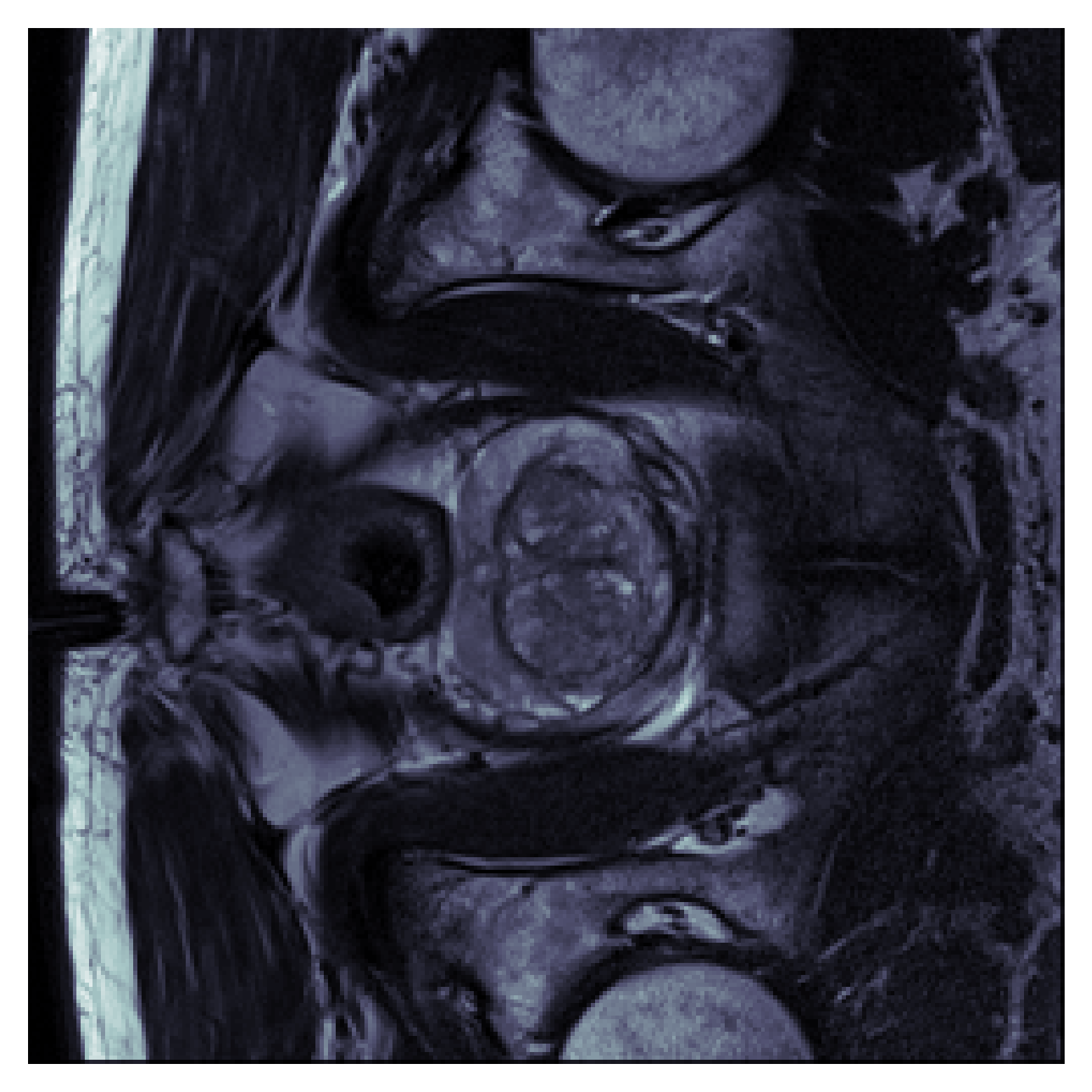}}
    \qquad
    \subfloat[\centering Ground truth]{\includegraphics[width=35mm]{predictions/MSD_gt.png}}
   \caption[example]{Worst predictions for a single image in MSD.}
  \label{fig:MSDWorstModels}
\end{figure}

The worst predictions according to the table are shown in Figure \ref{fig:MSDWorstModels}. Since MSE+hardtanh had convergence problems, we show BCE+hardtanh instead.

Hardtanh in combination with BCE loss results in poor segmentation with a high level of uncertainty. In contrast, the combination with dice loss does not produce any segmentation.  

\begin{figure}[H]
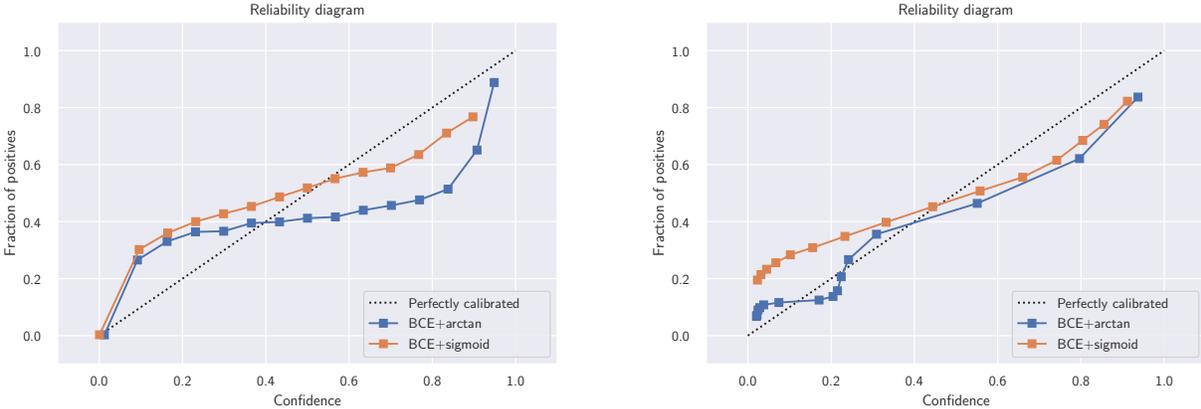
%
    \centering
    \subfloat[\centering Evenly spaced.]{{\scalebox{0.48}{\input{diagrams/calibration_plot_msd1.pgf}} }}
    \qquad
    \subfloat[\centering Adaptive.]{{\scalebox{0.48}{\input{diagrams/calibration_plot_msd1_adaptive2.pgf}} }}
   \caption[example]{Reliability diagrams on the MSD dataset for the best performing models compared to the sigmoid function as the output activation function.}
    \label{fig:reliabilityMSD}
\end{figure}

Figure \ref{fig:reliabilityMSD} shows the reliability diagrams of the best two models. In (a) the BCE+sigmoid combination seems closer to the perfectly calibrated line, but in (b) BCE+arctan is better. The adaptive strategy leads to more bins at $0$ and $0.2$ for arctangent. The plot in (b) is more trustworthy, as adaptivity leads to a better distributed plot. 

Both combinations are not calibrated. The diagrams are similar to the previous calibration diagrams and show overestimation and underestimation of the probabilities. We see that arctan, which is an activation function with a wider effective domain, produces better calibrated predictions than sigmoid.

\section{DISCUSSION}
\label{sec:dis}

\paragraph{General Observations}

The results of the conducted analysis have shown that the output activation function in combination with the loss function has an important impact on the segmentation result in medical imaging.

\begin{figure}[H]
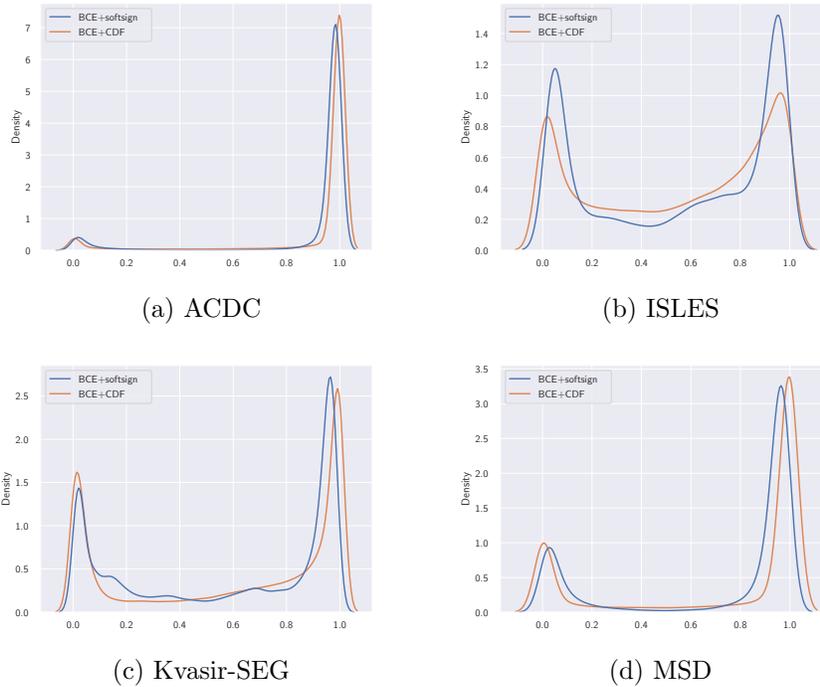

\centering
\begin{tabular}{cc}
  {\scalebox{0.35}{\input{diagrams/kde_plot_acac.pgf}}} &   {\scalebox{0.35}{\input{diagrams/kde_plot_isles.pgf}}}\\
  (a) ACDC & (b) ISLES \\
  {\scalebox{0.35}{\input{diagrams/kde_plot_kvasir.pgf}}} &   {\scalebox{0.35}{\input{diagrams/kde_plot_msd.pgf}}}\\
  (c) Kvasir-SEG & (d) MSD
\end{tabular}
\caption{Kernel density estimation for $P(\hat{Y} \mid Y = 1)$ of the first validation fold}
\label{fig:allKDE}
\end{figure}

We have hypothesized in section \ref{sec:act} that the rate of change of the activation function affects the range of the probabilities on the segmentation mask. From the metrics, reliability diagrams, and examples, we see that the activation function has a strong influence on the prediction results. However, the effect on the range of the probabilities is not always obvious. For this reason, we plot in Figure \ref{fig:allKDE} the probability density function of two extreme examples: CDF (small effective domain) and softsign (large effective domain).

For all datasets, the peaks at 0.0 and 1.0 when considering BCE+softsign are slightly shifted to the right and left. This means that the network tends to make fewer certain predictions. By choosing a different activation, we can thus affect the distribution of the probabilities.

However, a less certain prediction is not always better, as we saw in the tables from the last section. The softsign function did not do as well as CDF or sigmoid in terms of segmentation errors. A smaller effective domain tends to improve the segmentation error. One explanation is that we are restricting the network's freedom of action. When we reduce the domain from $[-6, 6]$ to $[-3, 3]$, all the weights in the neural network must decrease. This then also affects the backpropagation during training and can lead to better local minima.

The advantage of having a larger effective domain is that it can improve the reliability of the predictions. If we just look at the BCE loss function in the graphs in the last section, we can see that a larger effective domain tends to increase reliability. For the ISLES dataset, sigmoid is better than CDF. For the MSD dataset, arctangent is better than sigmoid.

Two special cases are the linear activation and the hardtanh function. The linear activation adapts the domain to the values in the image. For this reason, it also gives good results, although it approaches the values 0 and 1 linearly. The hardtanh function has a small effective domain and should also give good results. However, because the function has discontinuities and not a good derivative, it has not always been able to converge or has had difficulty doing so. This could also explain why hardtanh rarely performed well.

Another aspect related to the effective domain is the loss function. We have seen that a greater effective domain can improve the probabilities for dice loss. Table \ref{tab:diceEffect} shows the reverse effect of what we observed for MSE/BCE loss. Furthermore, this decreasing effect is even clearer than the increasing effect for MSE/BCE.

\begin{table}[H]
    \centering
    \begin{tabular}{llllll}
      \toprule
      Dataset & normal CDF & sigmoid & inverse sqrt & arctangent & softsign\\ \hline
      ACDC & $0.2033$ & $0.1409$ & $0.1131$ & $0.0992$ & $0.0965$\\
      ISLES & $0.1084$ & $0.0905$ & $0.0712$ & $0.0608$ & $\underline{0.0644}$\\ 
      Kvasir-SEG & $1.3776$ & $1.1301$ & $0.7643$ & $0.5441$ & $0.5072$\\
      MSD & $0.1727$ & $0.1123$ & $0.077$ & $0.0655$ & $\underline{0.0708}$\\
      \bottomrule
    \end{tabular}
\centering
\caption{NLL error for models trained with dice loss. Activation functions that do not show a decreasing behavior were underlined.}
\label{tab:diceEffect}
\end{table}

The effect is reversed because dice loss does not show any uncertainties and optimizes the overlap. When we decrease the rate of change, it improves the probabilistic estimates and the segmentation error. Since dice loss takes into account the sum of many pixels at the same time, a change in the activation function enables better adaptation of the error of individual pixels.

Apart from the tables and the reliability diagrams, we also analyzed the results visually. We have seen that good predictions have a high probability in the core area and low probabilities in the border area. If the uncertainty for the whole area is high, the dice coefficient tends to be low. Thus, we observed that the uncertainty also correlates with the dice coefficient.

Furthermore, we have seen that small differences in the dice coefficient are not always obvious when looking at a single image. If the dice coefficient differs only slightly, there are no major changes in the predictions. However, if a function performs significantly worse, this is also reflected in the predicted output.

\paragraph{Interpretation of Test Results.}

We aggregate the results from Section \ref{sec:eval} to compare them against sigmoid. The best activation+loss function combination is contrasted with the best sigmoid+loss function combination, as shown in Table \ref{tab:betterSigmoid}. 

\begin{table}[H]
    \centering
    \begin{tabular}{lll|l}
      \toprule
      Activation function & Won NLL & Won Dice & Sum\\ \hline
      Normal CDF & \textbf{2} & \textbf{3} & \textbf{5}\\
      Inverse Square Root & 0 & 2 & 2\\ 
      Arctangent & 0 & \textbf{3} & 3\\
      Softsign & 0 & 2 & 2\\ 
      \midrule
      HardTanh & 0 & 2 & 2\\ 
      Linear & \textbf{2} & 2 & 4\\
      \bottomrule
    \end{tabular}
\centering
\caption{Number of models with output activation functions that perform better than sigmoid output activation function across all datasets. Except for hardtanh and linear, all functions were ordered by their effective domain.}
\label{tab:betterSigmoid}
\end{table}

Of a total of $4$ datasets, the normal CDF won 3 times against sigmoid with respect to the optimal dice coefficient. Regarding the negative log-likelihood, it is a tie. Hence, the normal CDF activation is at least as good as sigmoid for the given datasets with respect to the segmentation error.

From Table \ref{tab:betterSigmoid}, we also see that the linear activation fared well against sigmoid. Here, we can determine that the sigmoid or CDF activation function should not be used exclusively. Instead, we suggest testing multiple output activation functions. In this paper, we have shown that there are several functions that can be used as output activation functions that outperform the sigmoid function. Table \ref{tab:betterSigmoid2} shows the results of a few activation functions that produced good results across all datasets.

\begin{table}[H]
    \centering
    \begin{tabular}{lll}
      \toprule
      Activation function & Loss function & Avg. dice coefficient \\ \hline
      Sigmoid & dice loss & 0.726575\\
      Arctangent & dice loss & 0.742\\
      Linear & MSE & 0.748875\\
      Sigmoid & BCE & 0.752425\\ 
      Normal CDF & BCE & \textbf{0.753675}\\
      \bottomrule
    \end{tabular}
\centering
\caption{Average dice coefficient across all datasets for a selected number of output activation and loss functions.}
\label{tab:betterSigmoid2}
\end{table}

Next, we count how many times each loss function won by dataset and activation function.

\begin{table}[H]
\begin{tabular}{lllH|l}
\toprule
Loss & Won NLL & Won Dice & Won $\text{Dice}_{avg}$ & Sum\\ \hline
BCE & \textbf{21} & \textbf{13} & 3 & \textbf{34}\\
Dice & 0 & 6 & \textbf{19} & 6\\ 
MSE & 7 & 9 & 6 & 16\\ 
\bottomrule
\end{tabular}
\centering
\caption{The number of models with the respective loss functions that have won the evaluation of the four datasets presented. The maximum of each of the first two columns is 7 activation functions times 4 datasets.}
\end{table}

The table shows that BCE was the best loss function for most activation functions and datasets. By using BCE as the loss function we optimize NLL, but in some cases, MSE was still the better loss function. Therefore, the metric does not always have to be the same as the loss function.

Given the optimal threshold, the best single dice coefficient is more frequently found by using BCE or MSE. For BCE/MSE, the threshold of the dice coefficient is an additional parameter that increases or reduces the predicted area. However, optimization of this parameter can also lead to overfitting to the validation set. Dice loss has not the same issue, as it is largely threshold-independent. One possible reason that BCE/MSE is better, then, is the lack of an additional parameter that controls the predicted area.

Another reason is that dice loss is only a surrogate loss function to the dice coefficient because it relaxes the constraint $\hat{y} \in \{0, 1\}$ to $\hat{y} \in [0, 1]$ where $\hat{y}$ is the predicted pixel. While this makes the loss function differentiable, it is no longer the dice coefficient. Then it is not surprising that other loss functions can be better.

Our empirical result that dice loss is inferior to cross entropy contradicts the findings by Bertels et al.\cite{DBLP:journals/corr/abs-1911-01685}. One possible reason is that we stopped training the network when the Dice coefficient stopped improving. Comparatively, the authors stopped the training when the validation loss did not reduce anymore. Another reason is that we used a pre-trained encoder and a 2D U-Net. We also obtained a 10-20\% higher dice coefficient than the authors on ISLES 2018 with 5-fold cross-validation. Our code is available on GitHub for further testing.

\paragraph{Limitations and Future Directions.}

A limitation of this paper is that we did not consider all possible loss functions due to resource constraints. The Lovász-Softmax loss\cite{DBLP:journals/corr/BermanB17} is another way to obtain a surrogate function of the dice coefficient. There are also many extensions to dice loss\cite{yeung2021unified,DBLP:journals/corr/SalehiEG17a} that could be considered in further work.

Apart from that, we treated multi-label segmentation with the datasets ACDC and MSD as multiple binary segmentation problems by applying the sigmoid instead of the softmax function. The difference is that the softmax function normalizes the sum of all classes to $1$. The softmax function corresponds to a categorical distribution, while our sigmoid function corresponds to multiple Bernoulli distributions. Hence, for proper multi-label classification, the proposed activation functions have to be extended to produce a categorical distribution.

We only considered the case of medical image segmentation. Further work could also evaluate the effect of the output activation in regular neural networks or other tasks like classification. Trainable activation functions could also be considered. Molina et al. \cite{Molina2020Pade} have shown with their Padé Activation Unit that trainable functions can surpass regular functions like ReLU.

Finally, another promising direction is to analyze more the effect of the activation function on the reliability of the probabilities.

\section{CONCLUSIONS}
\label{sec:con}

In this paper, we proposed alternative activation functions to the sigmoid function. We motivated this by examining the rate of change of these functions. Our results have shown that the activation function affects both the segmentation error and the calibration of the neural network. Functions with a high rate of change tend to produce better segmentation results, while functions with a lower rate of change can produce better calibration. In particular, we saw that the normal CDF is a viable alternative to sigmoid in binary segmentation. Furthermore, we showed that the loss function has also an impact on which activation function is best suited for segmentation. When changing the loss function to dice loss, arctangent gave the best results. We encourage further research to find better output activation functions.

In future research, we aim to extend our analysis to other application areas of deep learning, such as mechanical engineering or natural language processing. Further, we want to provide visual assistance tools that make it possible to communicate the effectiveness of a chosen output activation function.

\section{CONFLICT OF INTEREST}

The authors declare that there is no conflict of interest.

\newpage

\section{APPENDIX}
\label{sec:appendix}

\subsection{Automated Cardiac Diagnosis Challenge (ACDC)}

\begin{figure}[H]
\centering
\scalebox{.77}{\begin{tabular}{cccc}
  \includegraphics[width=35mm]{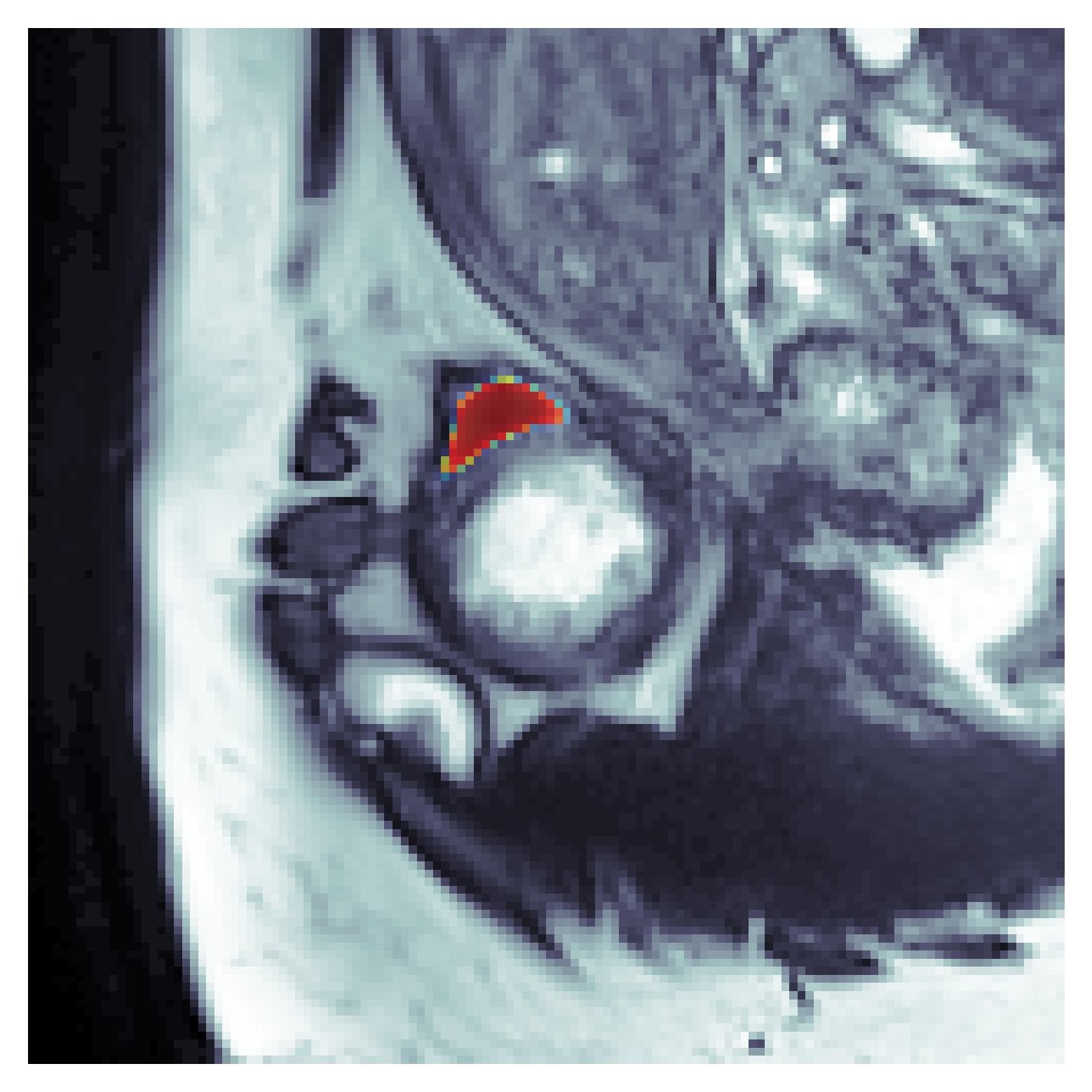} &   \includegraphics[width=35mm]{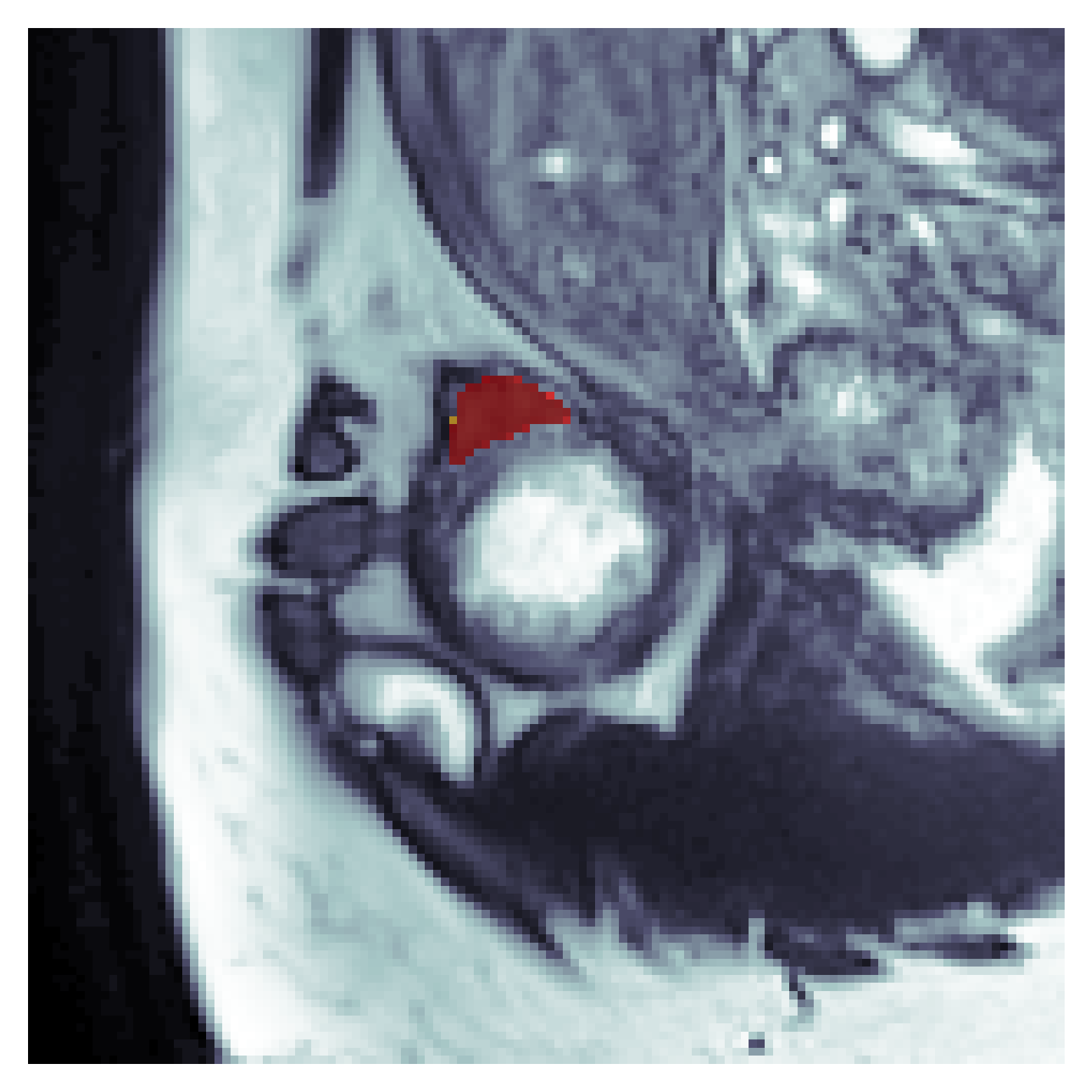} & \includegraphics[width=35mm]{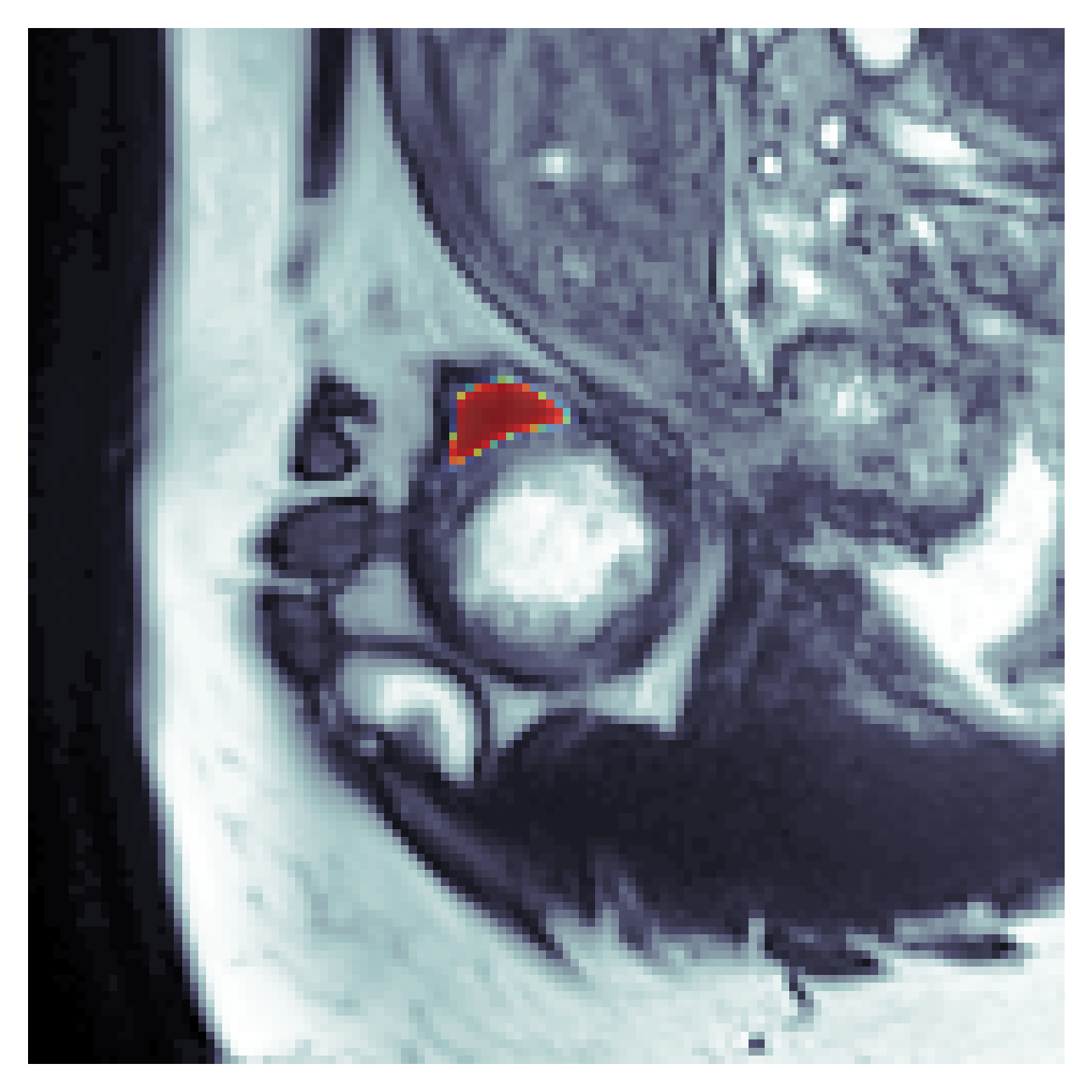} &   \includegraphics[width=35mm]{predictions/ACAC-BCELoss-cdf_activation-resnet34-Unet.png} \\
  (a) BCE+Arctangent & (b) Dice+Arctangent & (c) MSE+Arctangent & (d) BCE+CDF \\[6pt]

  \includegraphics[width=35mm]{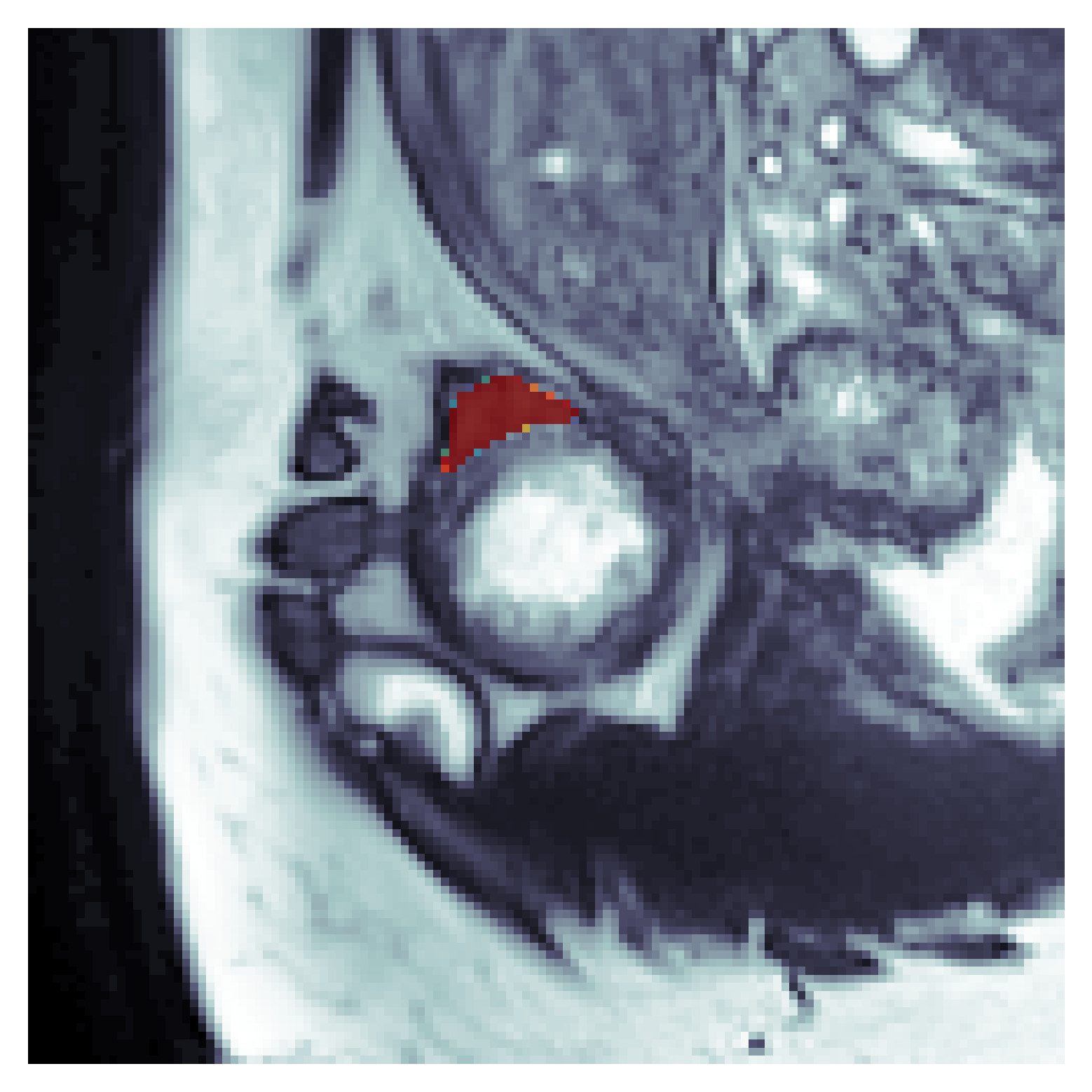} &   \includegraphics[width=35mm]{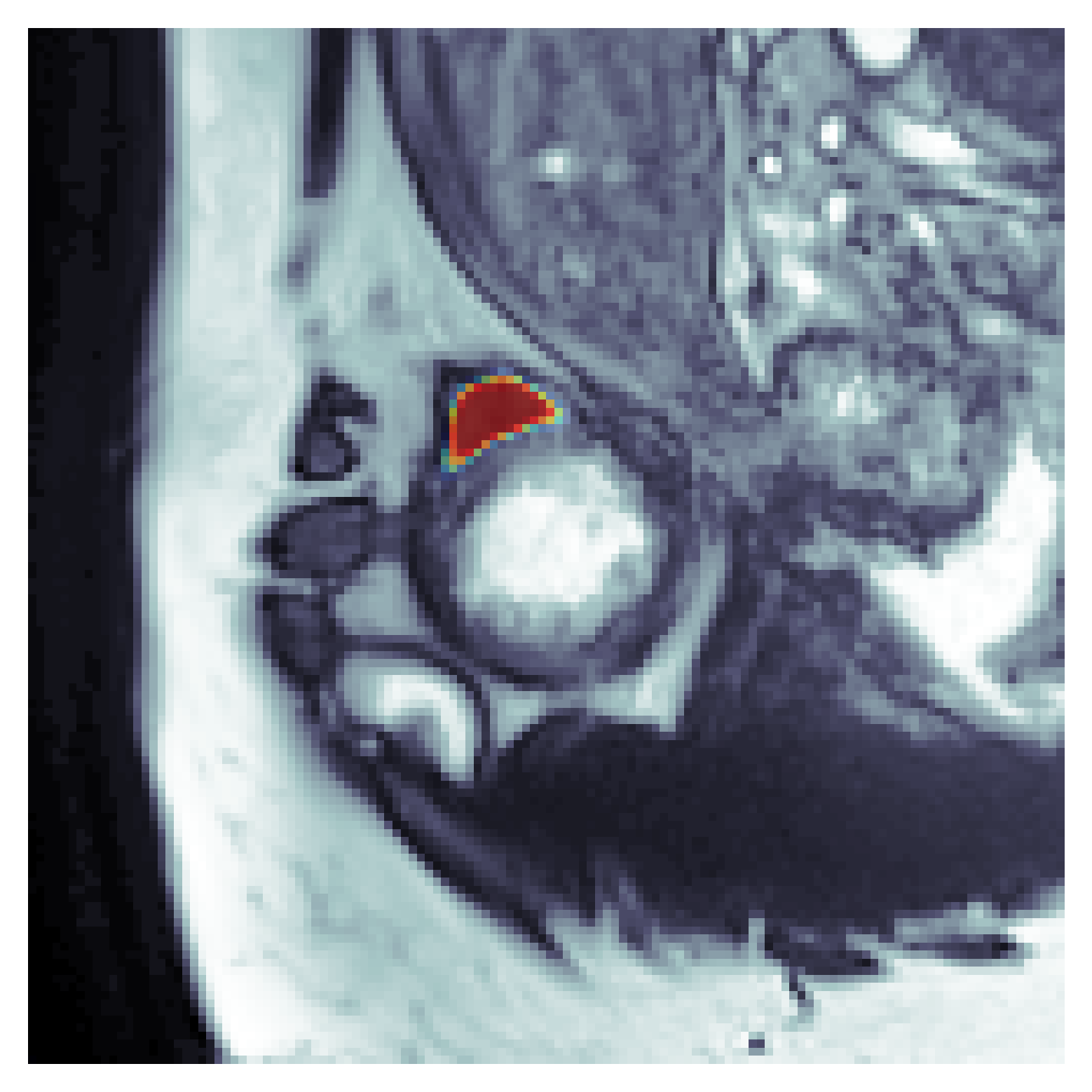} & \includegraphics[width=35mm]{predictions/ACAC-BCELoss-hardtanh_activation-resnet34-Unet.png} &   \includegraphics[width=35mm]{predictions/ACAC-DiceLoss-hardtanh_activation-resnet34-Unet.png} \\
  (e) Dice+CDF & (f) MSE+CDF & (g) BCE+HardTanh & (h) Dice+HardTanh \\[6pt]

  \includegraphics[width=35mm]{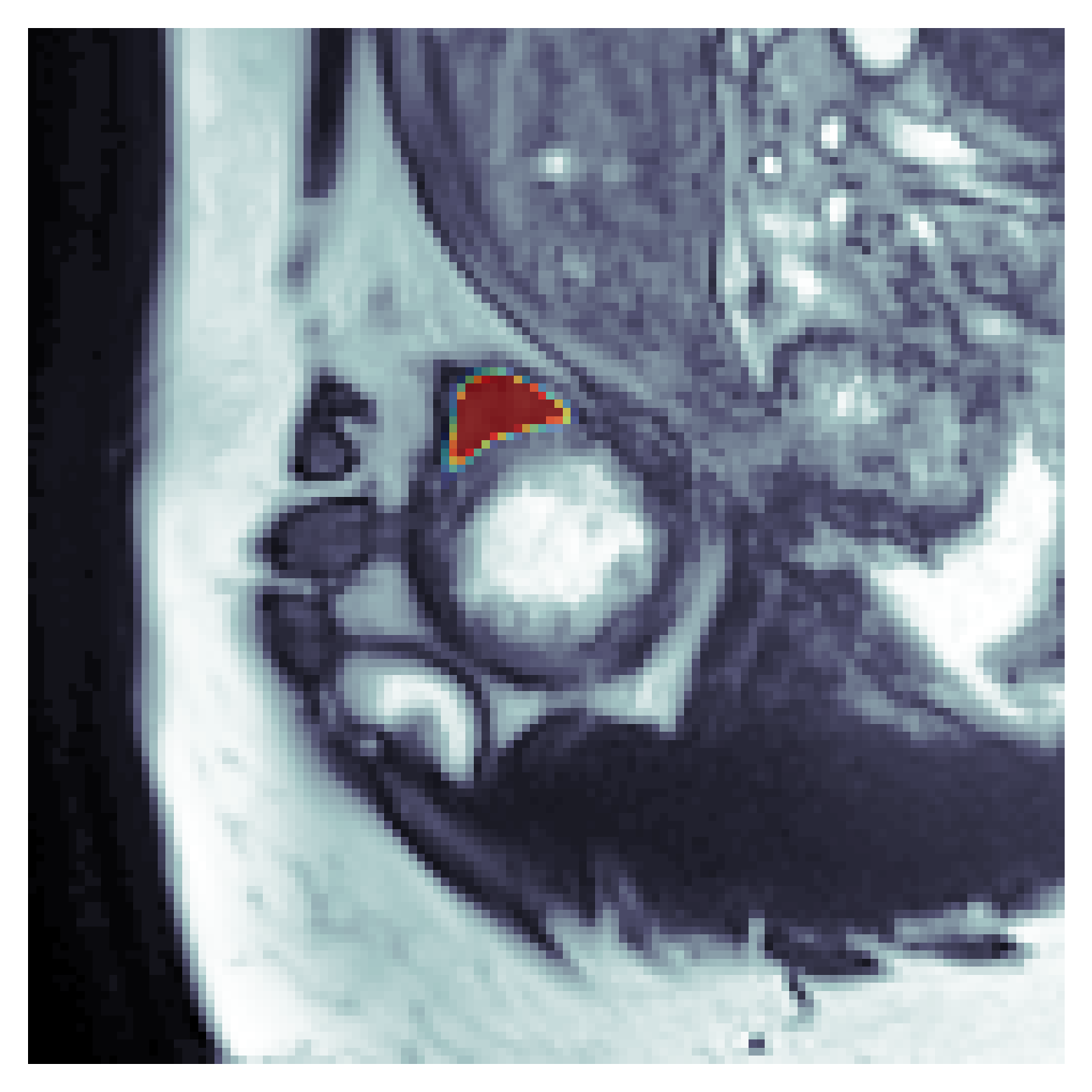} &   \includegraphics[width=35mm]{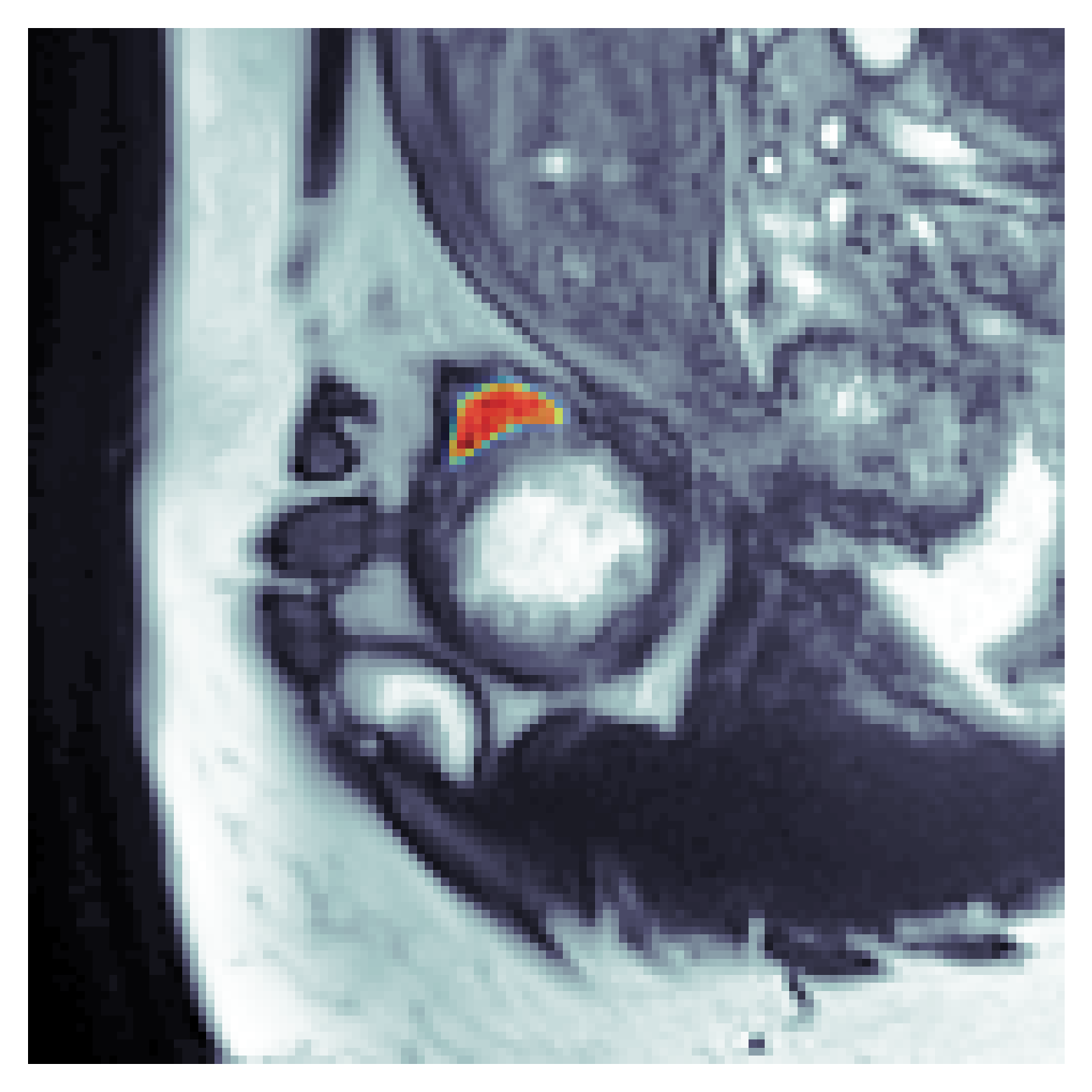} & \includegraphics[width=35mm]{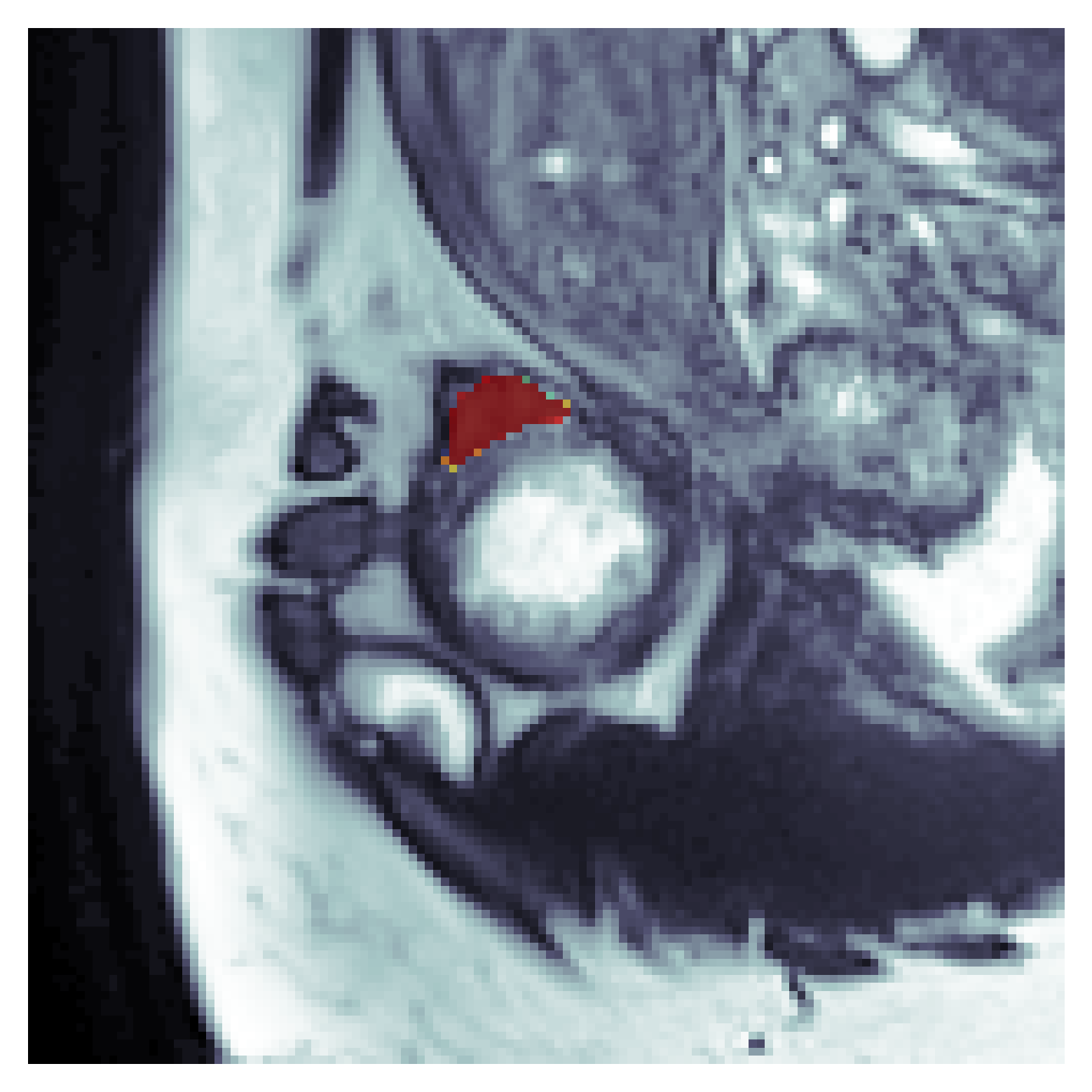} &   \includegraphics[width=35mm]{predictions/ACAC-MSELoss-inv_square_root_activation-resnet34-Unet.png} \\
  (i) MSE+HardTanh & (j) BCE+InvSquareRoot & (k) Dice+InvSquareRoot & (l) MSE+InvSquareRoot \\[6pt]

  \includegraphics[width=35mm]{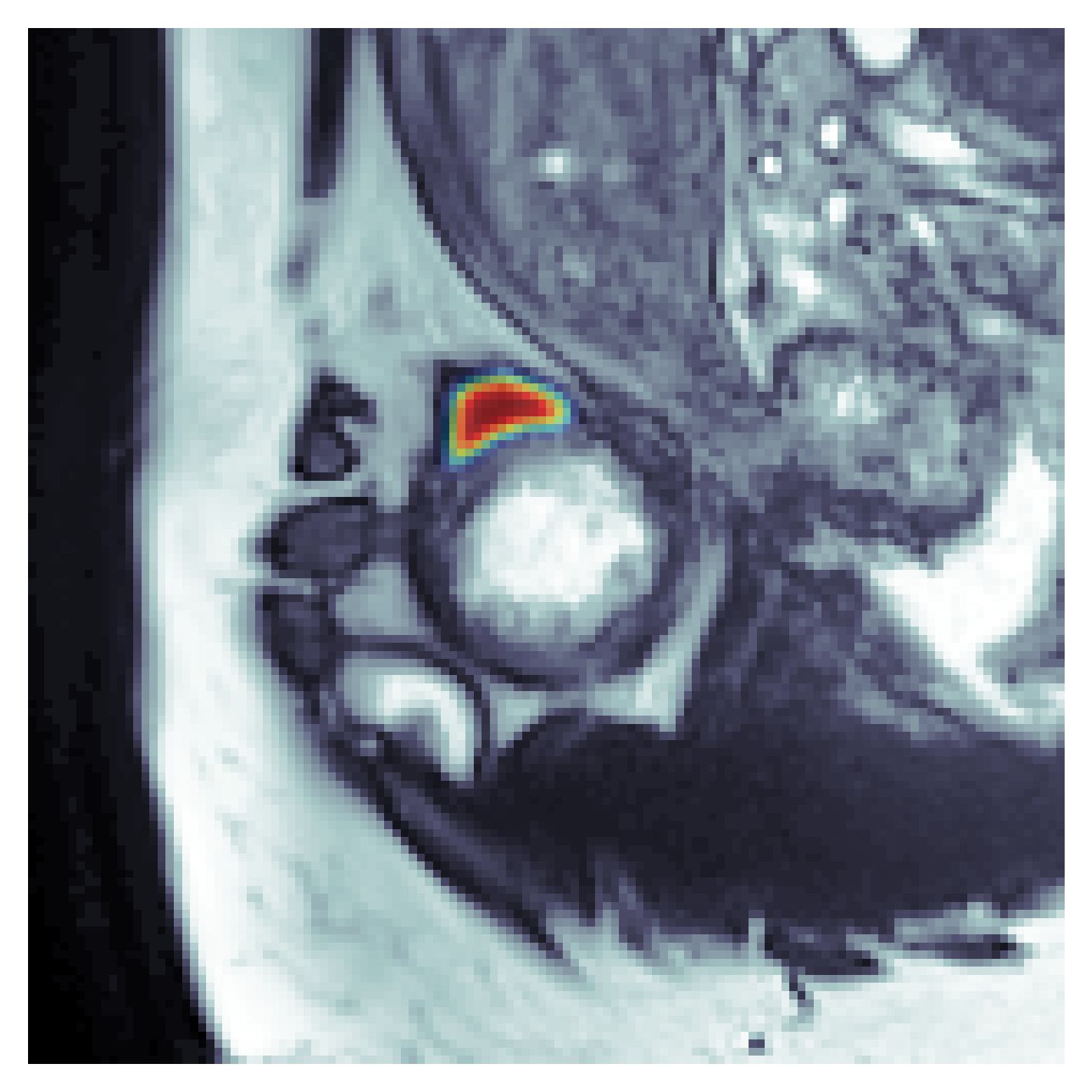} &   \includegraphics[width=35mm]{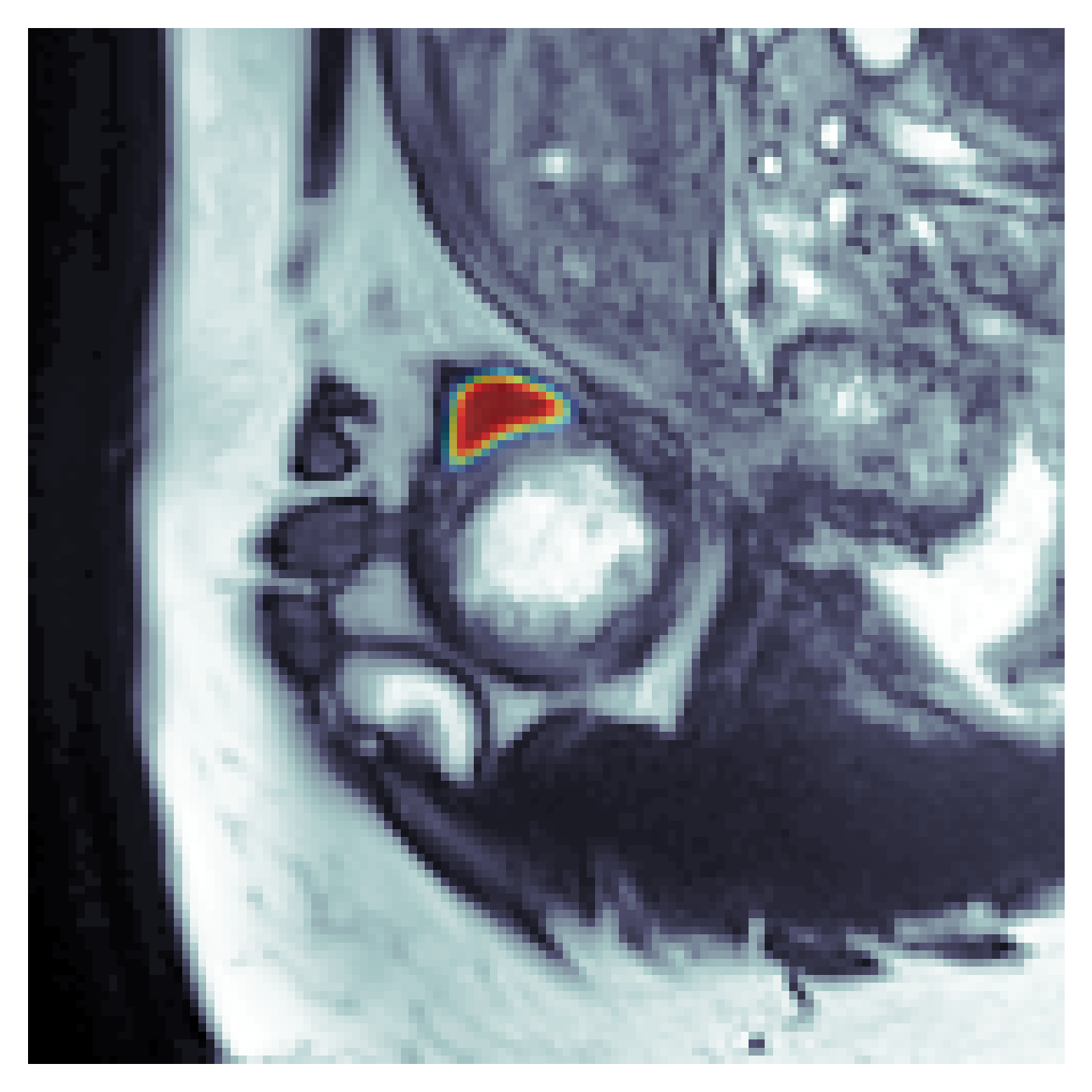} & \includegraphics[width=35mm]{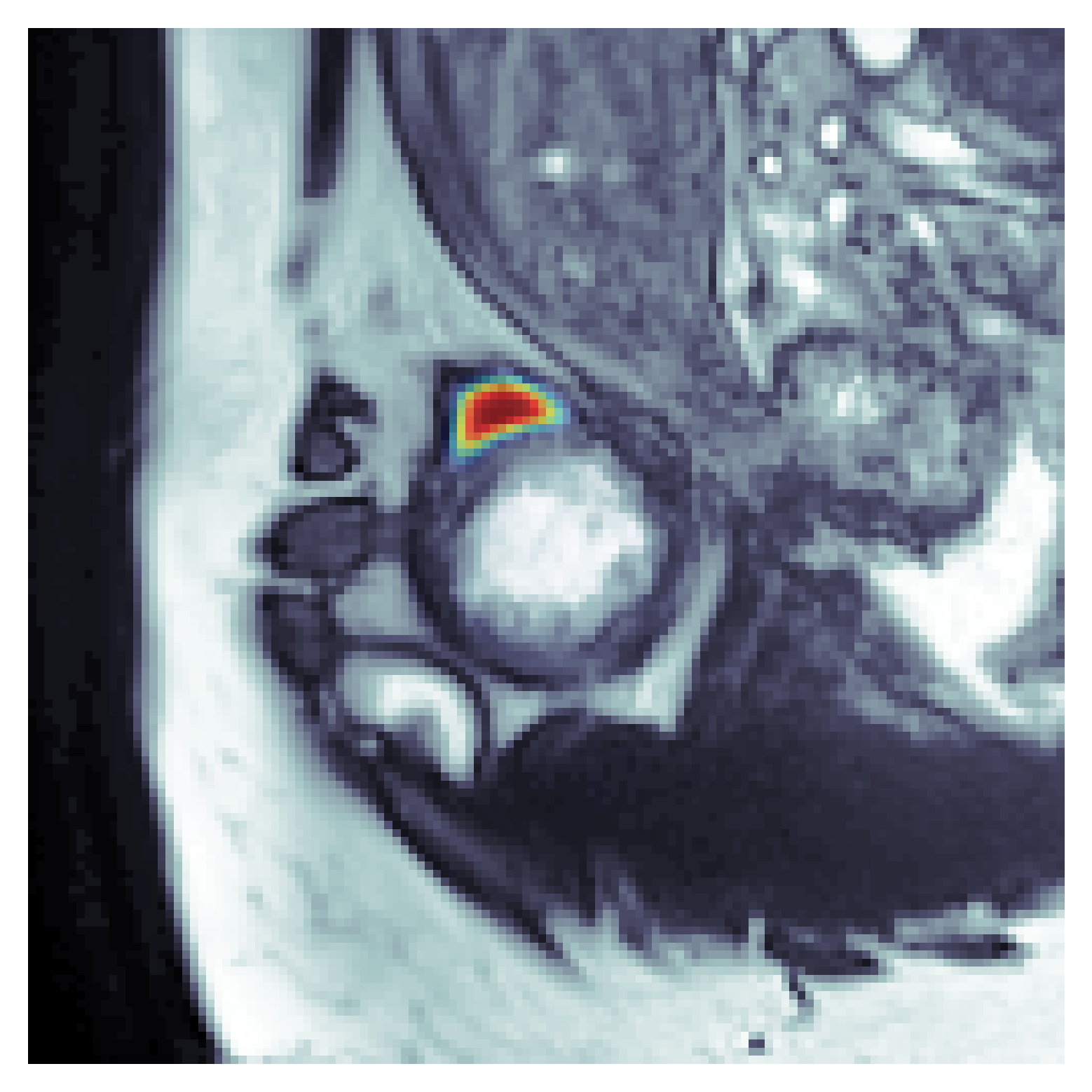} &   \includegraphics[width=35mm]{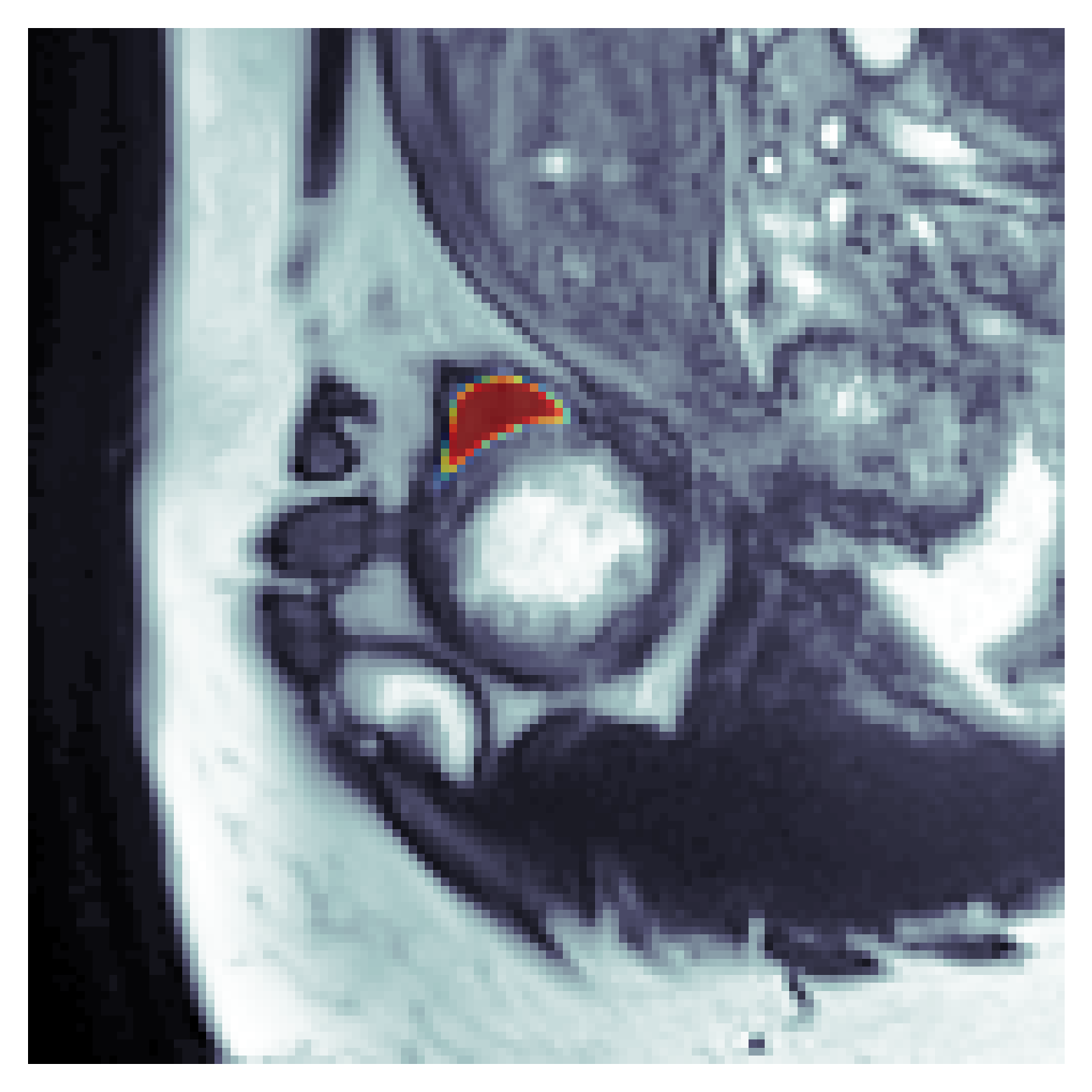} \\
  (m) BCE+Linear & (n) Dice+Linear & (o) MSE+Linear & (p) BCE+Sigmoid \\[6pt]

  \includegraphics[width=35mm]{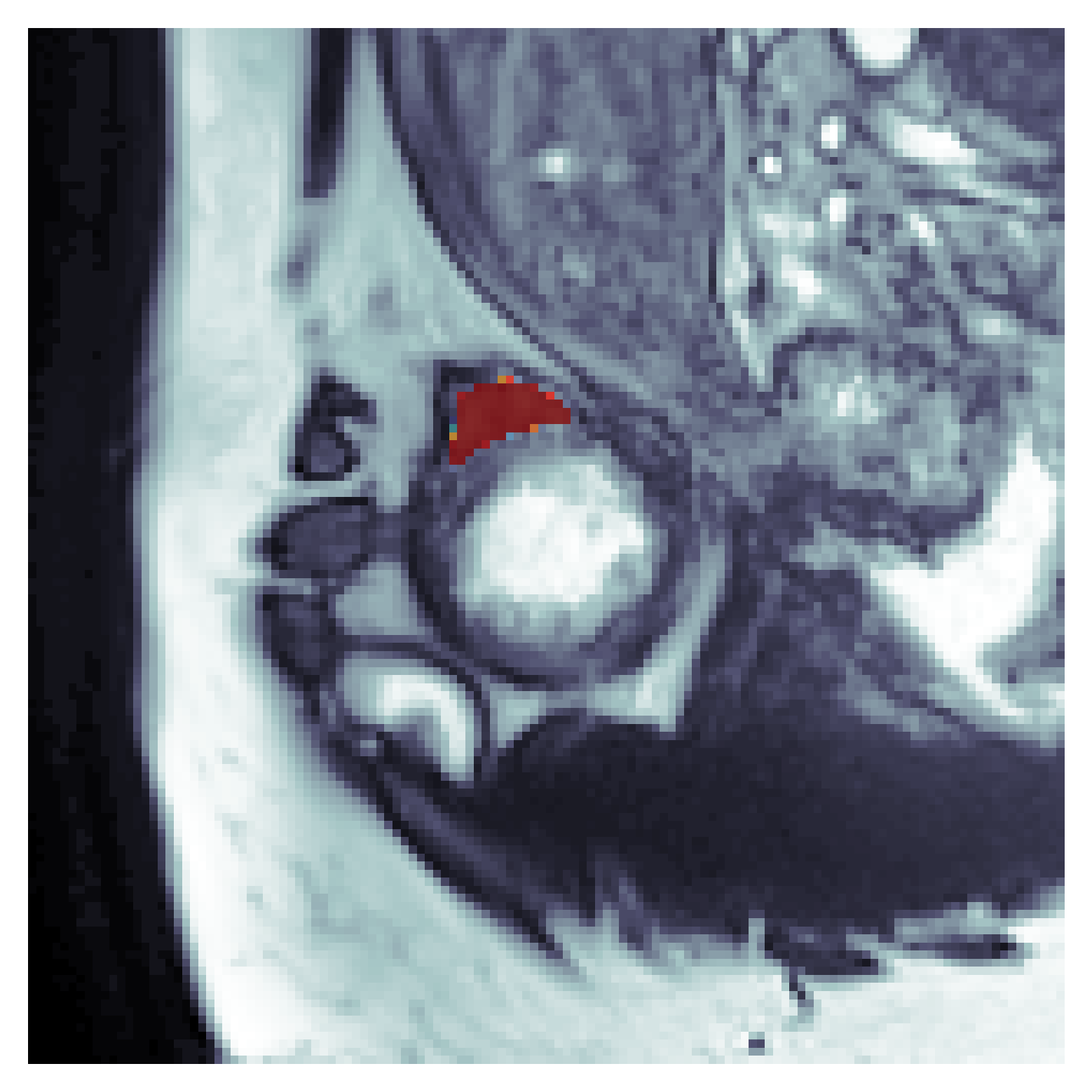} &   \includegraphics[width=35mm]{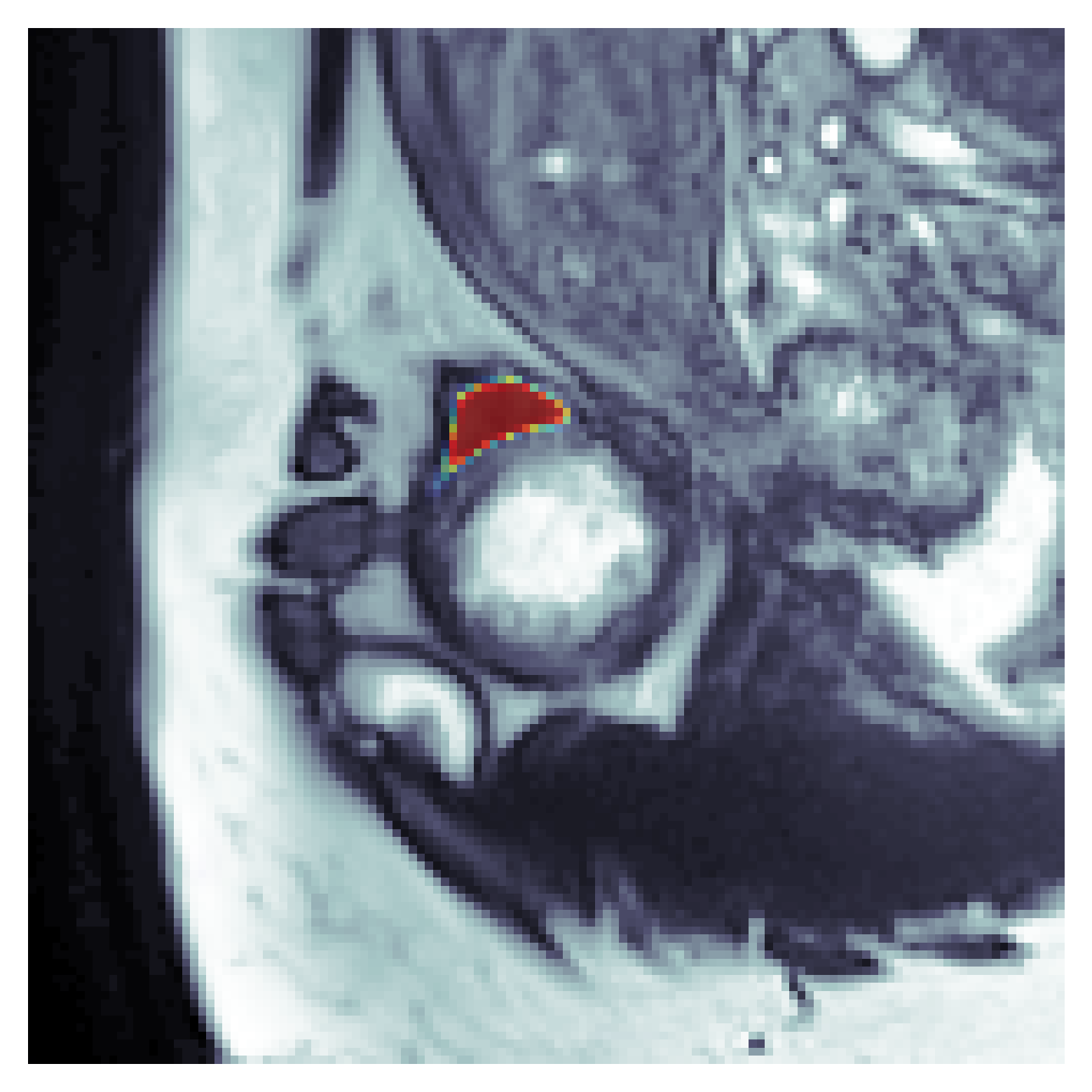} & \includegraphics[width=35mm]{predictions/ACAC-BCELoss-softsign_activation-resnet34-Unet.png} &   \includegraphics[width=35mm]{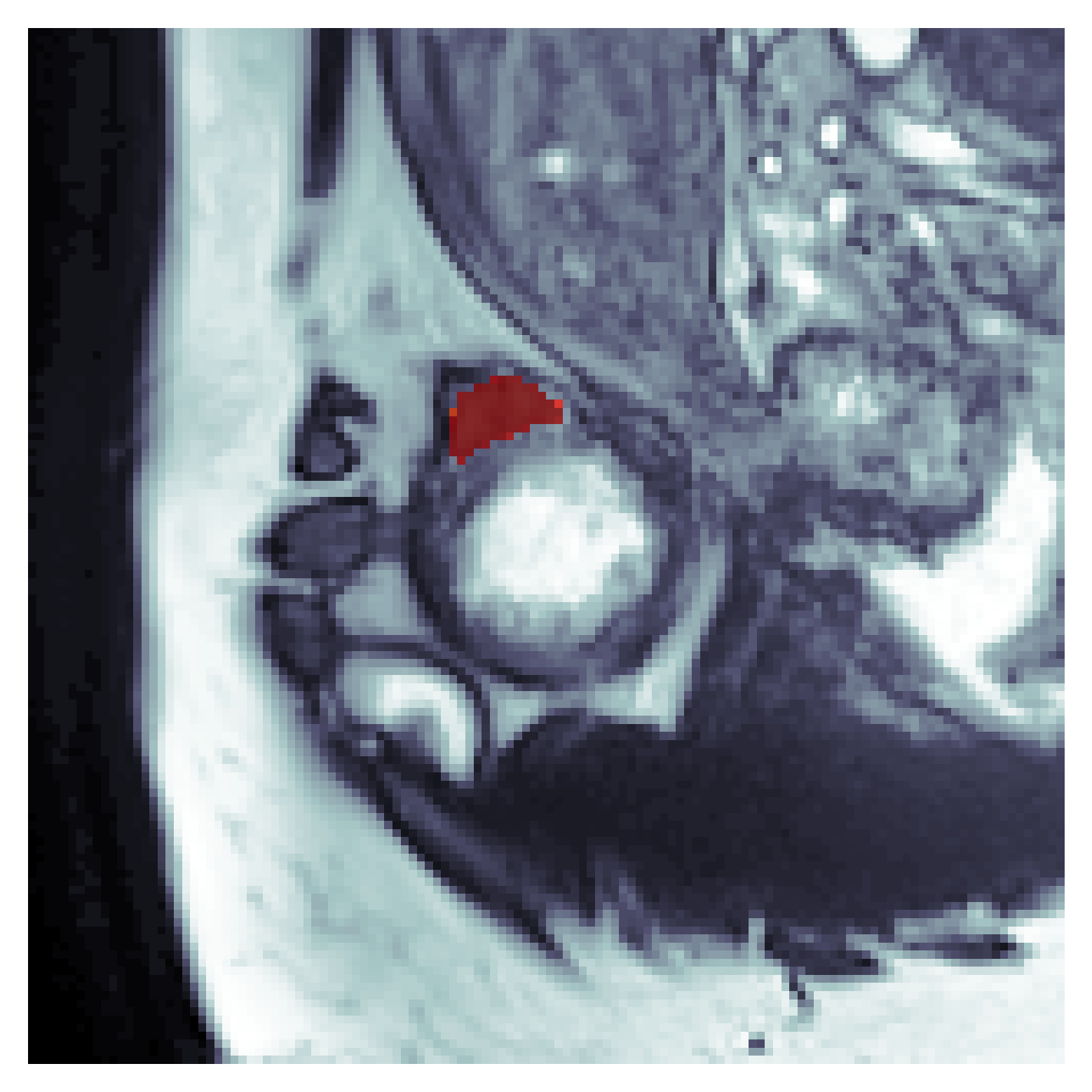} \\
  (q) Dice+Sigmoid & (r) MSE+Sigmoid & (s) BCE+Softsign & (t) Dice+Softsign \\[6pt]
  \includegraphics[width=35mm]{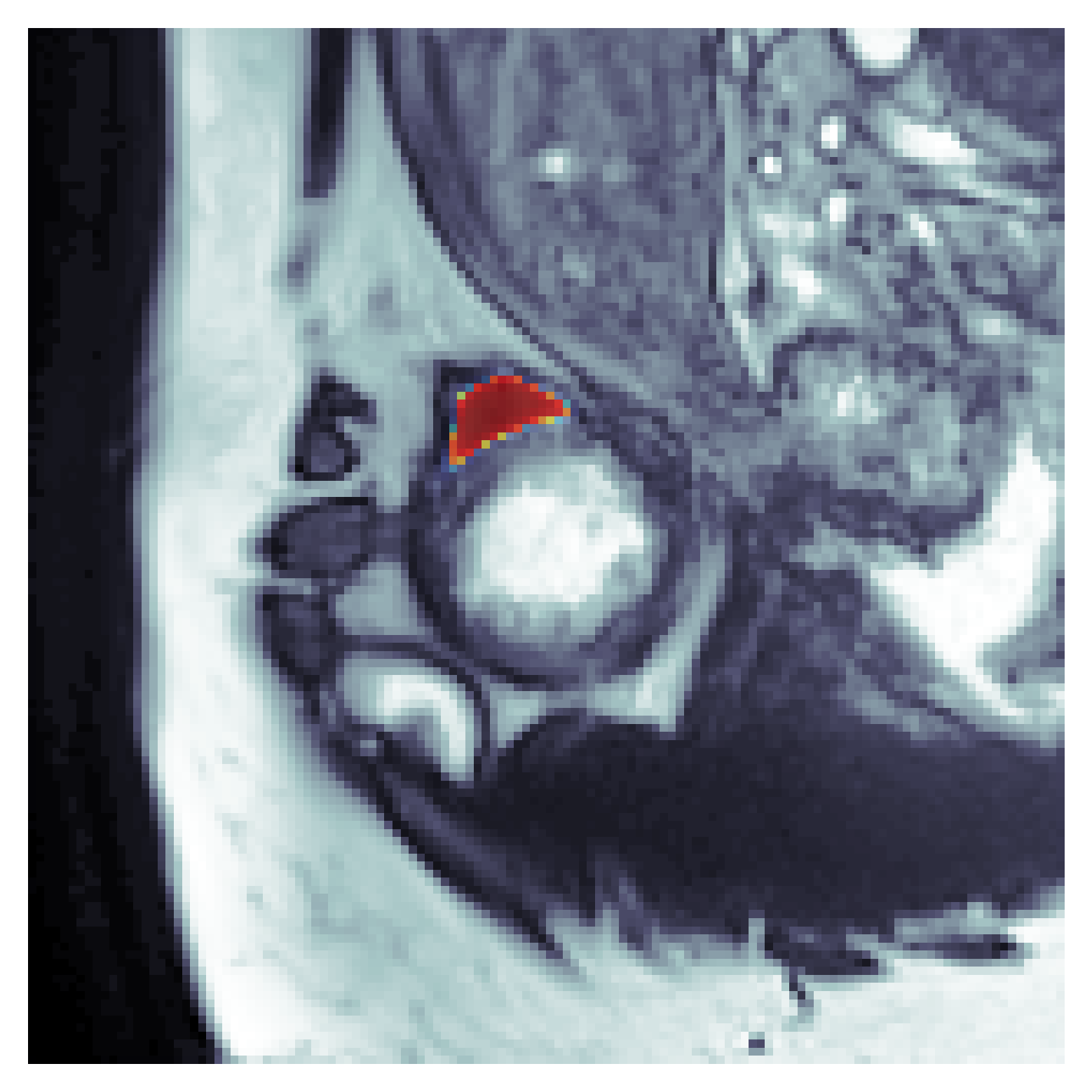} & \includegraphics[width=35mm] {predictions/ACAC_gt.png} \\ (u) MSE+Softsign & (v) Ground truth
\end{tabular}}
\label{fig:allpred4}
\caption{All $21$ predictions for a single image of the ACDC dataset, with the last image being the ground truth.}
\end{figure}

\subsection{Ischemic Stroke Lesion Segmentation Challenge 2018 (ISLES)}

\begin{figure}[H]
\centering
\scalebox{.77}{\begin{tabular}{cccc}
  \includegraphics[width=35mm]{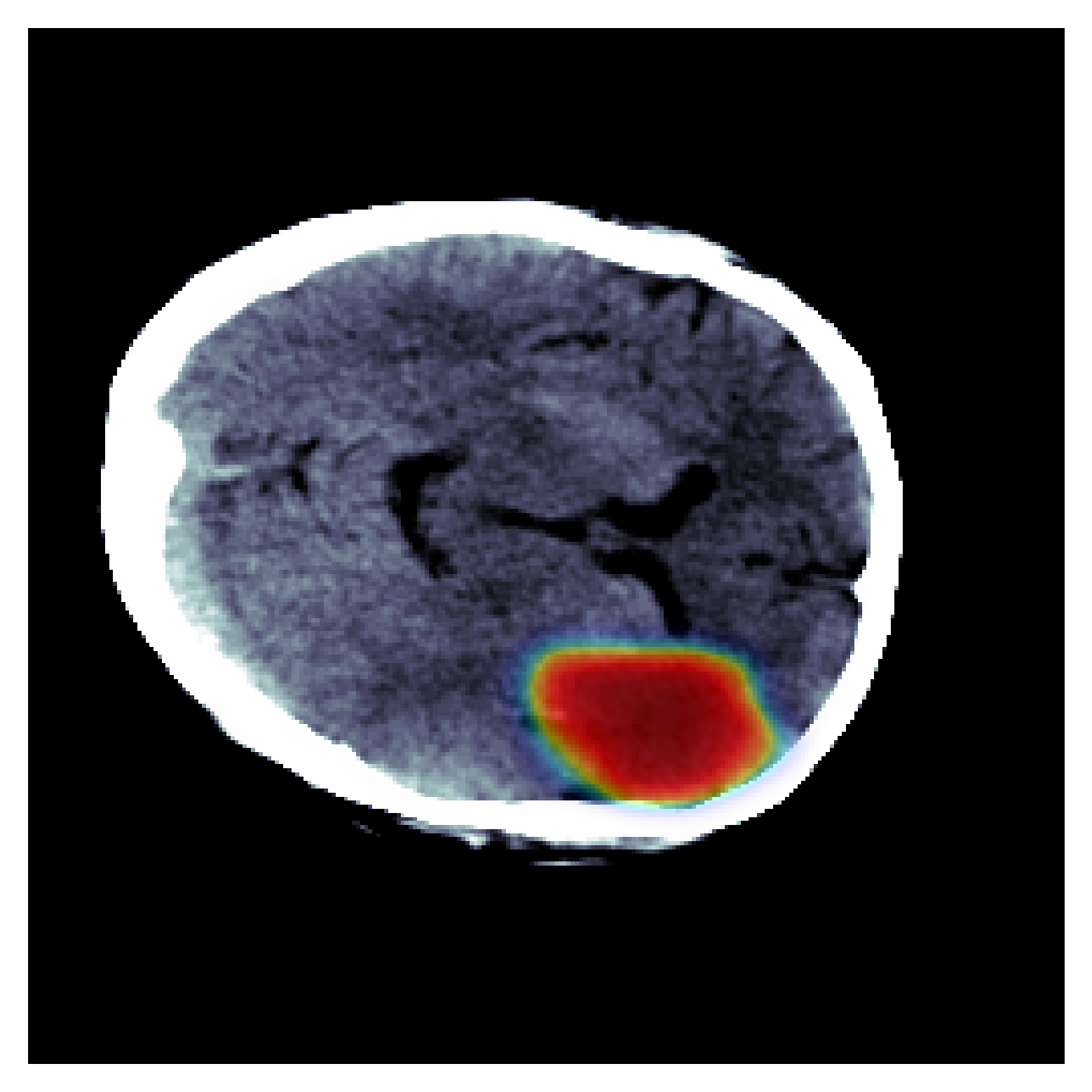} &   \includegraphics[width=35mm]{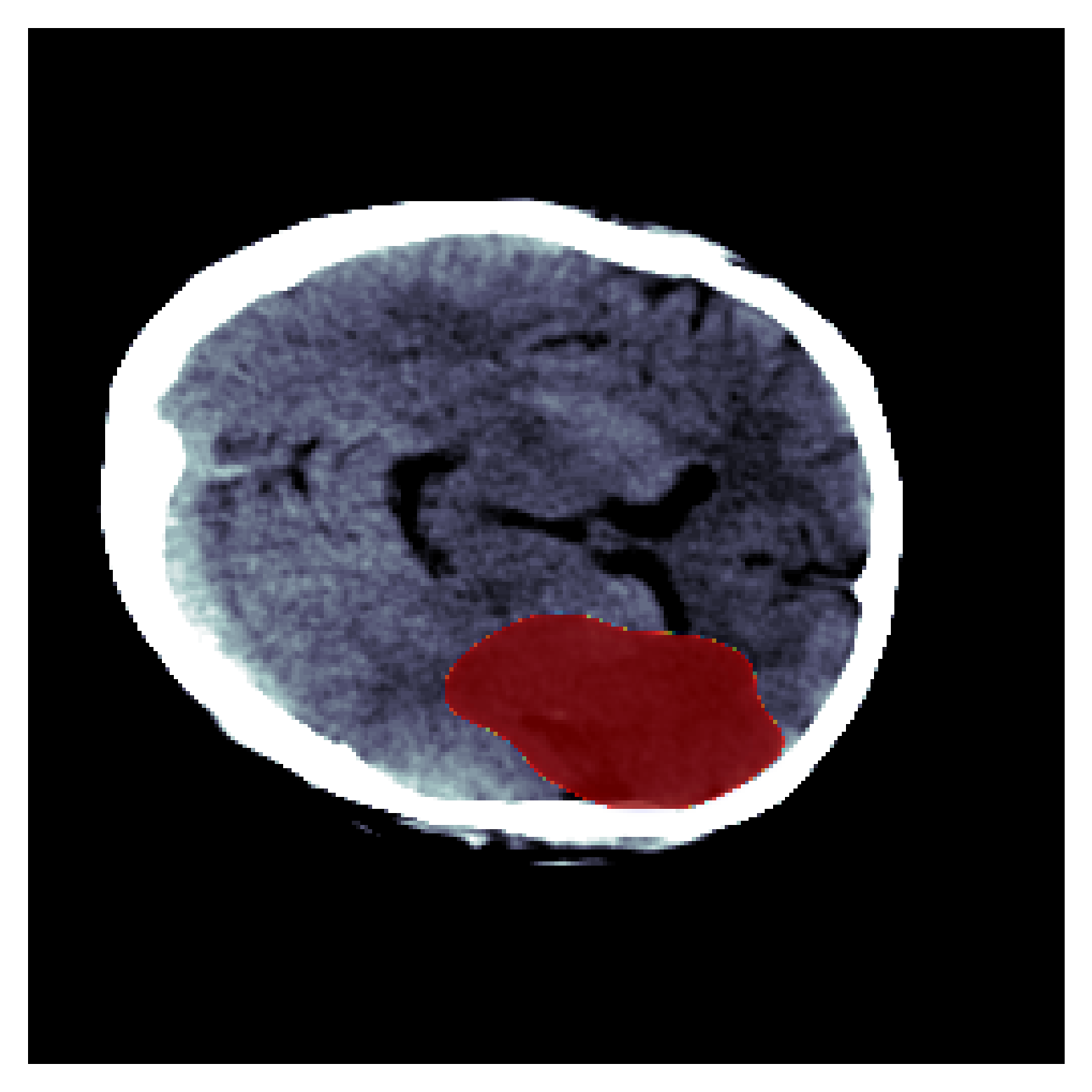} & \includegraphics[width=35mm]{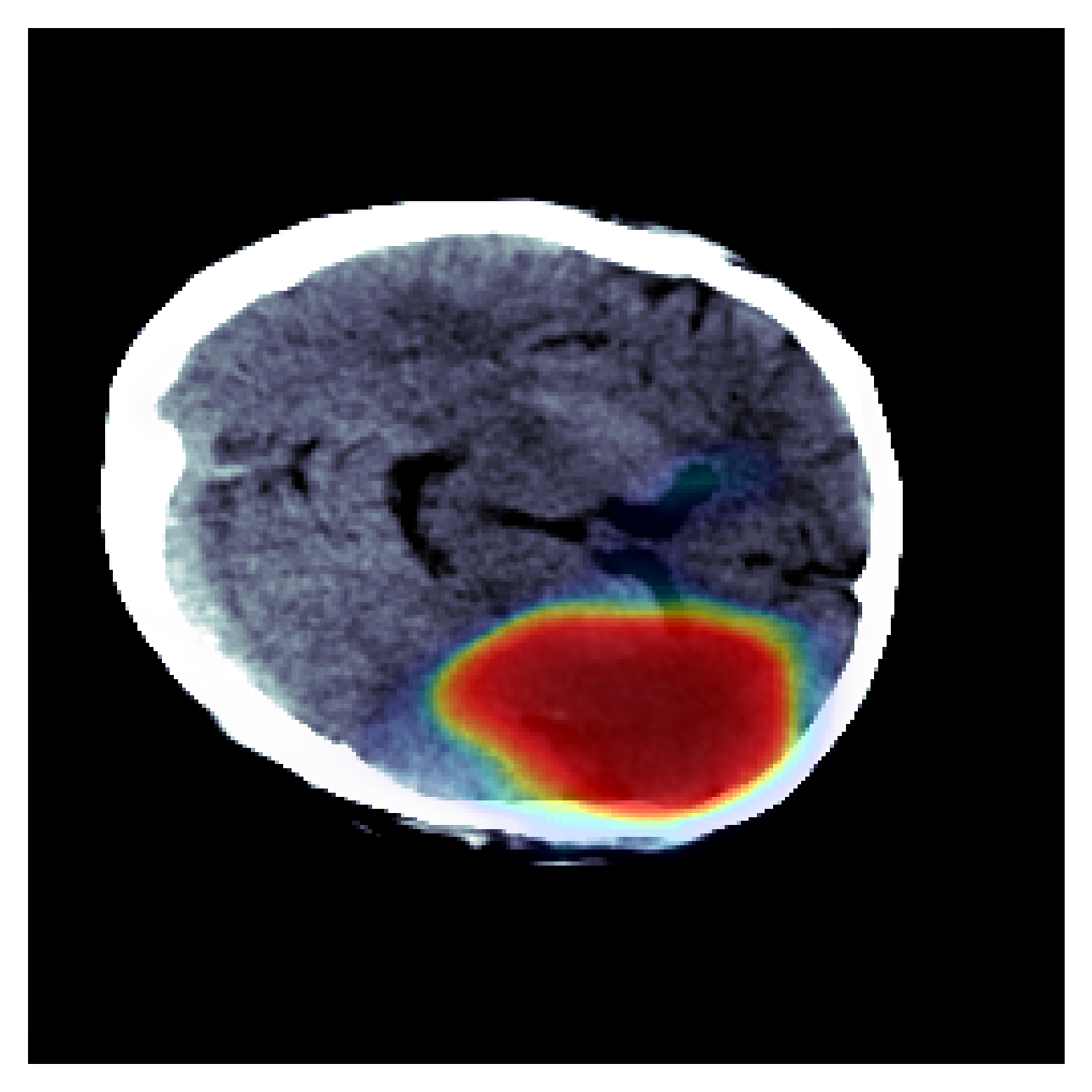} &   \includegraphics[width=35mm]{predictions/ISLES-BCELoss-cdf_activation-resnet34-Unet.png} \\
  (a) BCE+Arctangent & (b) Dice+Arctangent & (c) MSE+Arctangent & (d) BCE+CDF \\[6pt]

  \includegraphics[width=35mm]{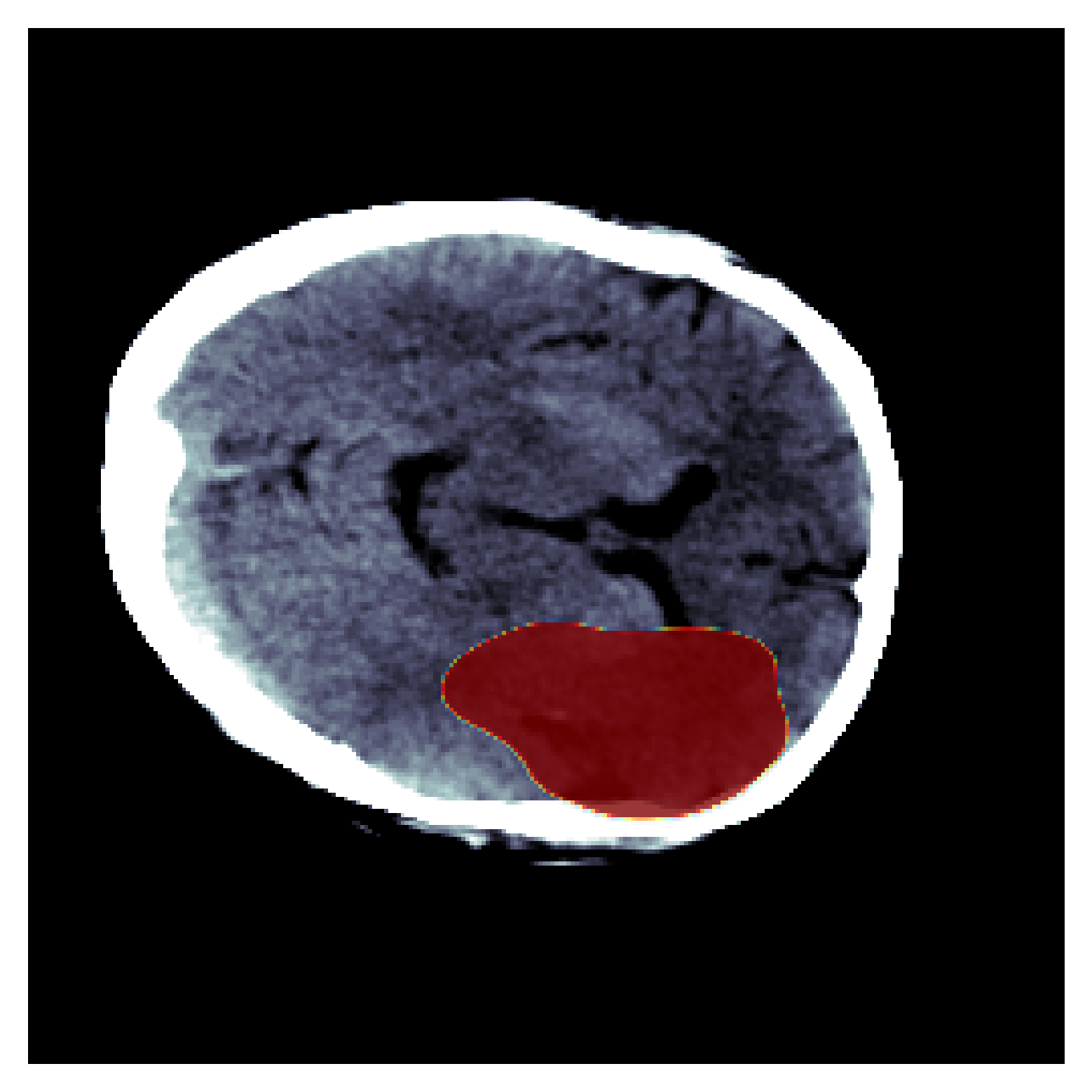} &   \includegraphics[width=35mm]{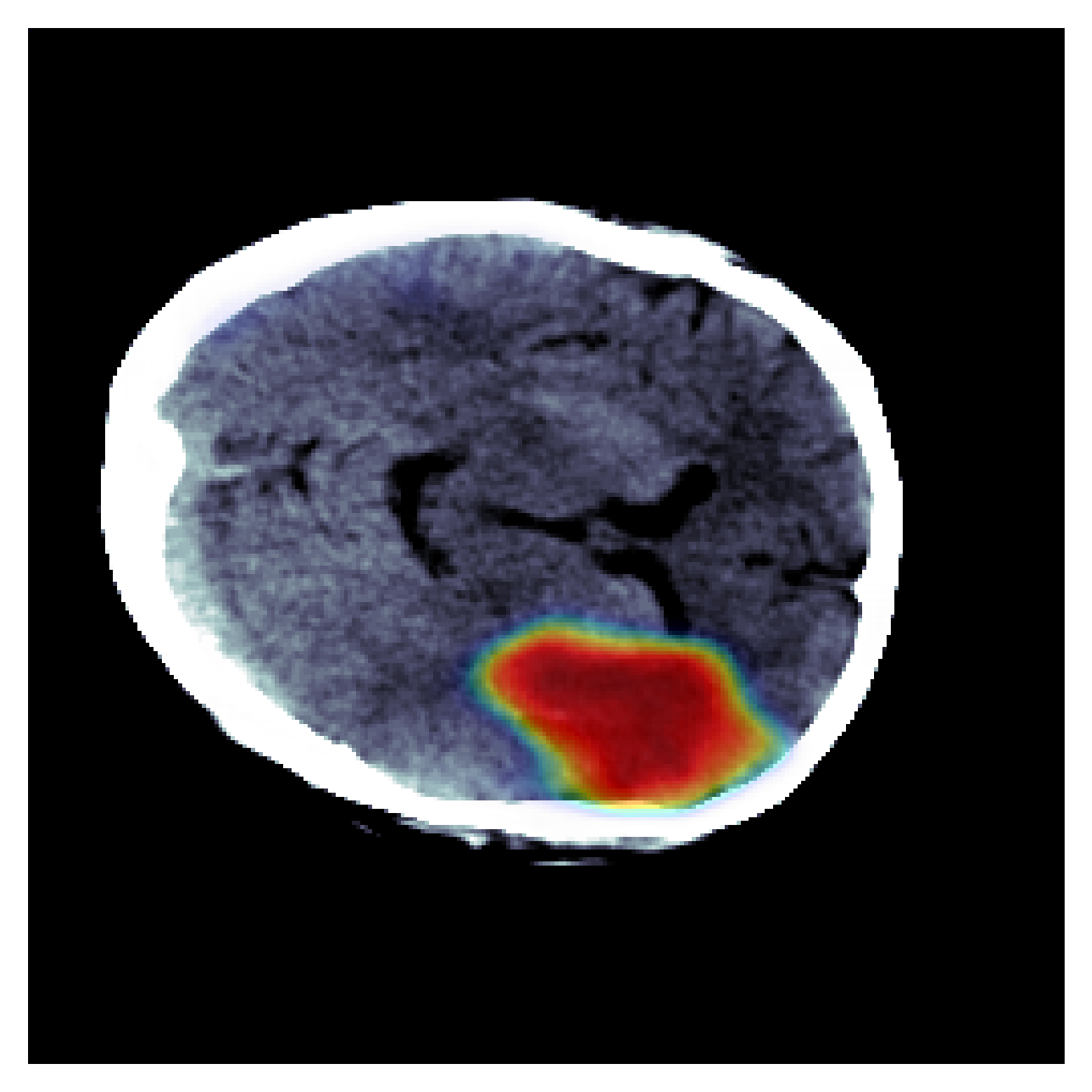} & \includegraphics[width=35mm]{predictions/ISLES-BCELoss-hardtanh_activation-resnet34-Unet.png} &   \includegraphics[width=35mm]{predictions/ISLES-DiceLoss-hardtanh_activation-resnet34-Unet.png} \\
  (e) Dice+CDF & (f) MSE+CDF & (g) BCE+HardTanh & (h) Dice+HardTanh \\[6pt]

  \includegraphics[width=35mm]{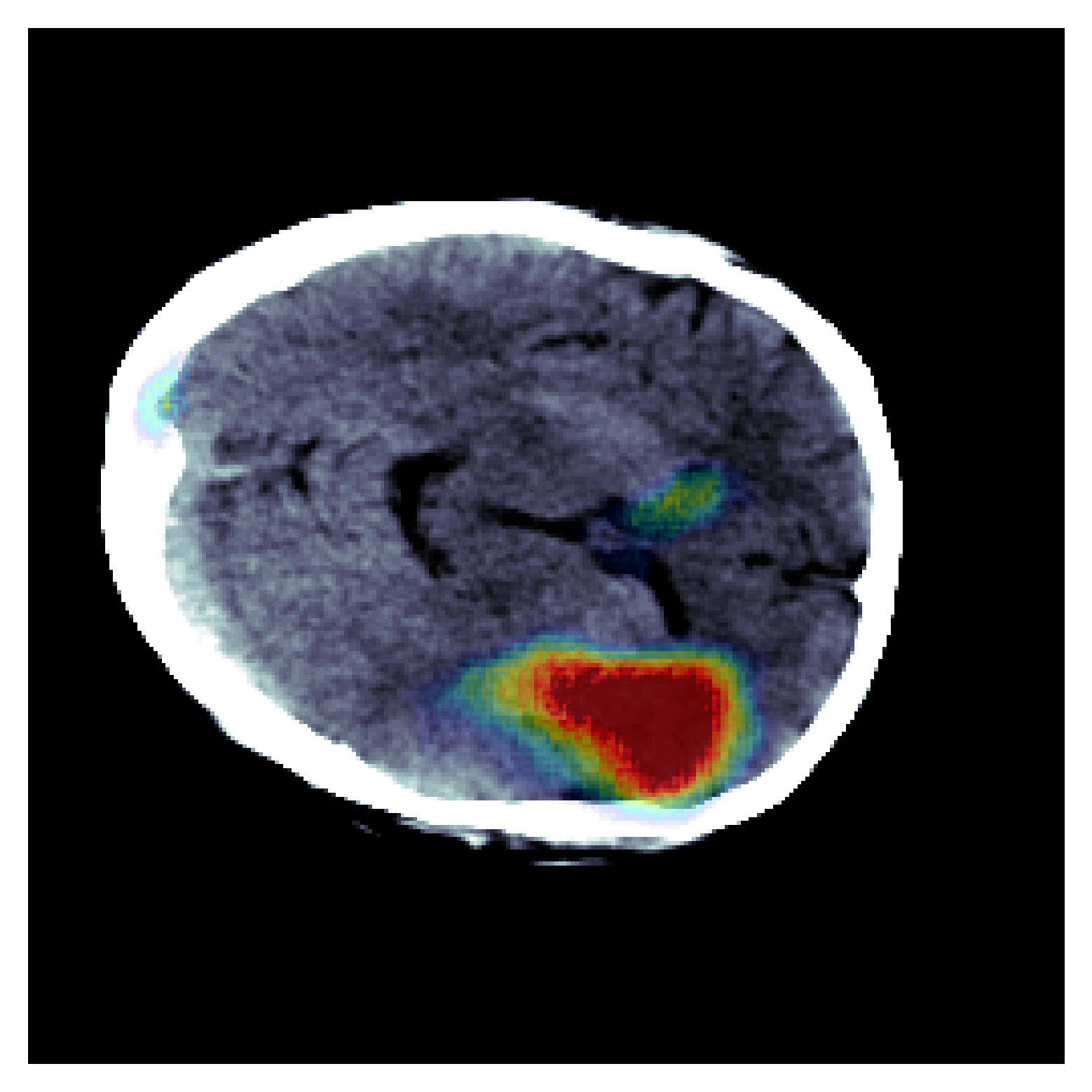} &   \includegraphics[width=35mm]{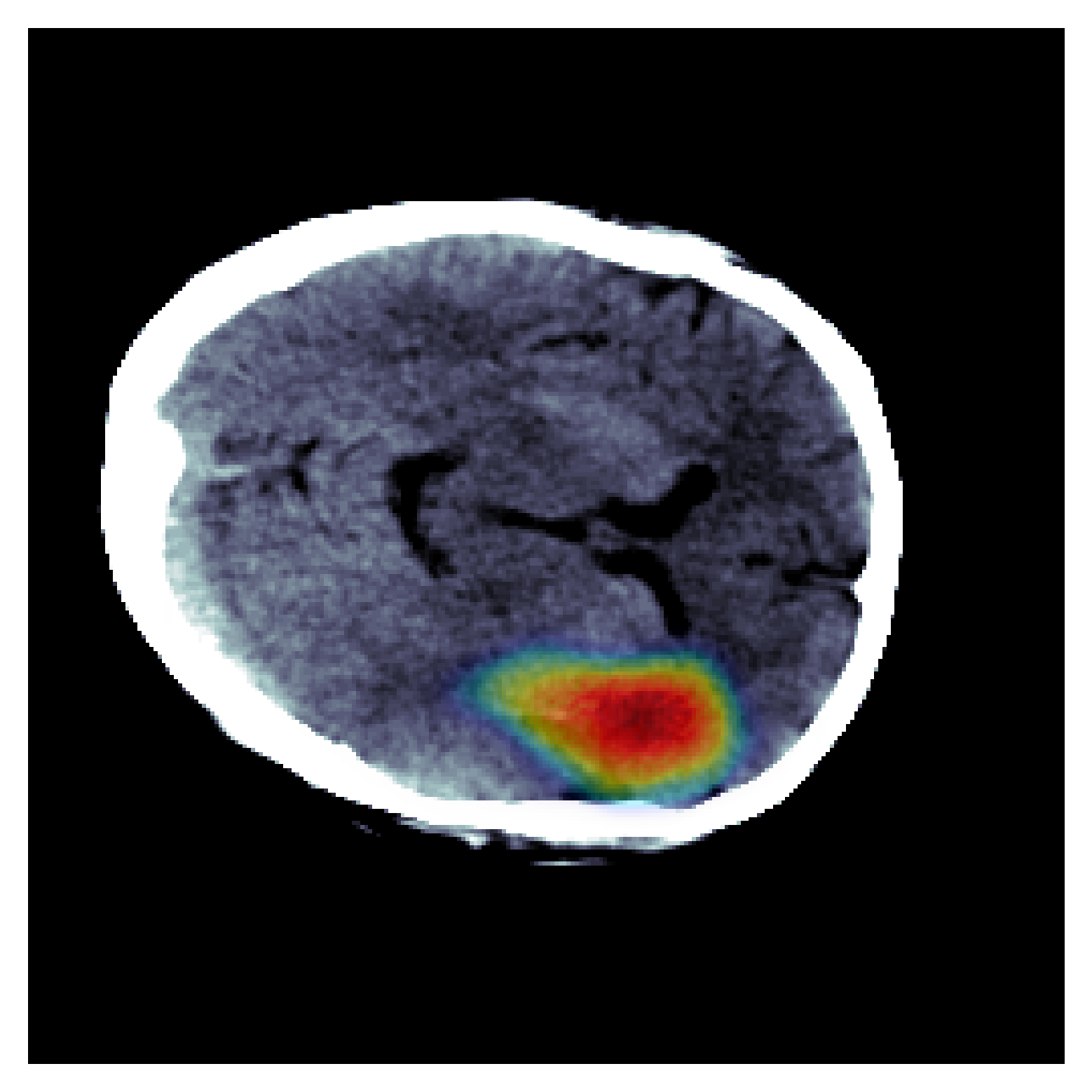} & \includegraphics[width=35mm]{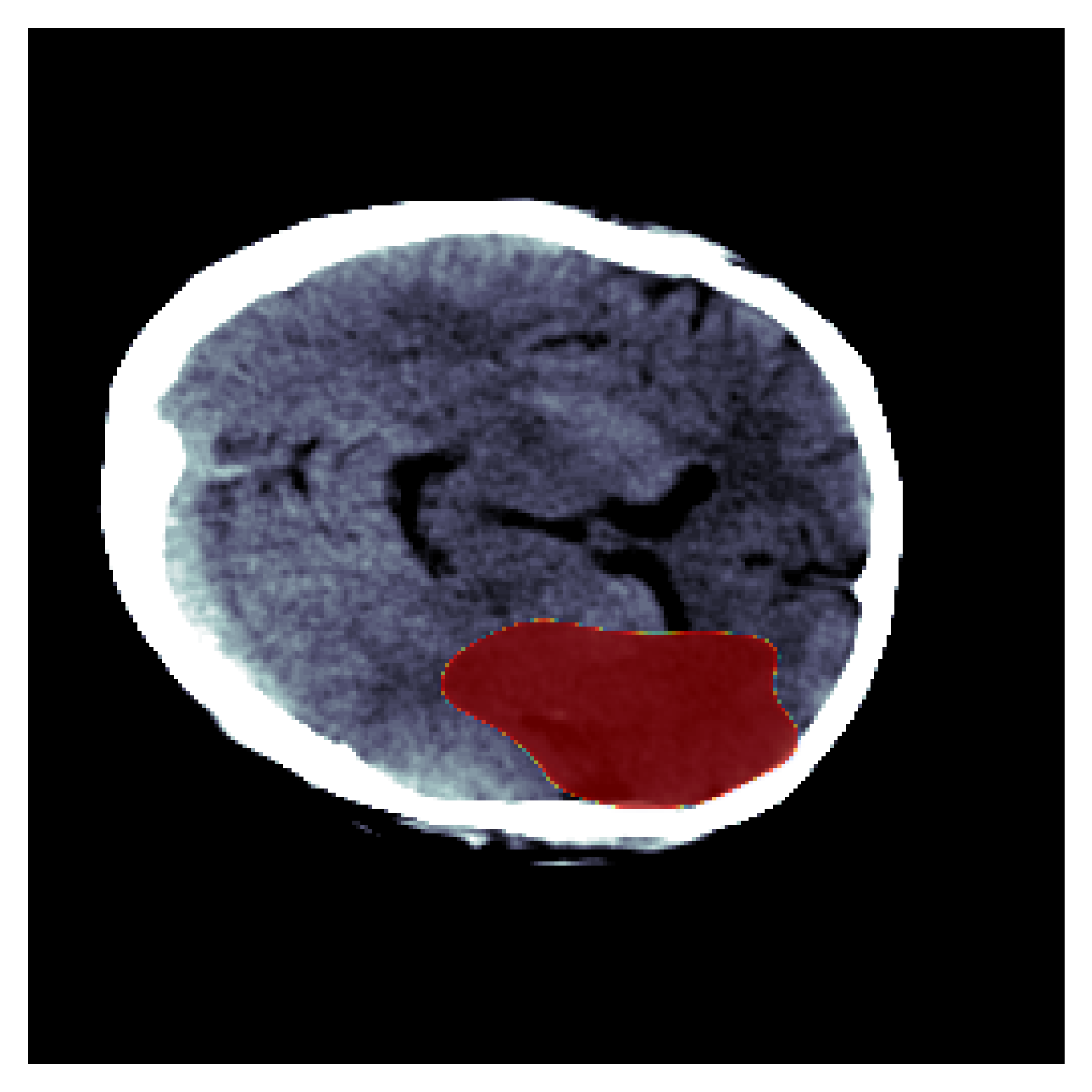} &   \includegraphics[width=35mm]{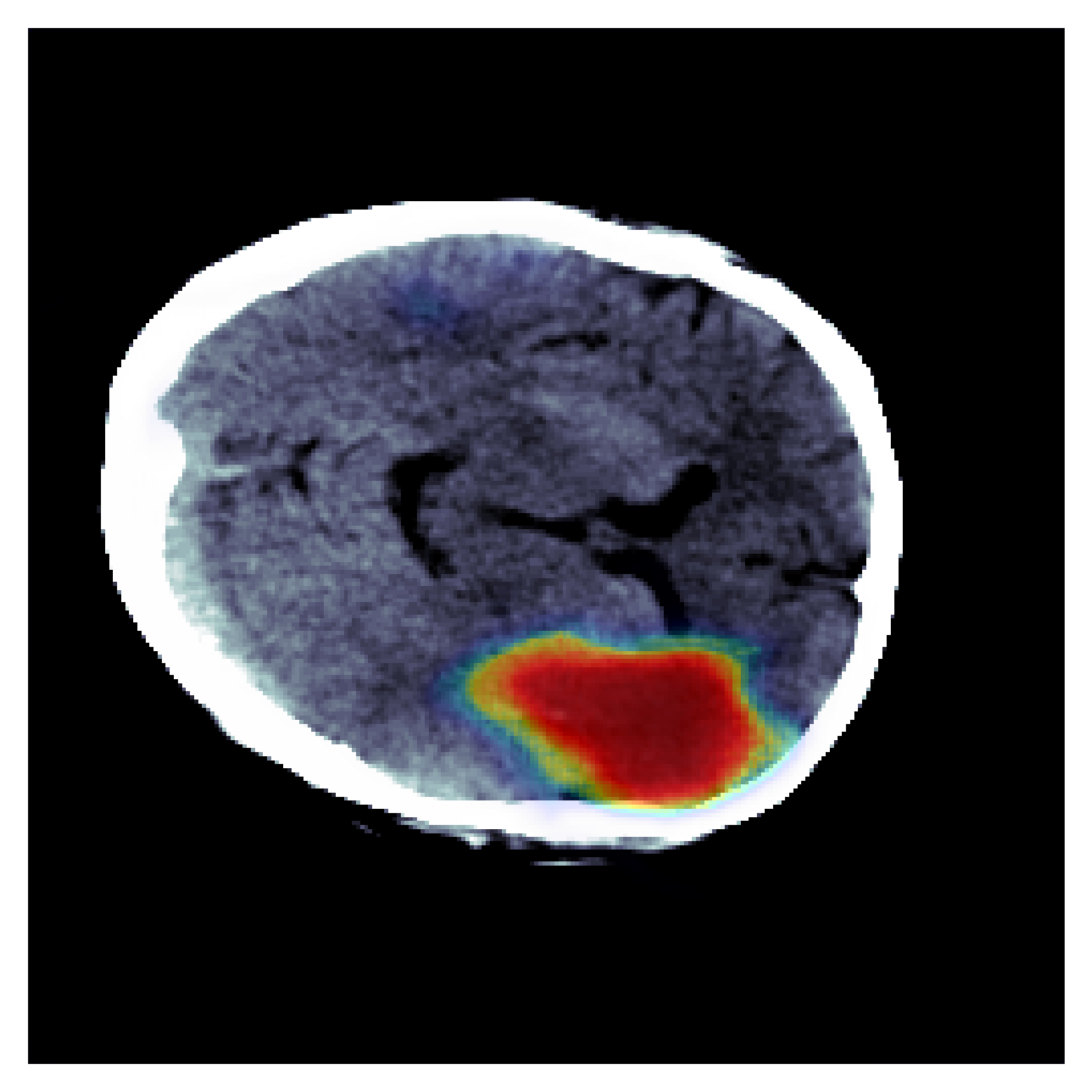} \\
  (i) MSE+HardTanh & (j) BCE+InvSquareRoot & (k) Dice+InvSquareRoot & (l) MSE+InvSquareRoot \\[6pt]

  \includegraphics[width=35mm]{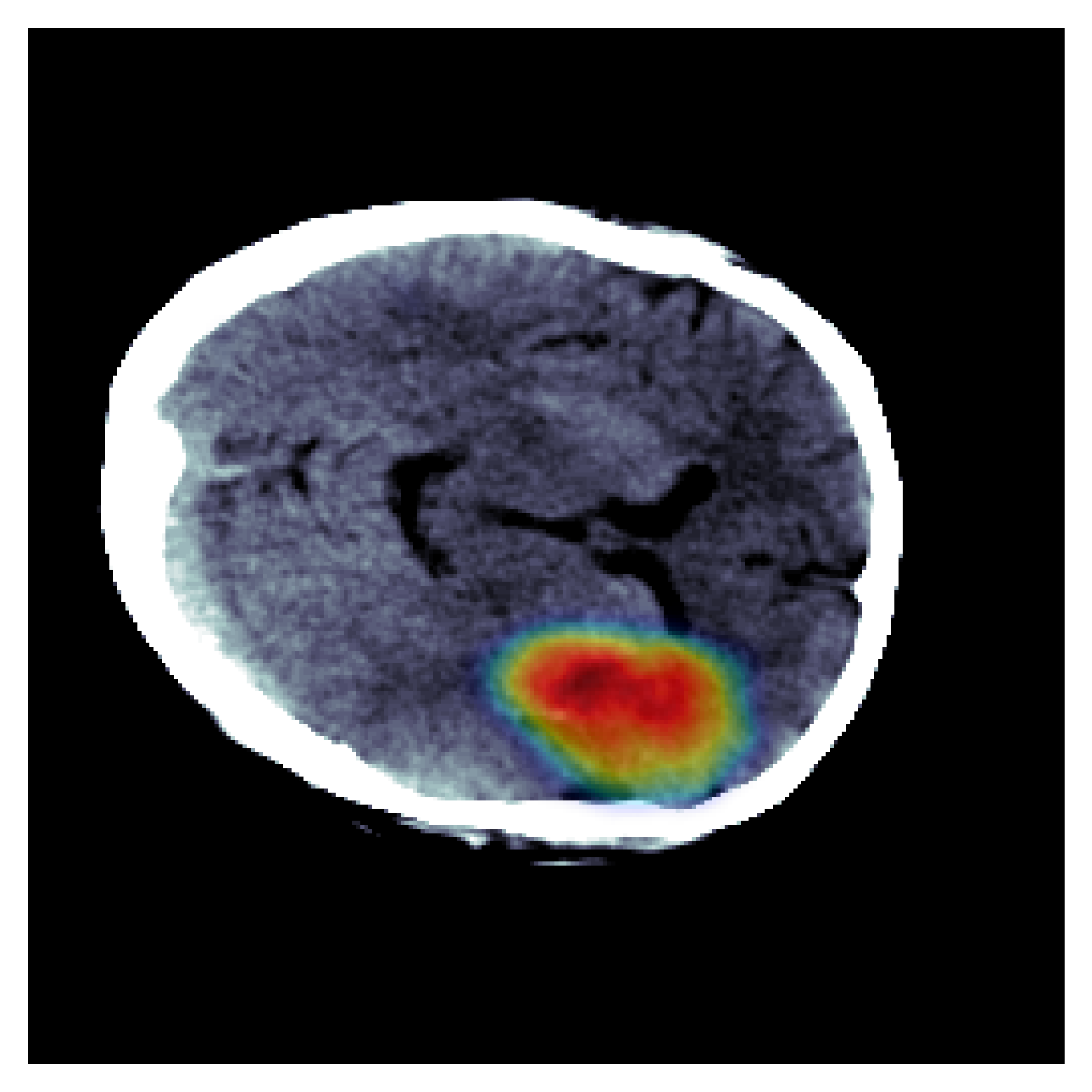} &   \includegraphics[width=35mm]{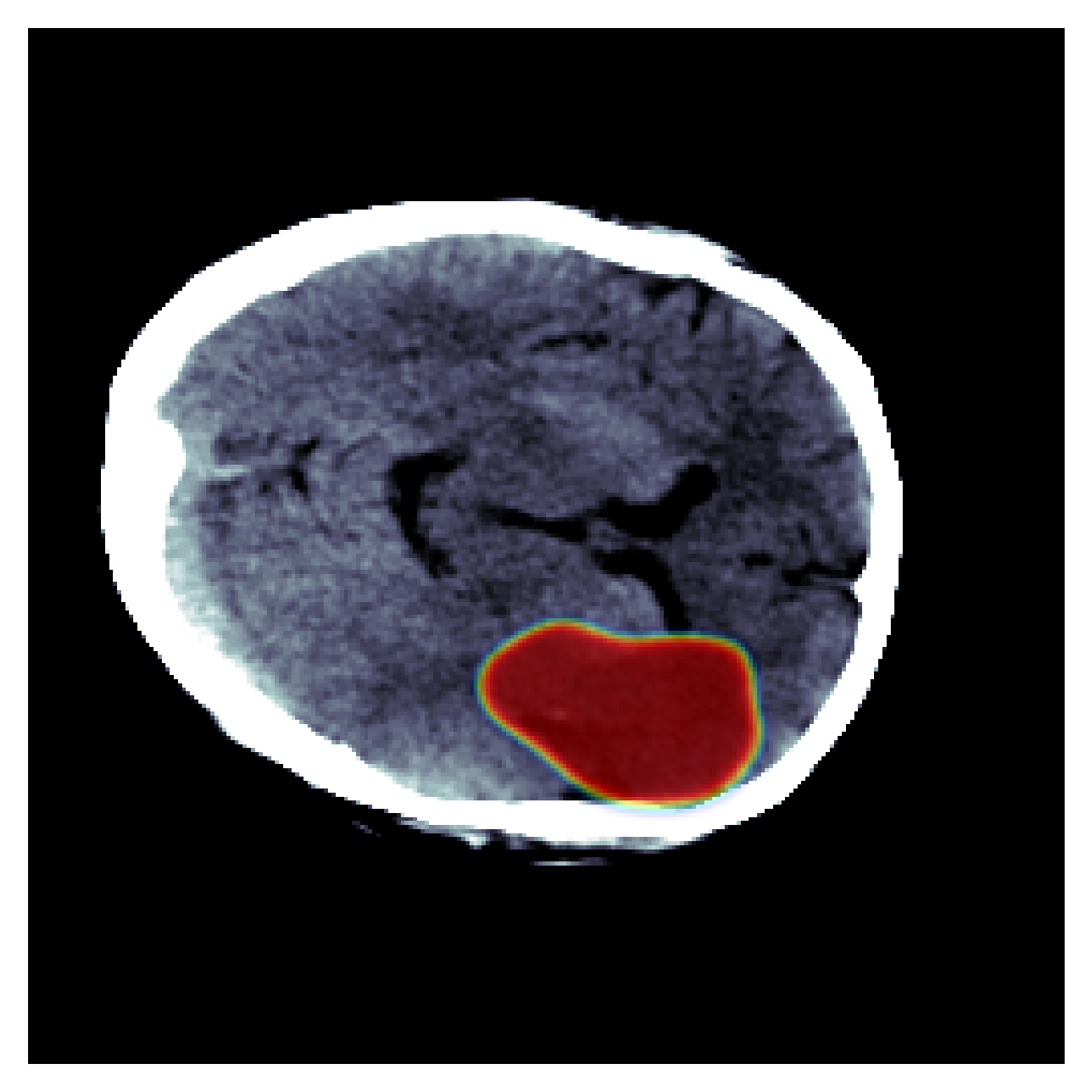} & \includegraphics[width=35mm]{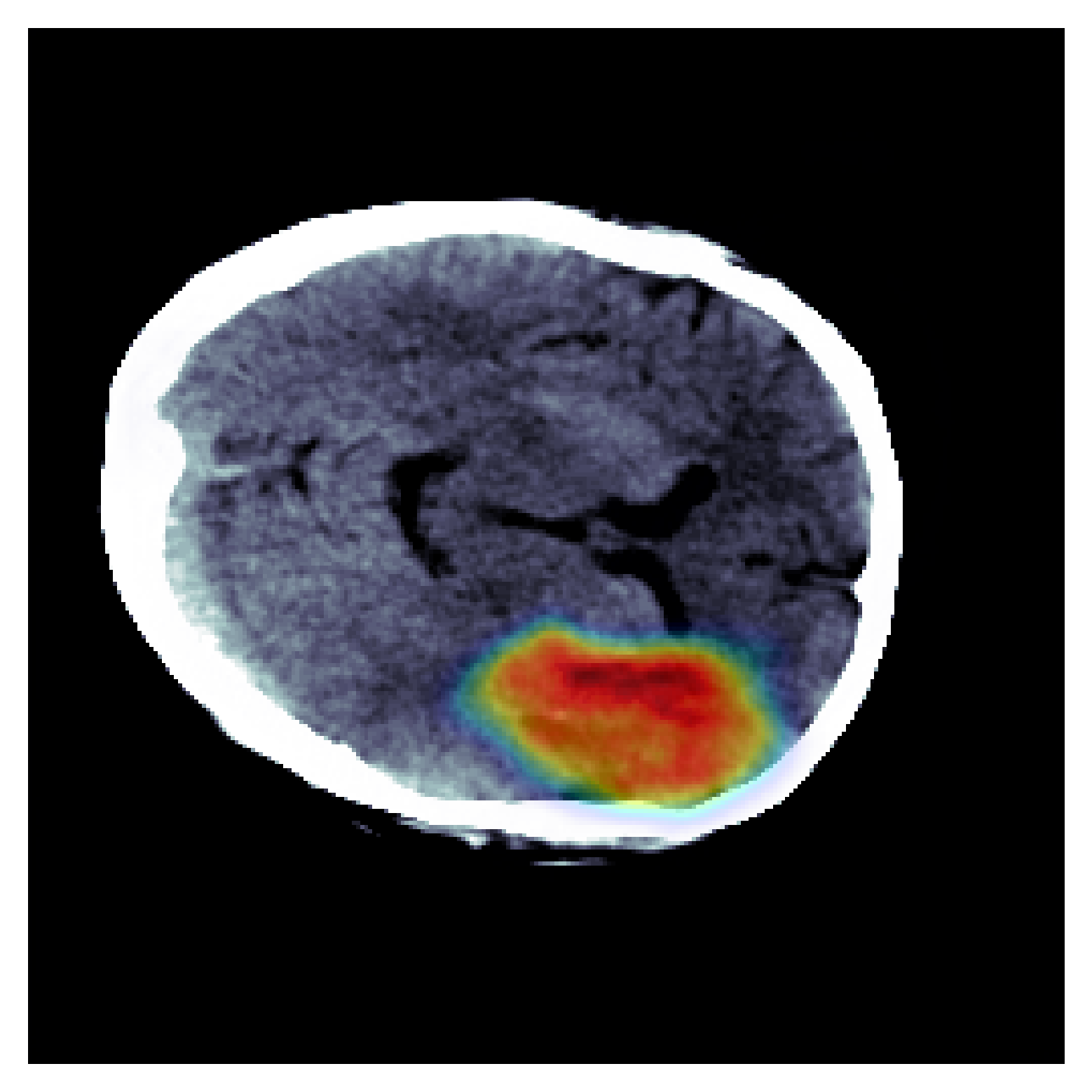} &   \includegraphics[width=35mm]{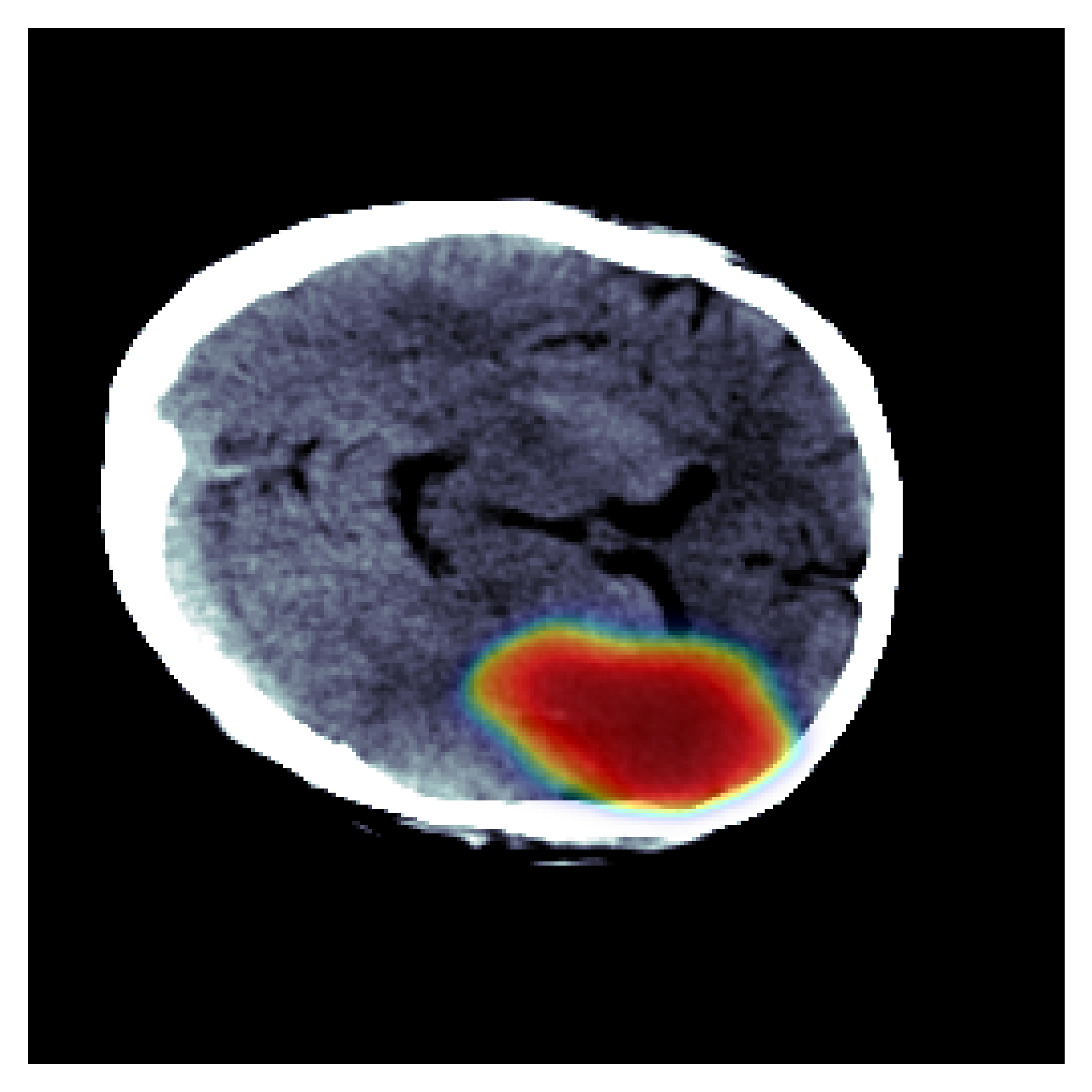} \\
  (m) BCE+Linear & (n) Dice+Linear & (o) MSE+Linear & (p) BCE+Sigmoid \\[6pt]

  \includegraphics[width=35mm]{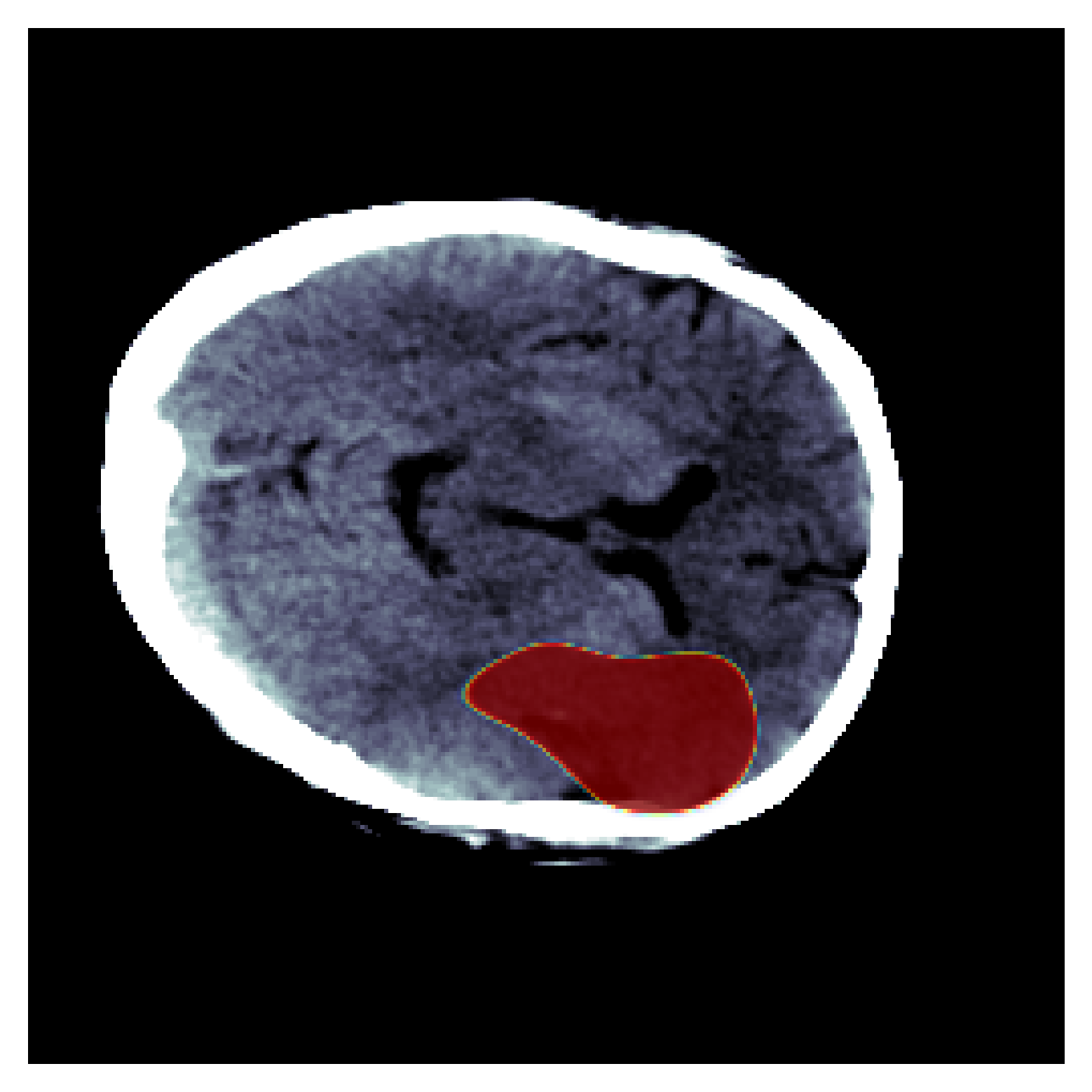} &   \includegraphics[width=35mm]{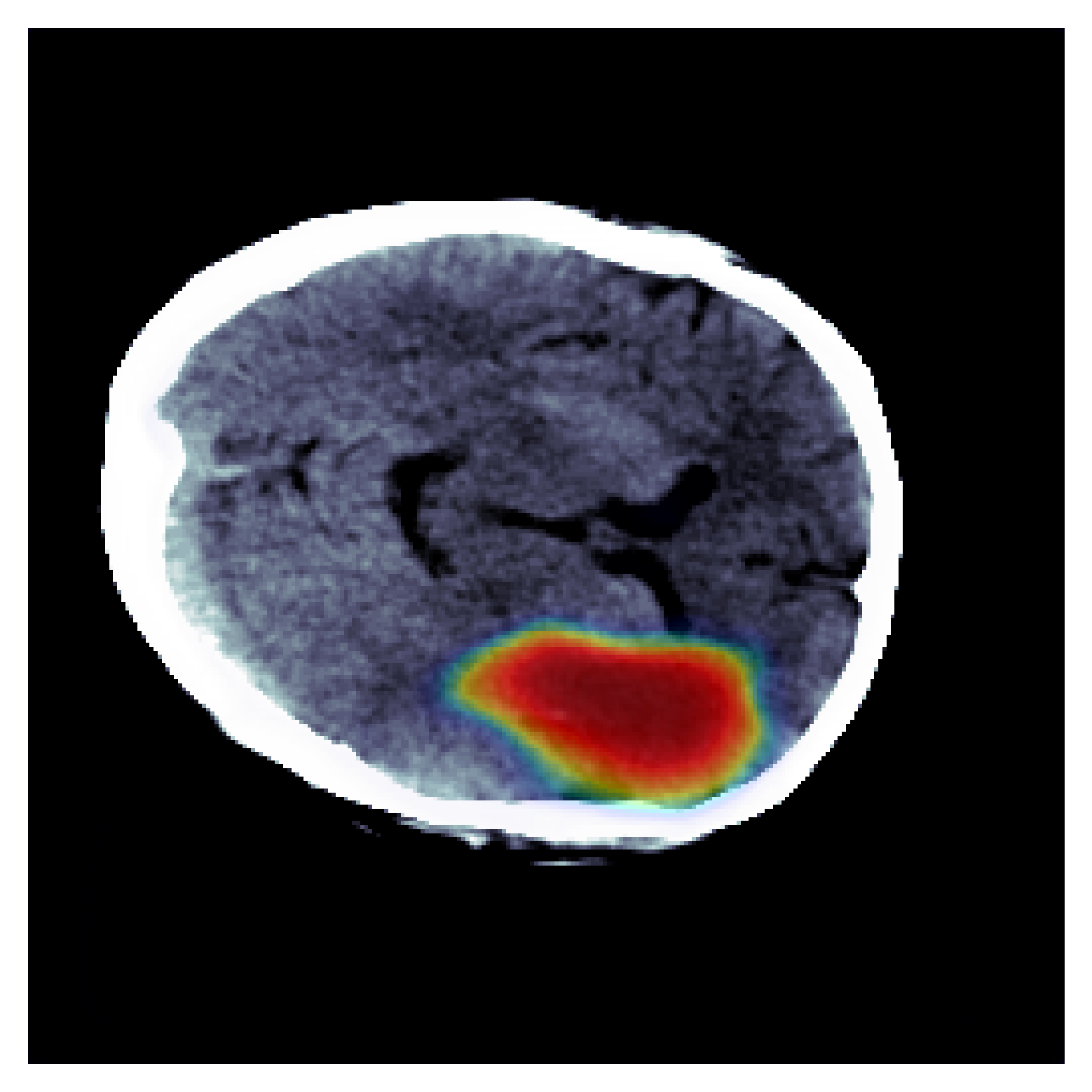} & \includegraphics[width=35mm]{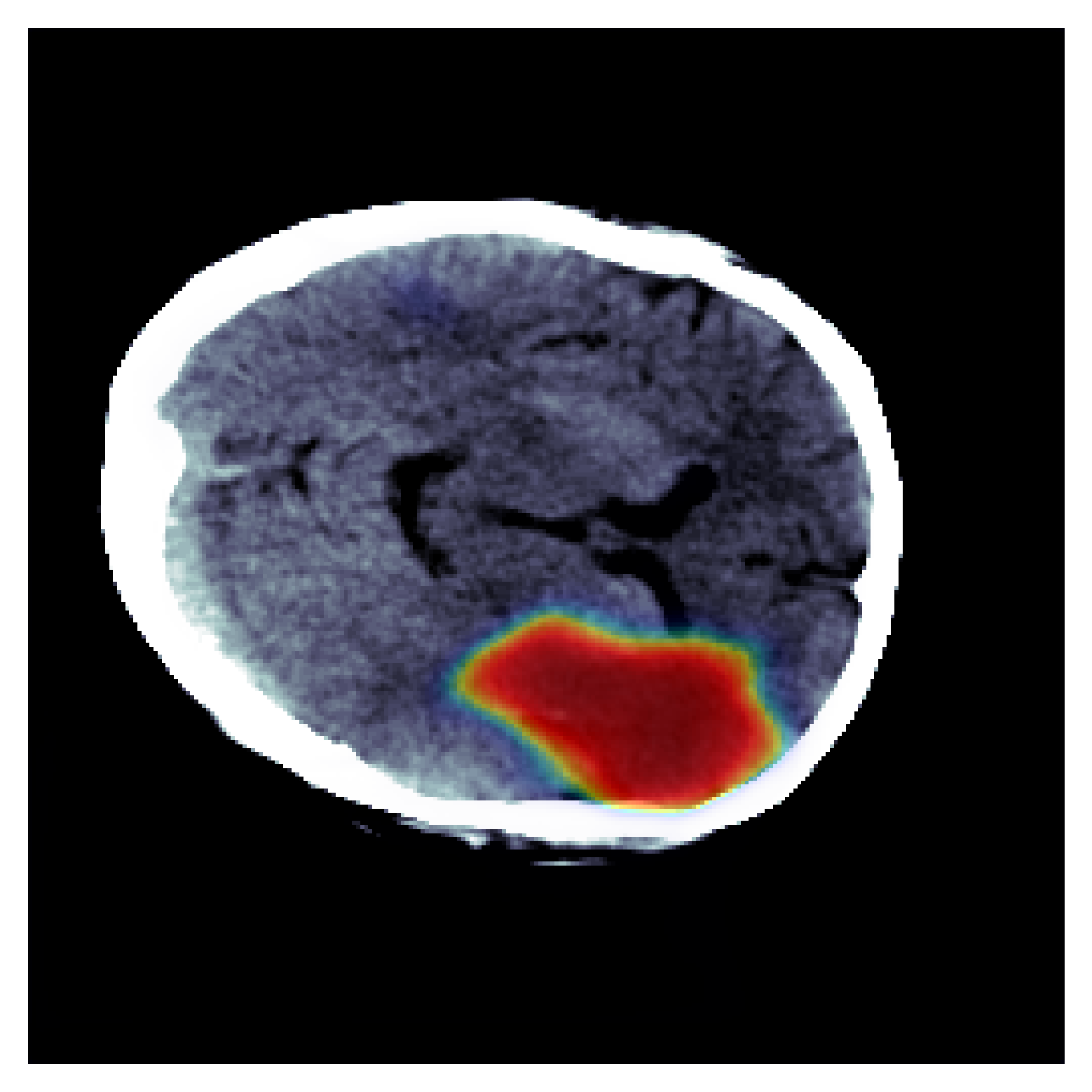} &   \includegraphics[width=35mm]{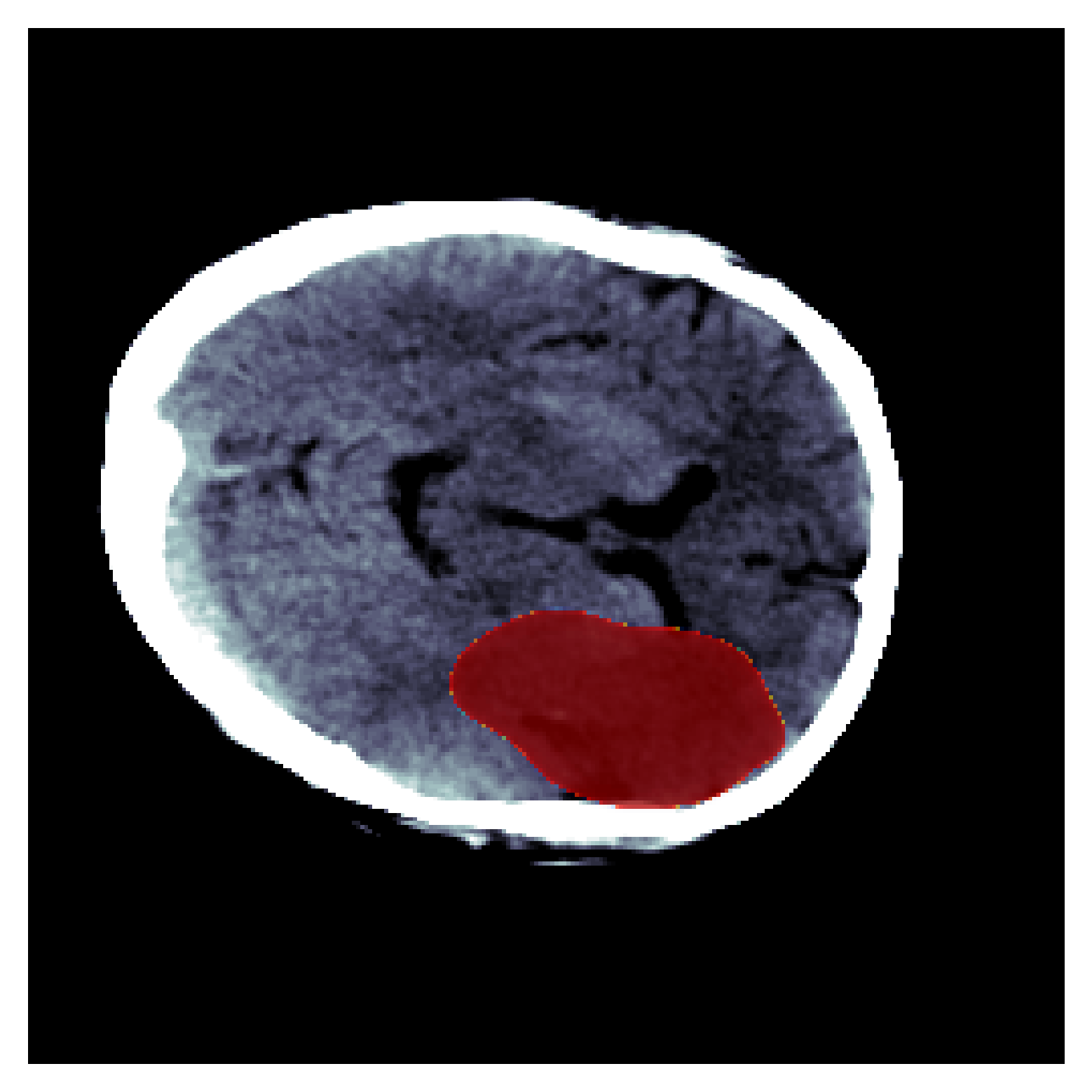} \\
  (q) Dice+Sigmoid & (r) MSE+Sigmoid & (s) BCE+Softsign & (t) Dice+Softsign \\[6pt]
  \includegraphics[width=35mm]{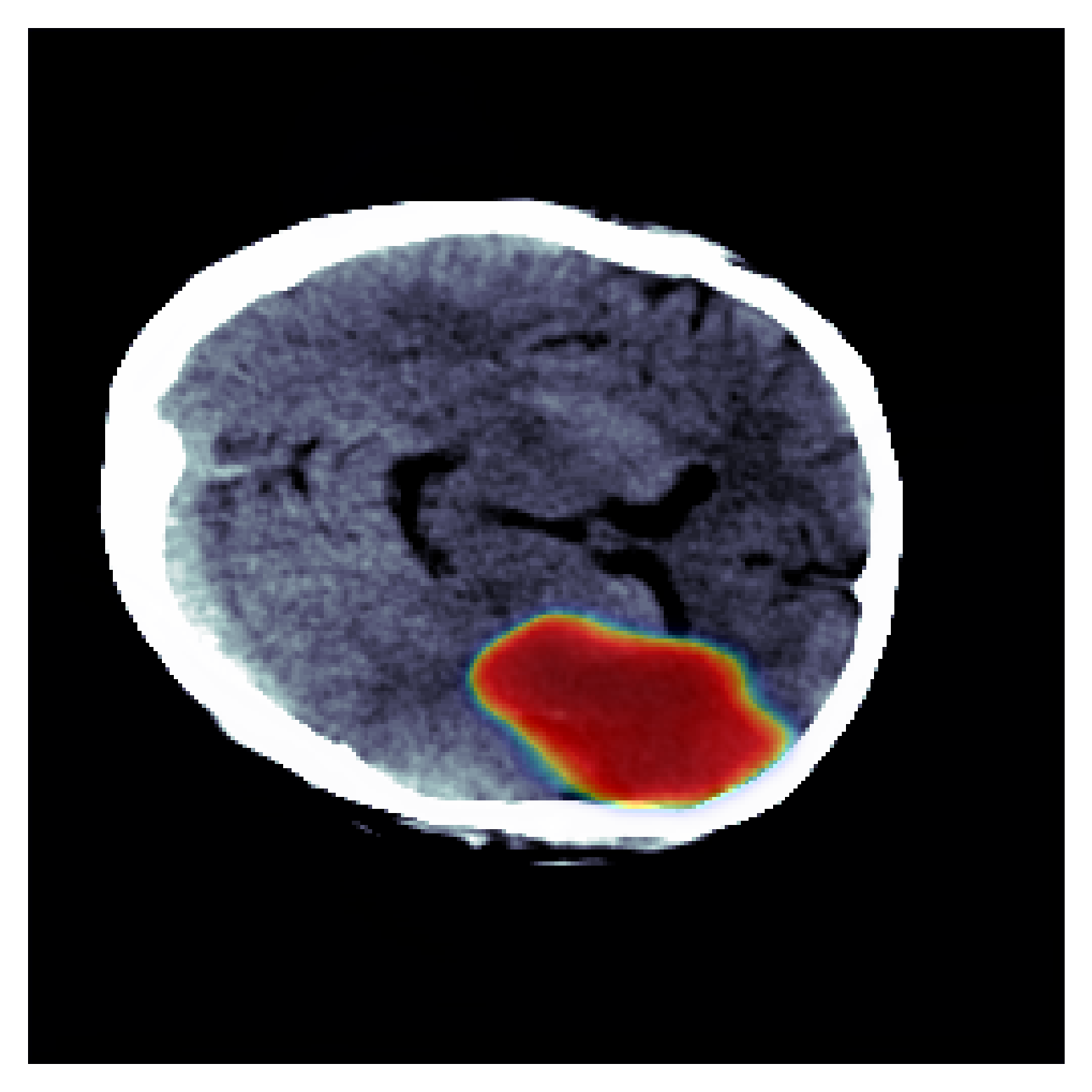} & \includegraphics[width=35mm] {predictions/ISLES_gt.png} \\ (u) MSE+Softsign & (v) Ground truth
\end{tabular}}
\label{fig:allpred3}
\caption{All $21$ predictions for a single image of the ISLES dataset, with the last image being the ground truth.}
\end{figure}

\subsection{Kvasir-SEG}

\begin{figure}[H]
\centering
\scalebox{.77}{\begin{tabular}{cccc}
  \includegraphics[width=35mm]{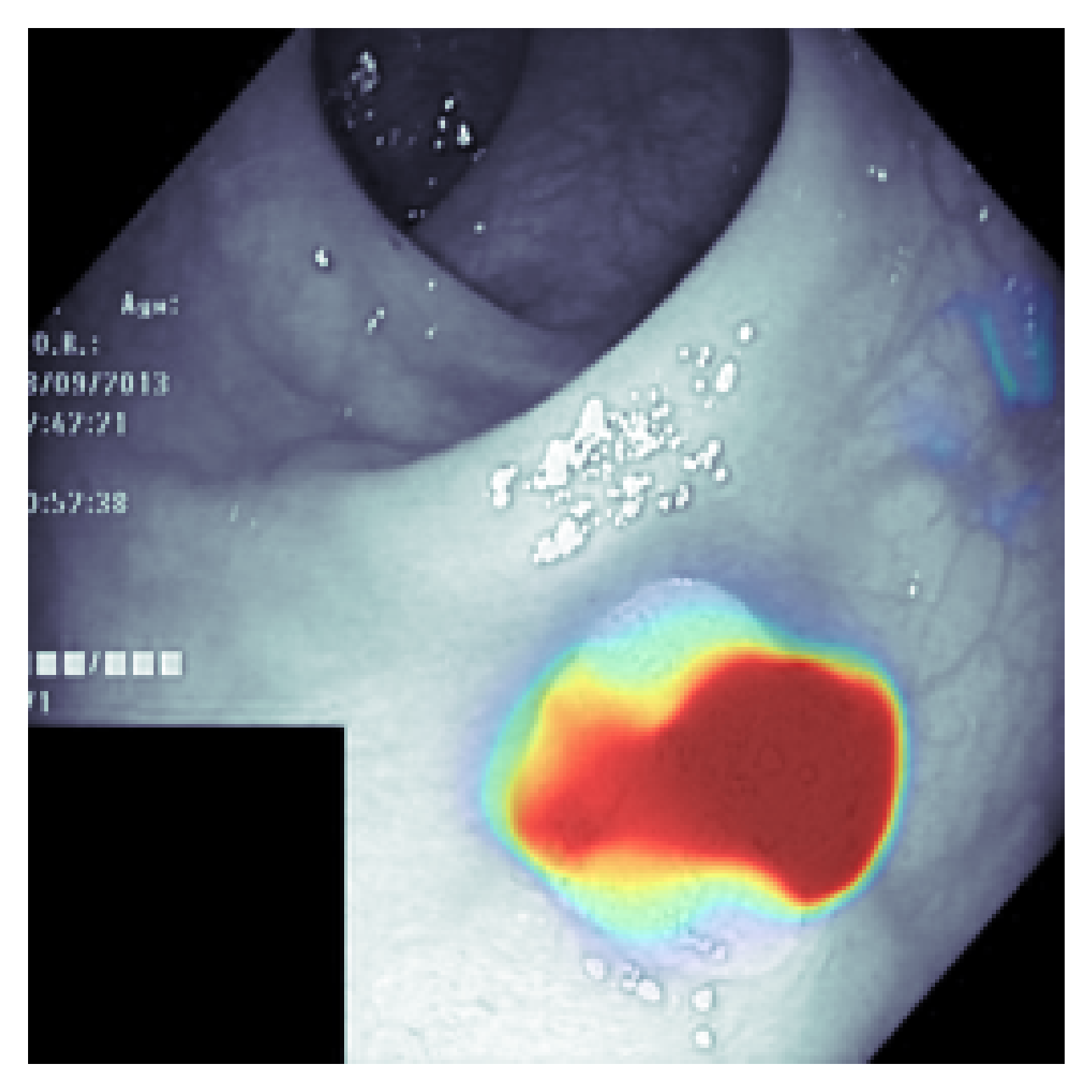} &   \includegraphics[width=35mm]{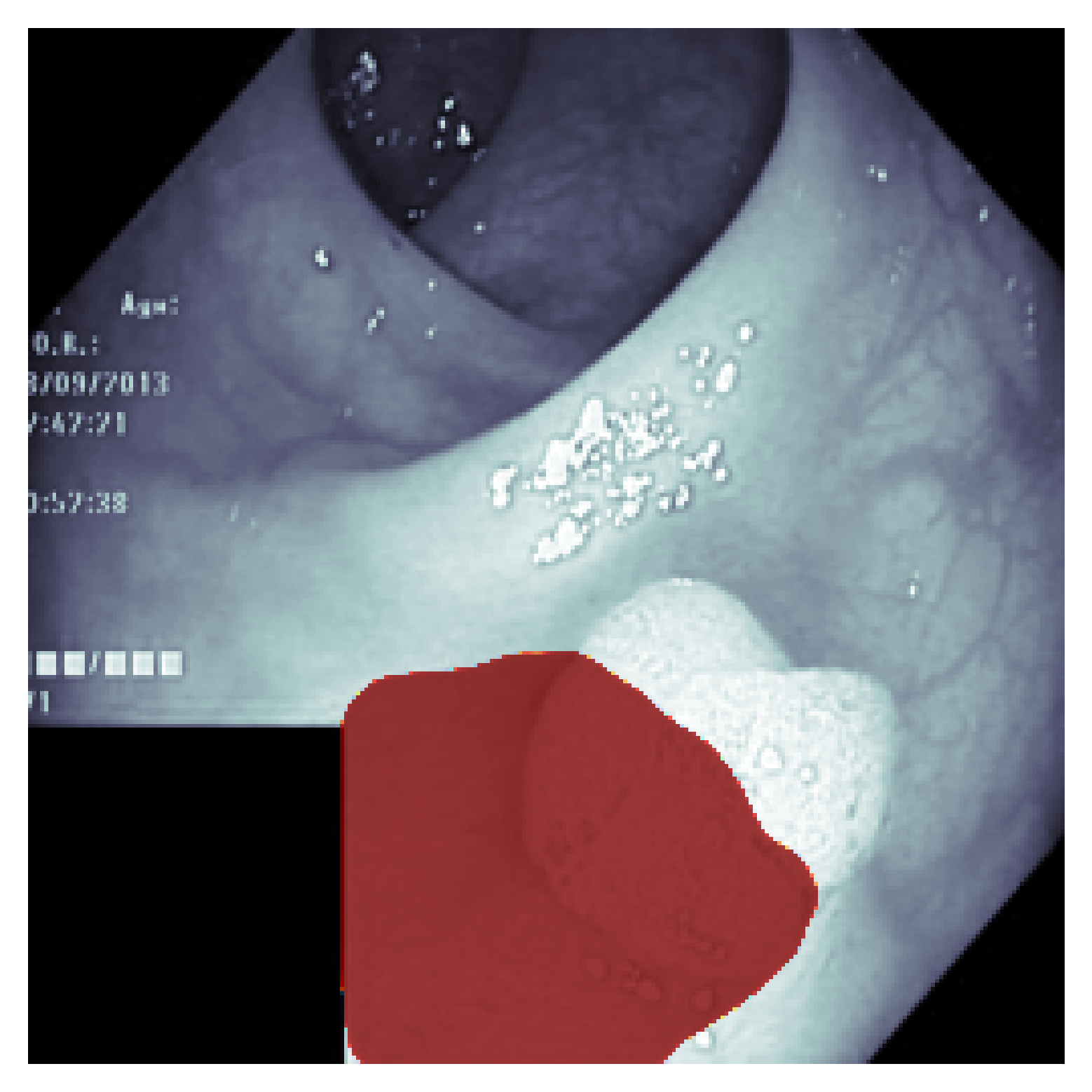} & \includegraphics[width=35mm]{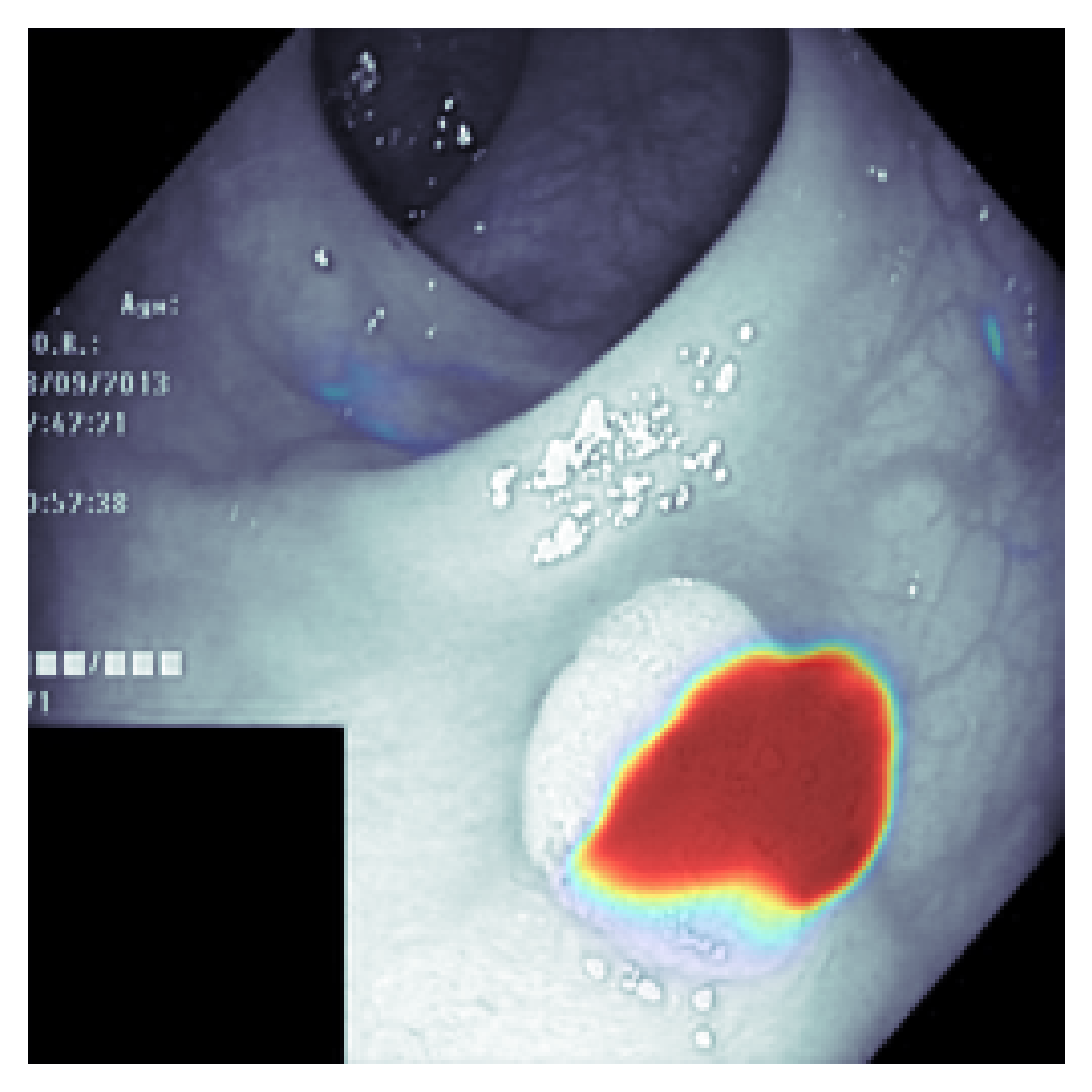} &   \includegraphics[width=35mm]{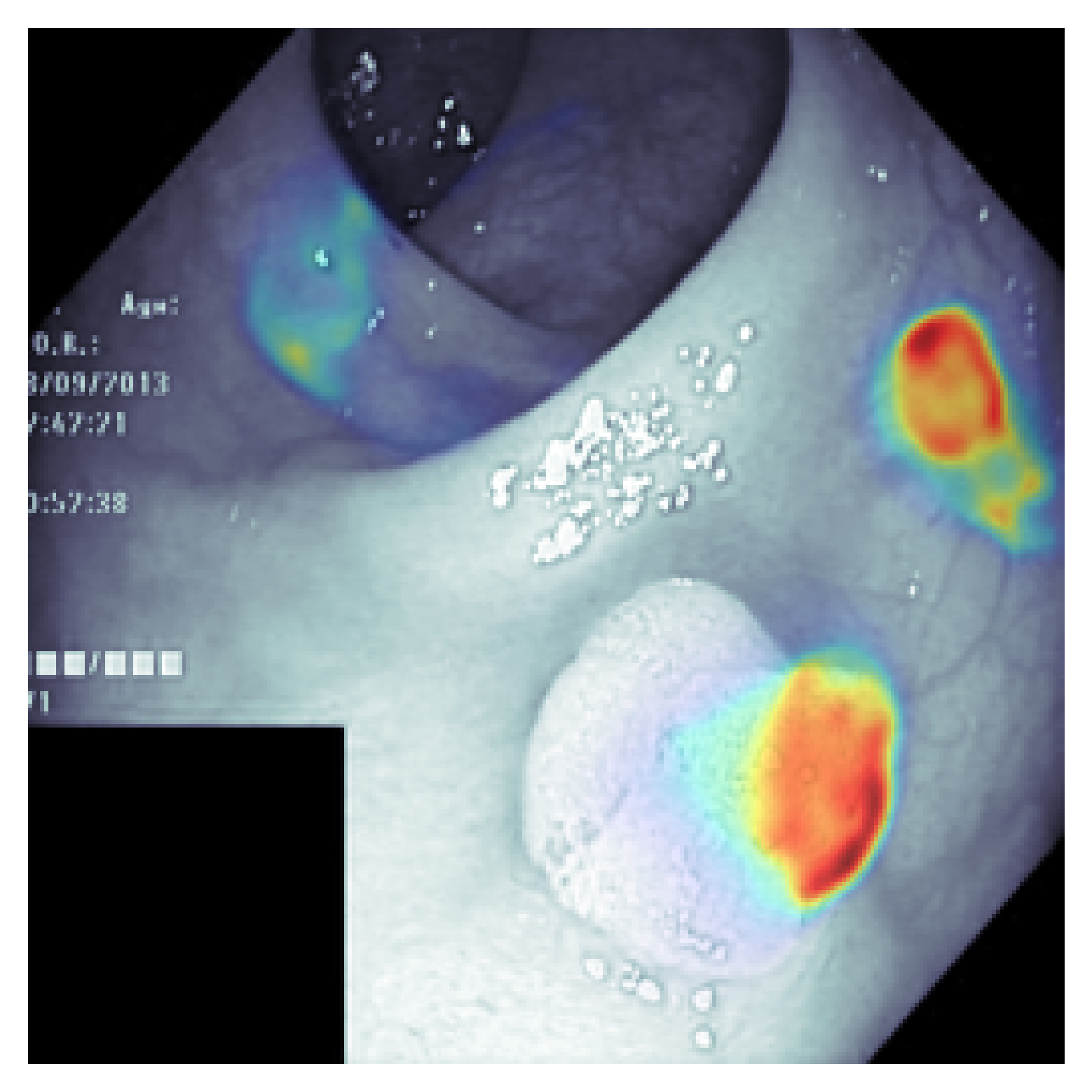} \\
  (a) BCE+Arctangent & (b) Dice+Arctangent & (c) MSE+Arctangent & (d) BCE+CDF \\[6pt]

  \includegraphics[width=35mm]{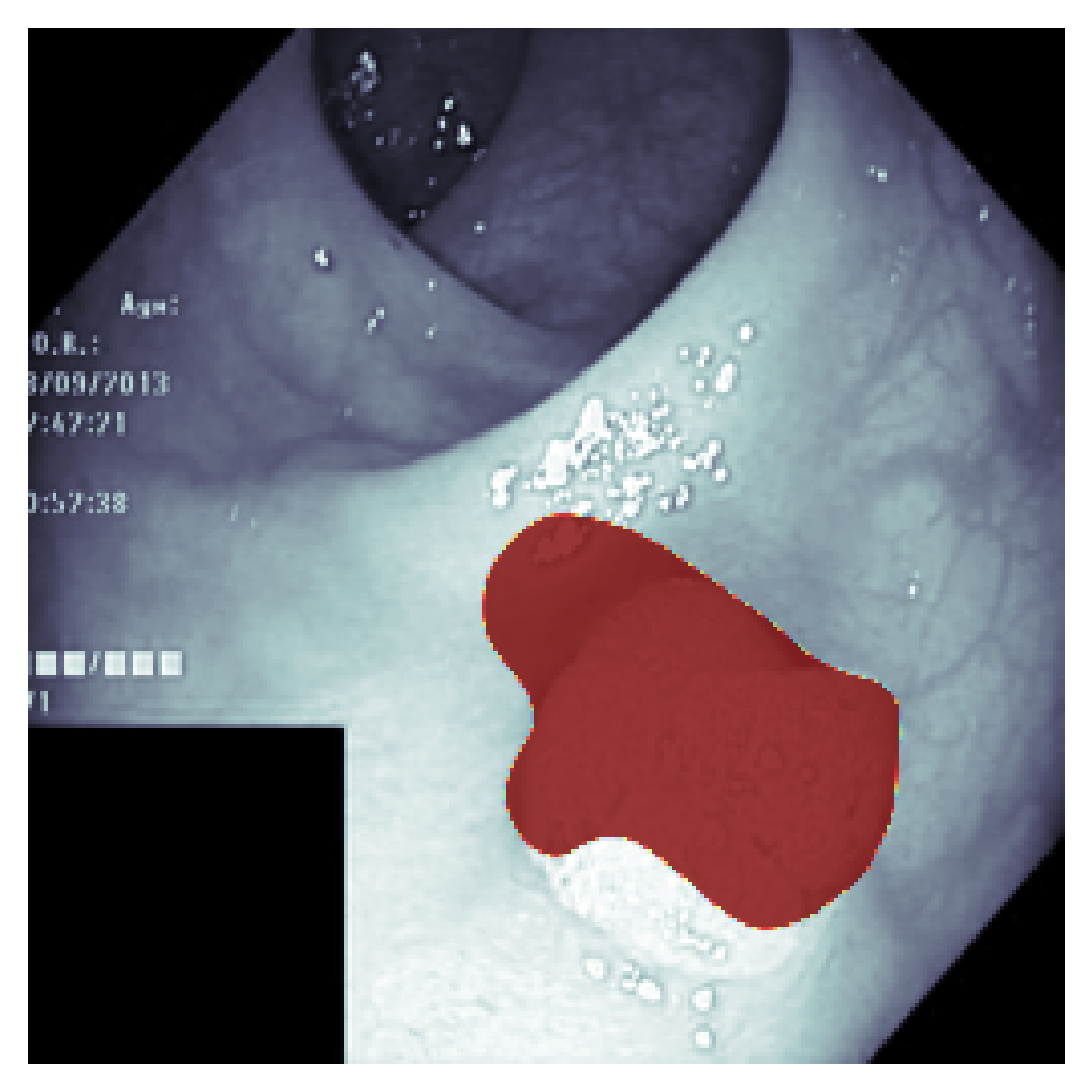} &   \includegraphics[width=35mm]{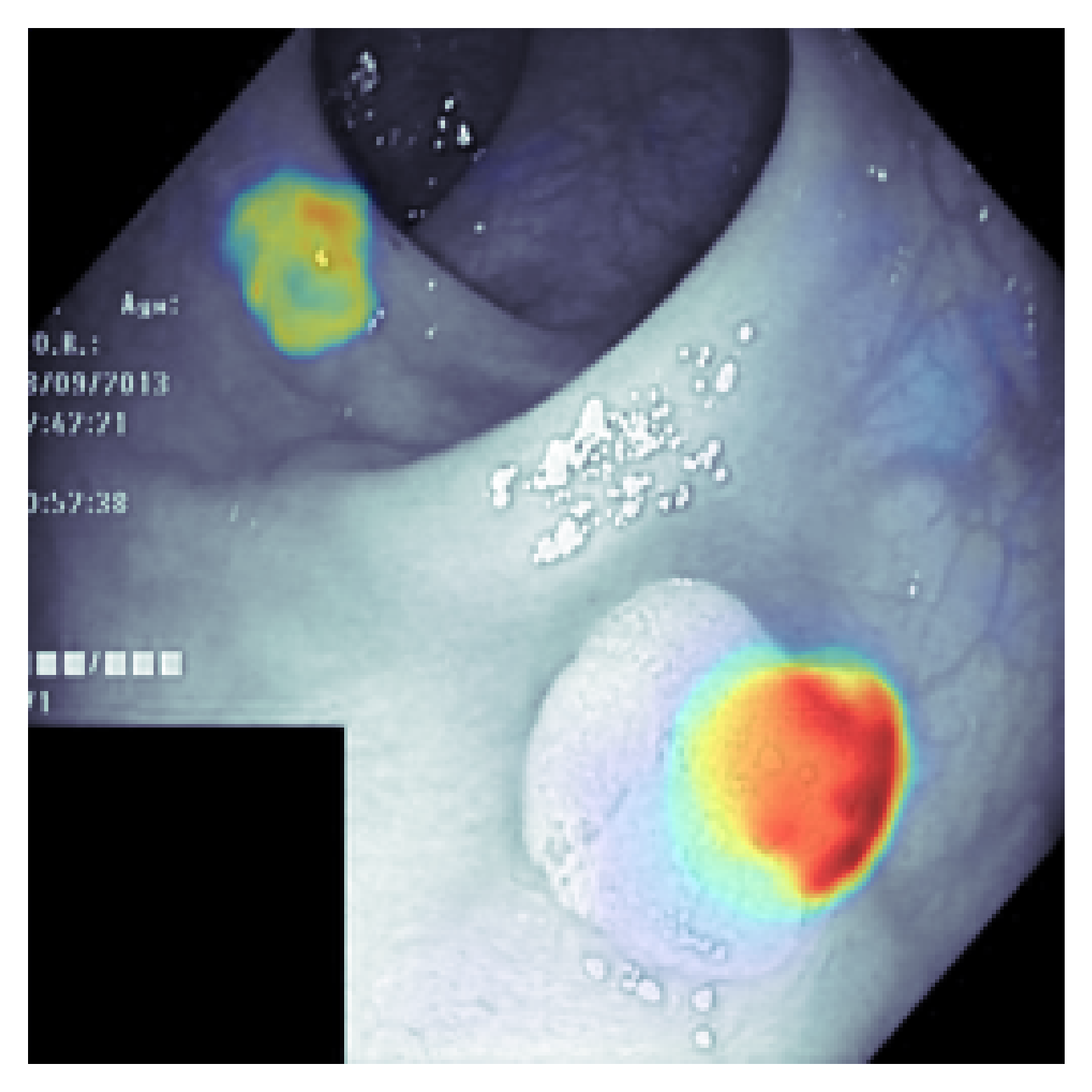} & \includegraphics[width=35mm]{predictions/Kvasir-BCELoss-hardtanh_activation-resnet34-Unet.png} &   \includegraphics[width=35mm]{predictions/Kvasir-DiceLoss-hardtanh_activation-resnet34-Unet.png} \\
  (e) Dice+CDF & (f) MSE+CDF & (g) BCE+HardTanh & (h) Dice+HardTanh \\[6pt]

  \includegraphics[width=35mm]{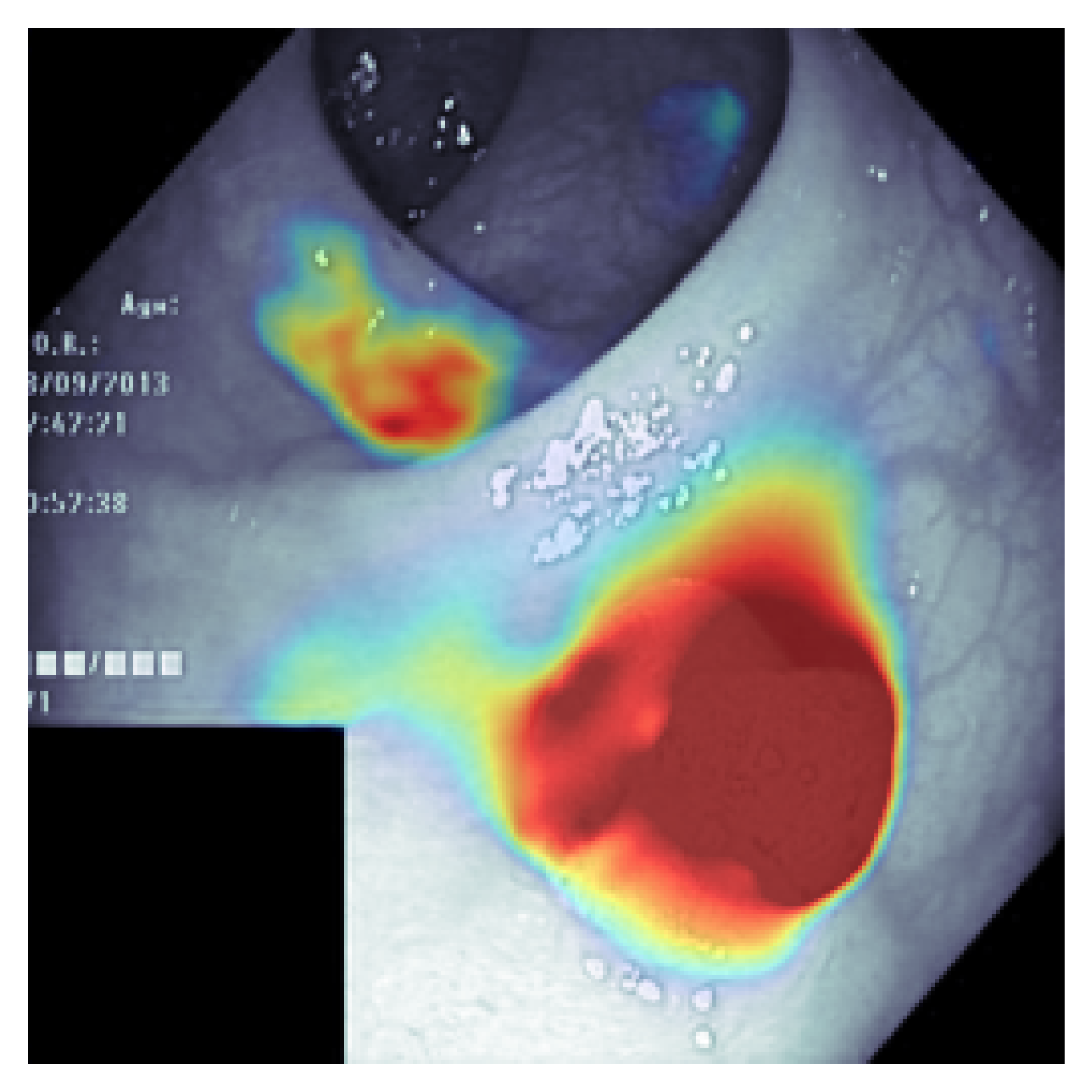} &   \includegraphics[width=35mm]{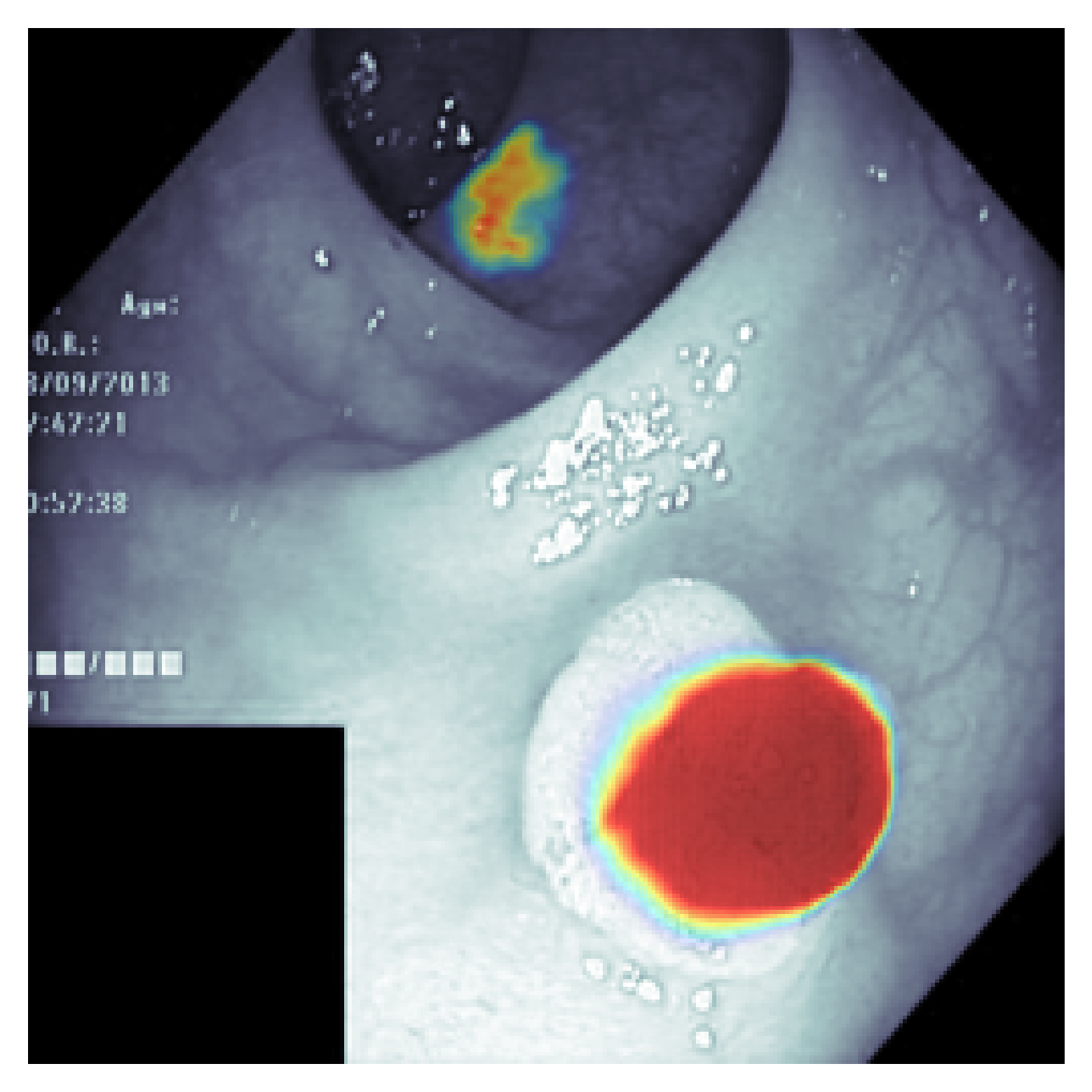} & \includegraphics[width=35mm]{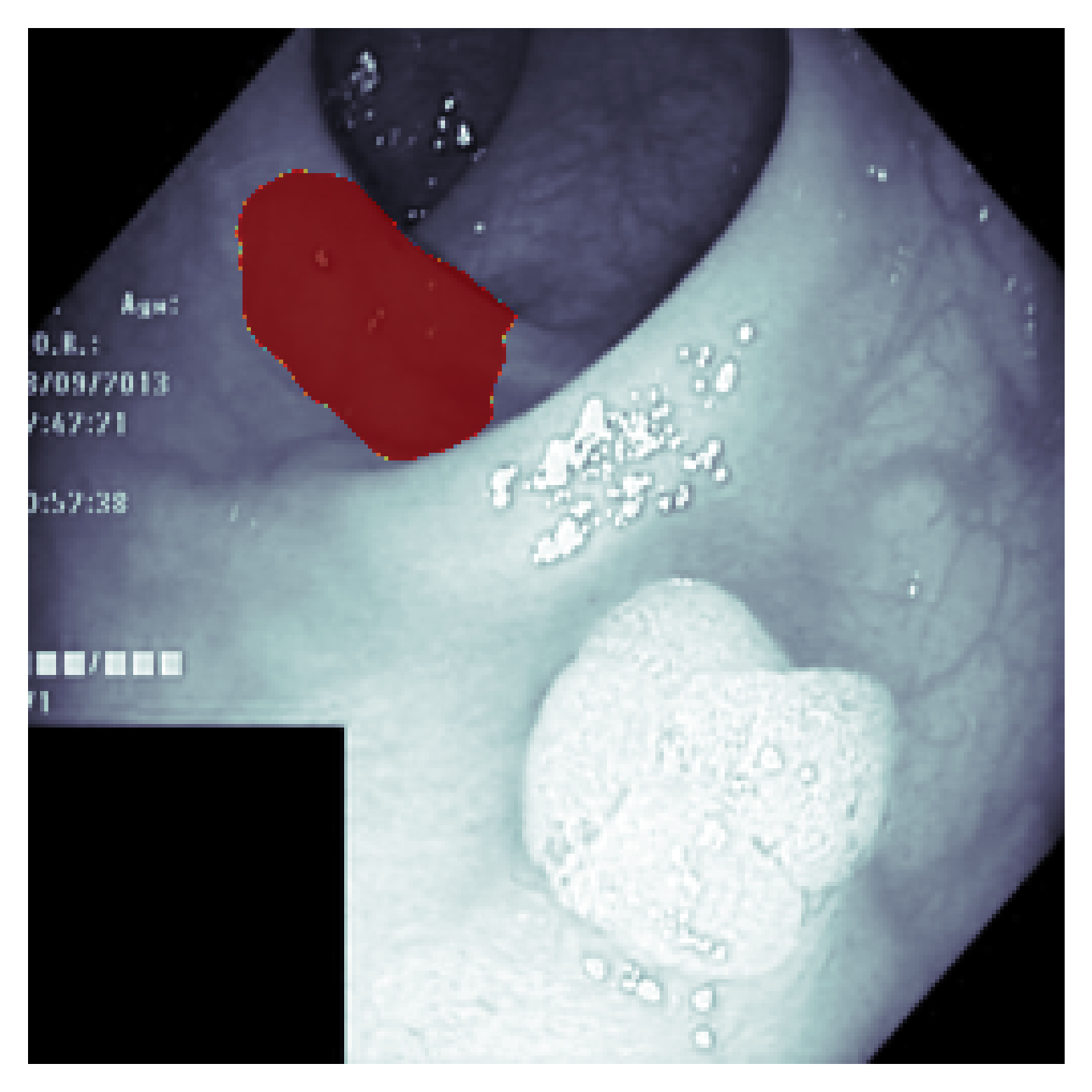} &   \includegraphics[width=35mm]{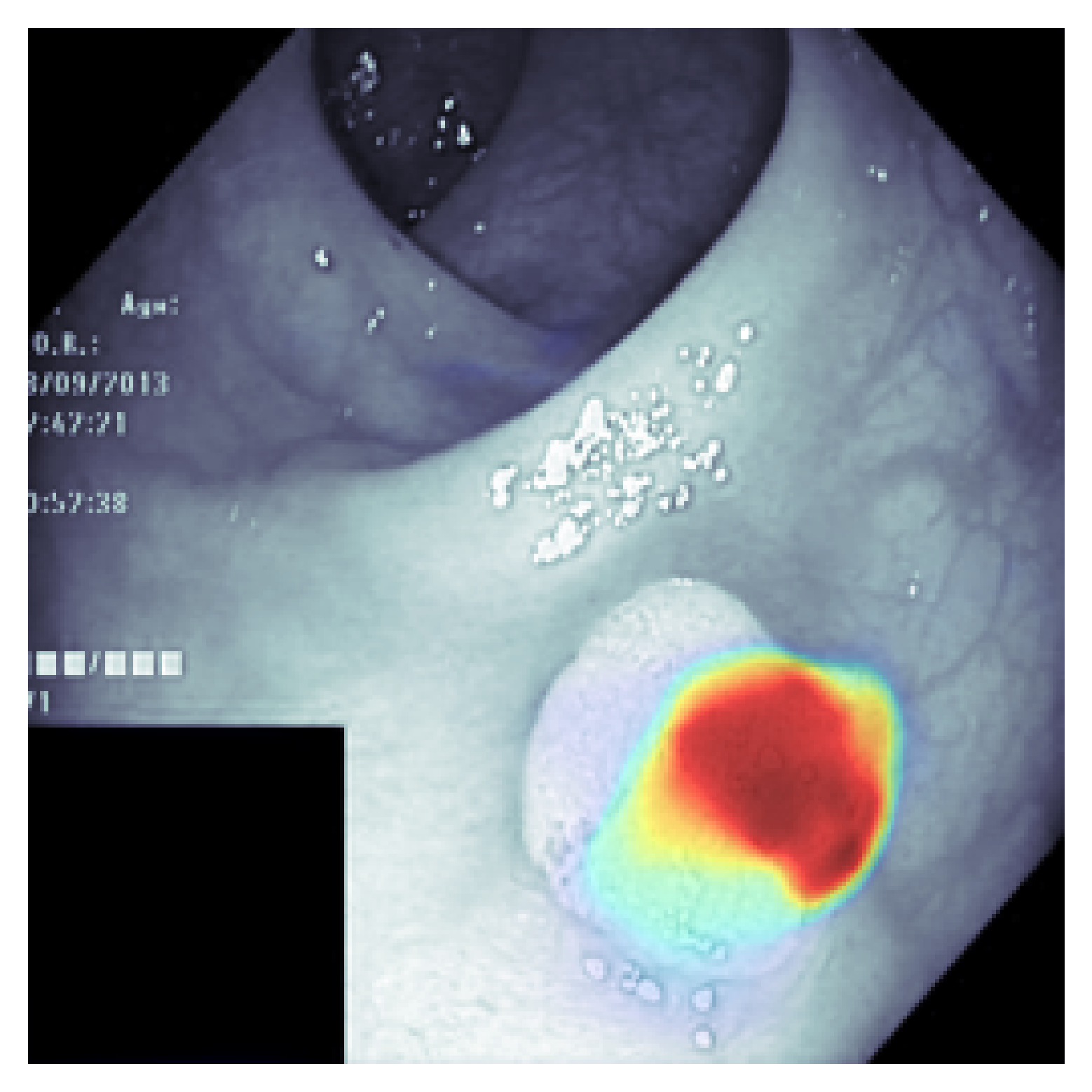} \\
  (i) MSE+HardTanh & (j) BCE+InvSquareRoot & (k) Dice+InvSquareRoot & (l) MSE+InvSquareRoot \\[6pt]

  \includegraphics[width=35mm]{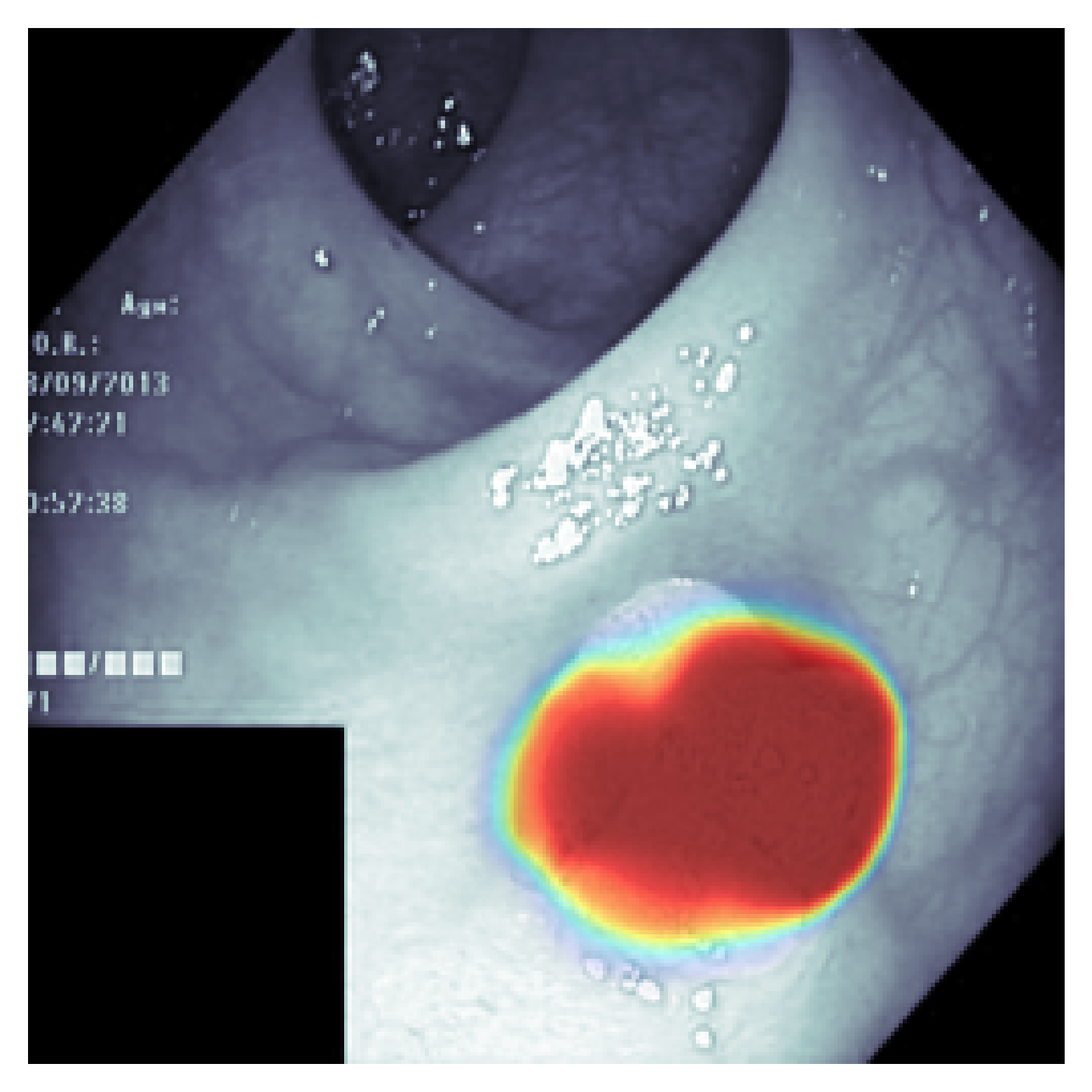} &   \includegraphics[width=35mm]{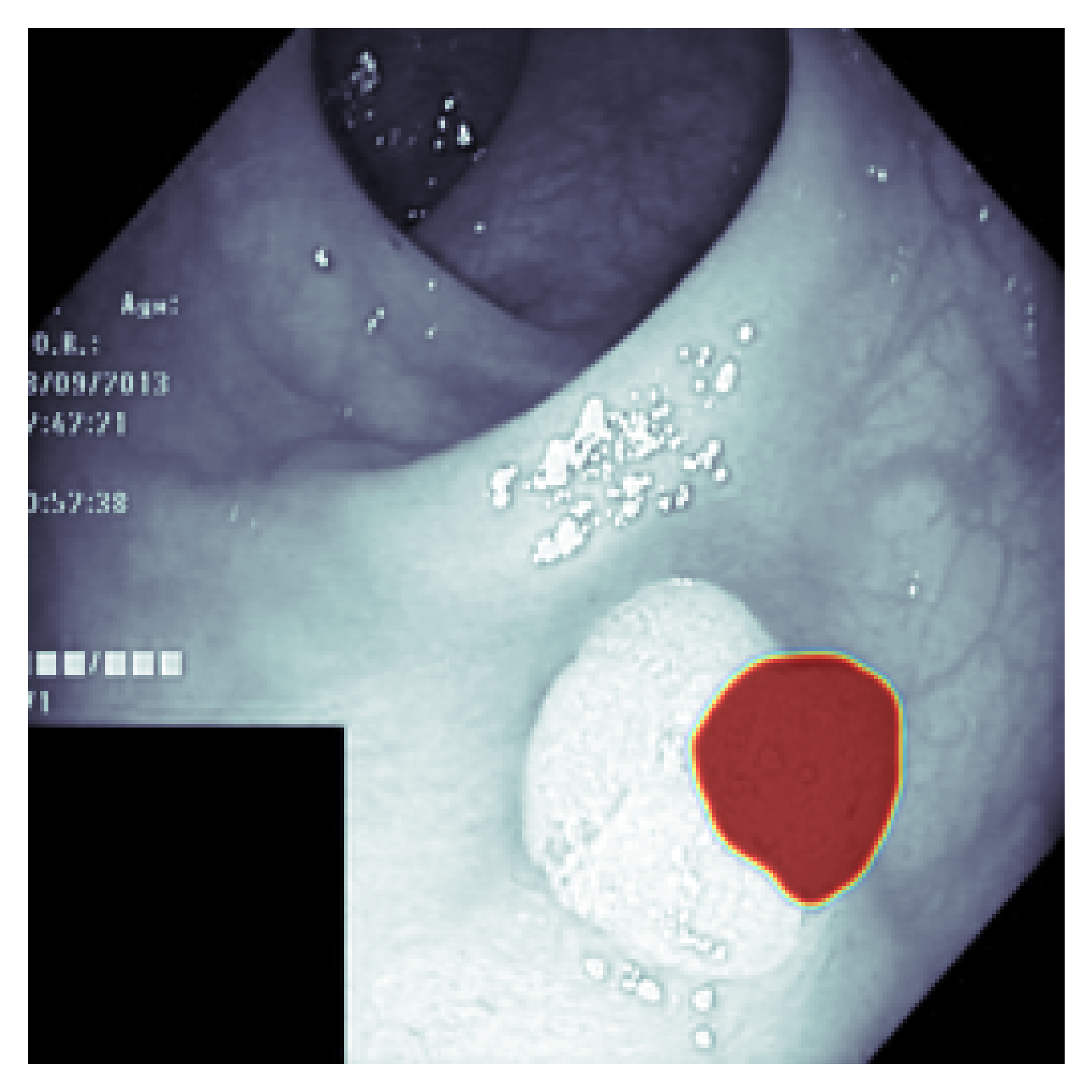} & \includegraphics[width=35mm]{predictions/Kvasir-MSELoss-linear_activation-resnet34-Unet.png} &   \includegraphics[width=35mm]{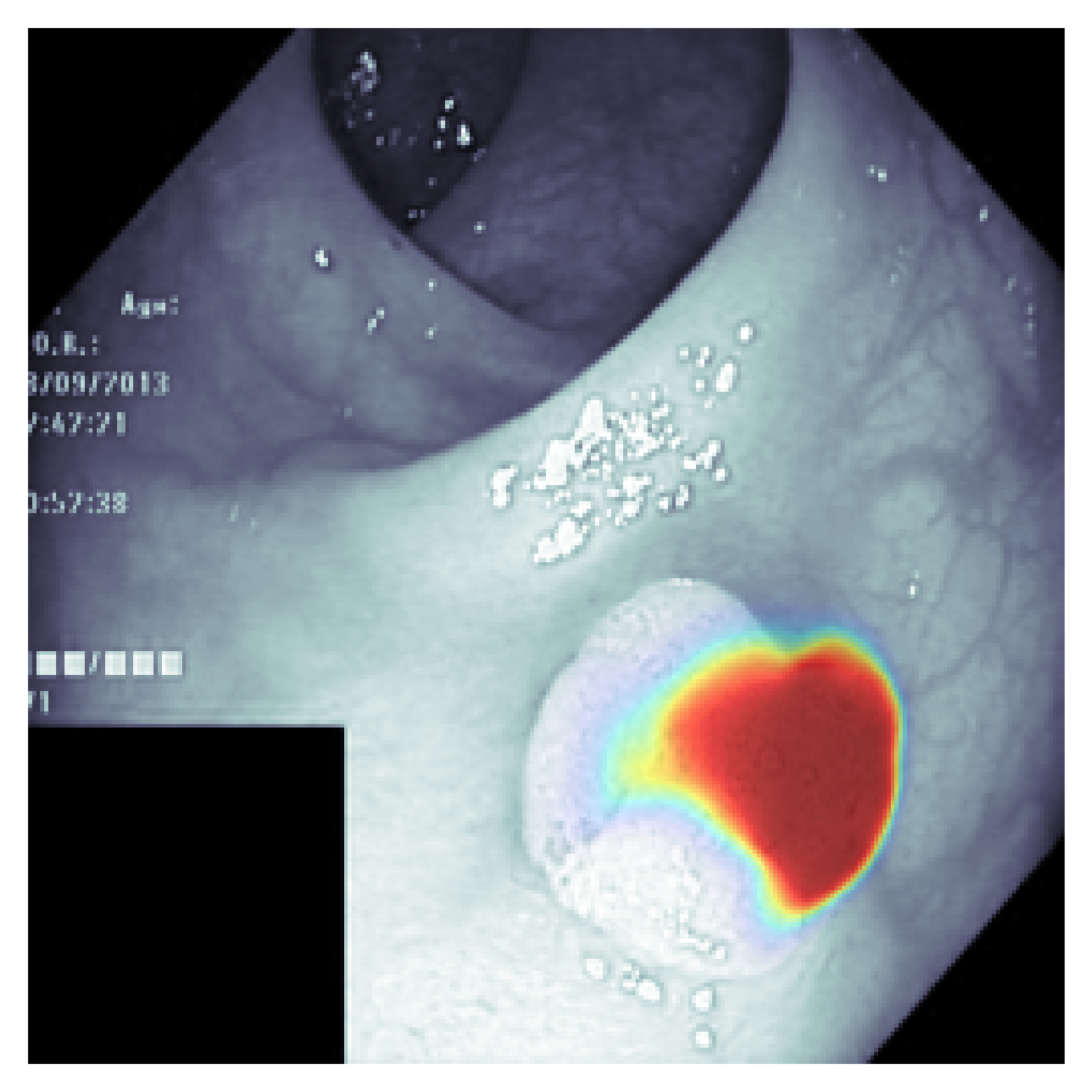} \\
  (m) BCE+Linear & (n) Dice+Linear & (o) MSE+Linear & (p) BCE+Sigmoid \\[6pt]

  \includegraphics[width=35mm]{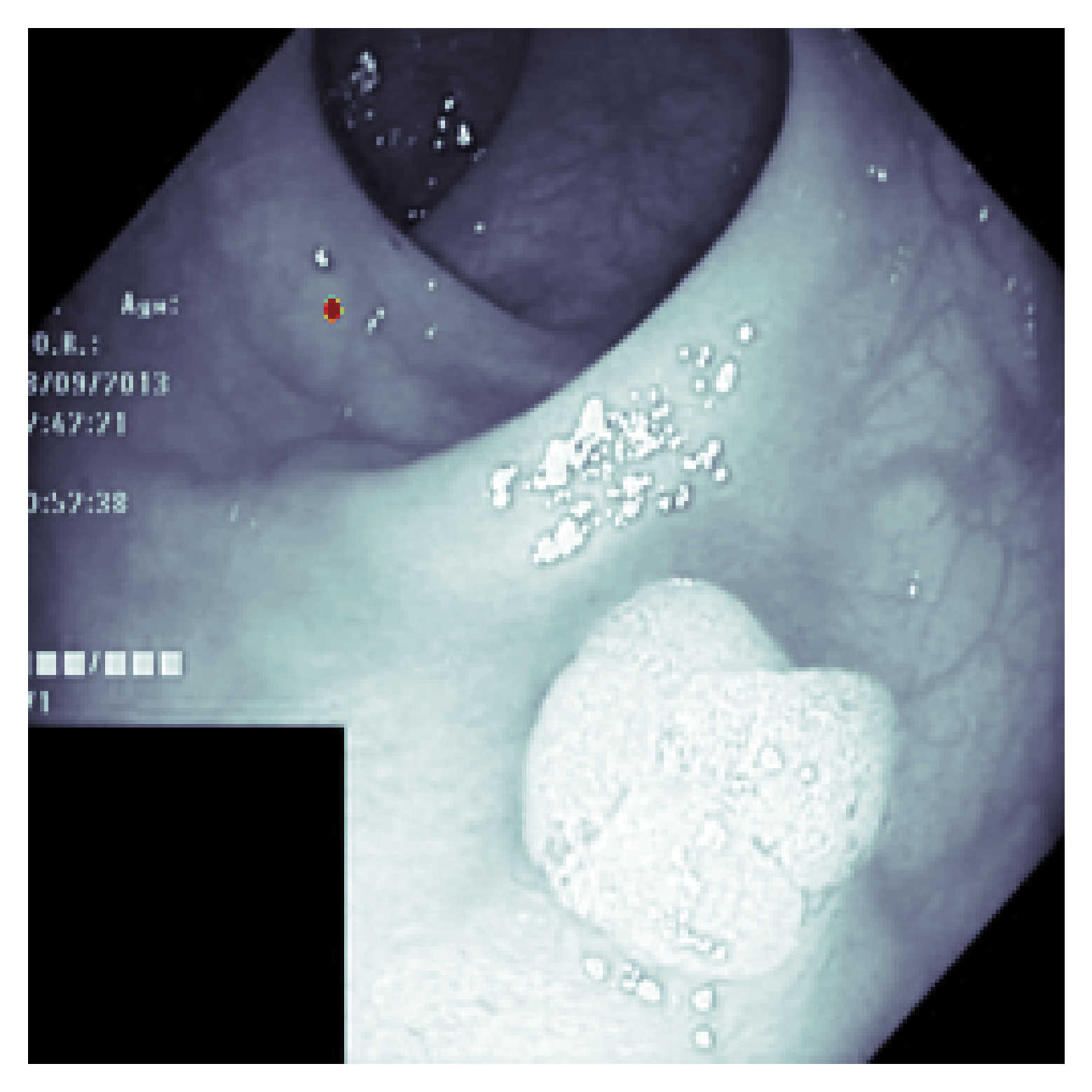} &   \includegraphics[width=35mm]{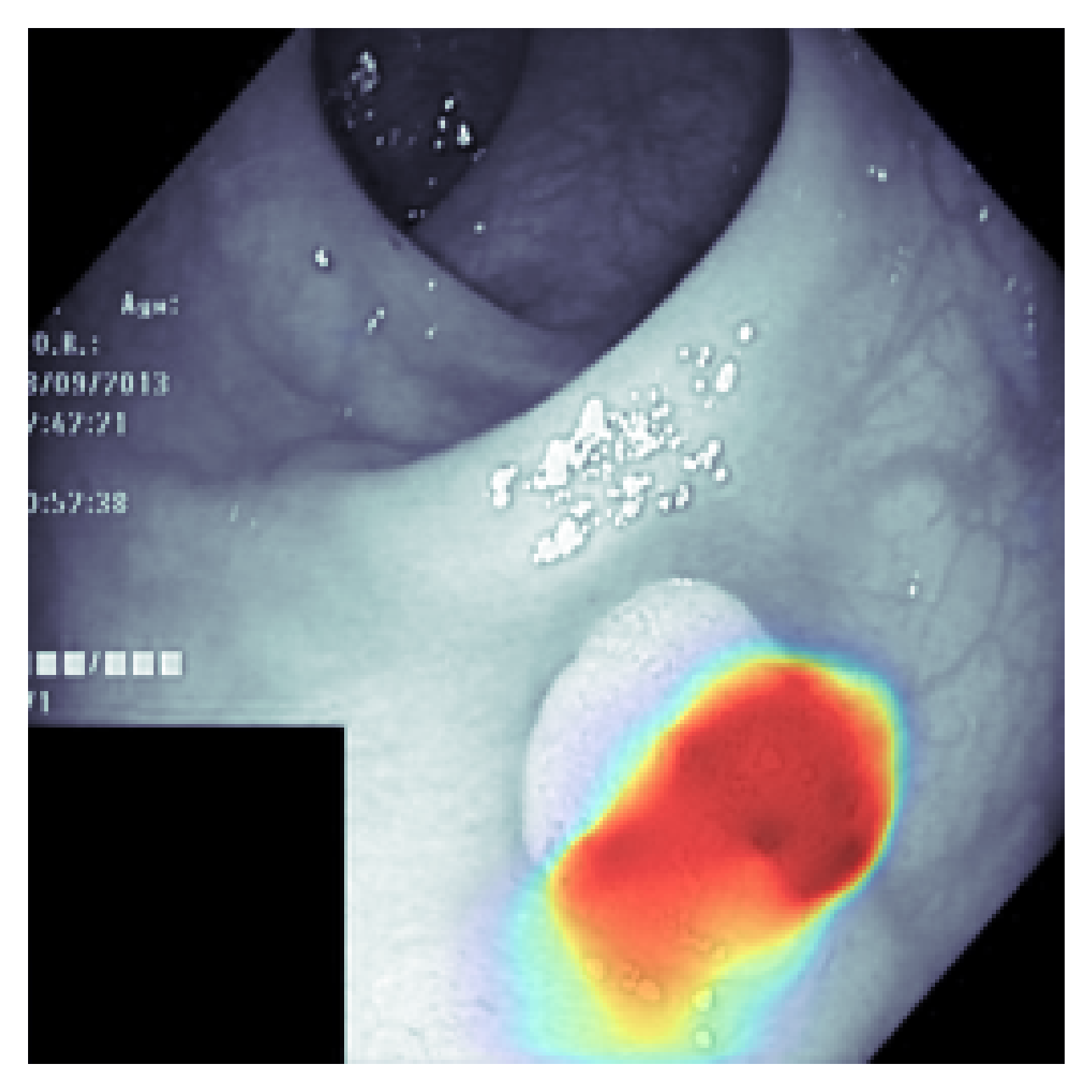} & \includegraphics[width=35mm]{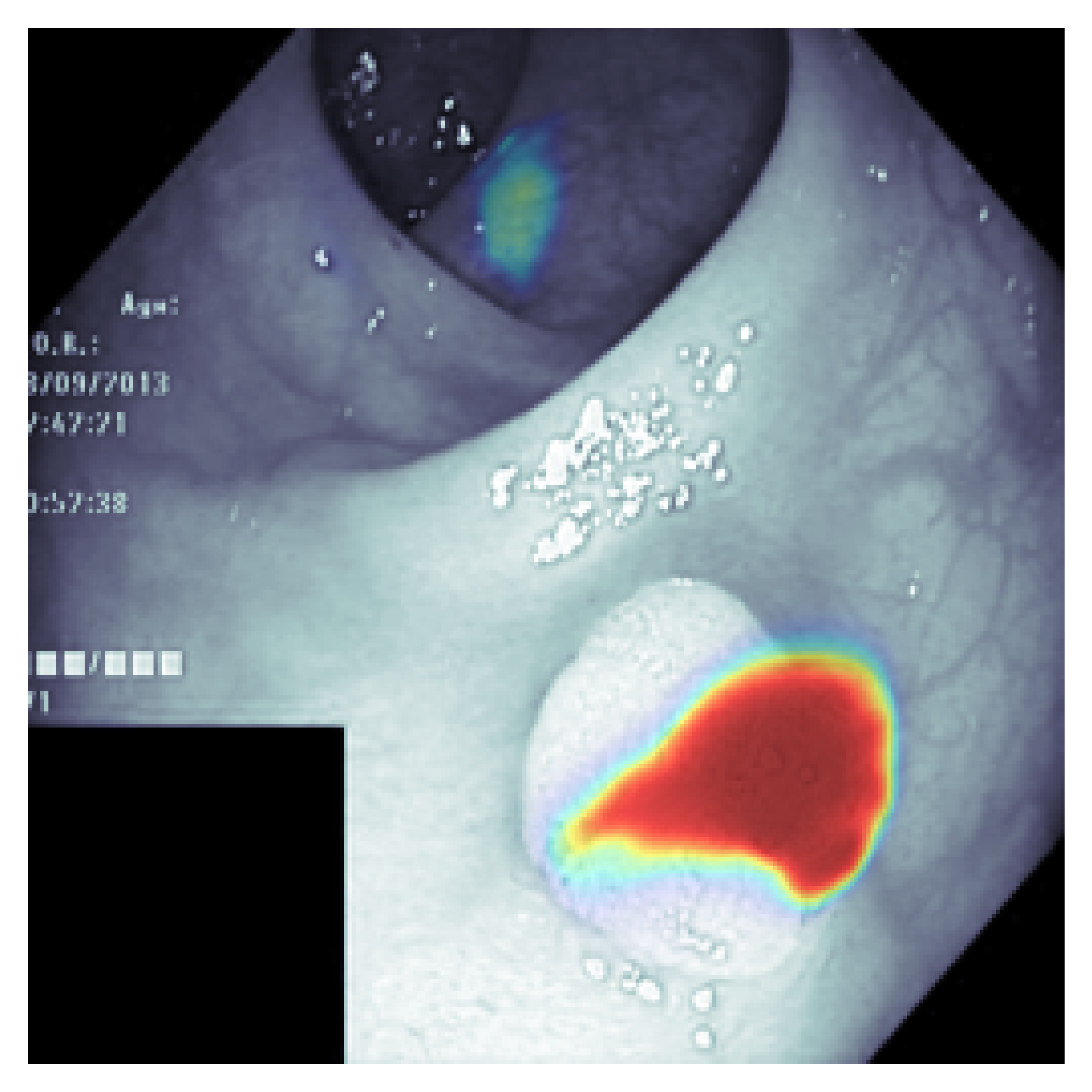} &   \includegraphics[width=35mm]{predictions/Kvasir-DiceLoss-softsign_activation-resnet34-Unet.png} \\
  (q) Dice+Sigmoid & (r) MSE+Sigmoid & (s) BCE+Softsign & (t) Dice+Softsign \\[6pt]
  \includegraphics[width=35mm]{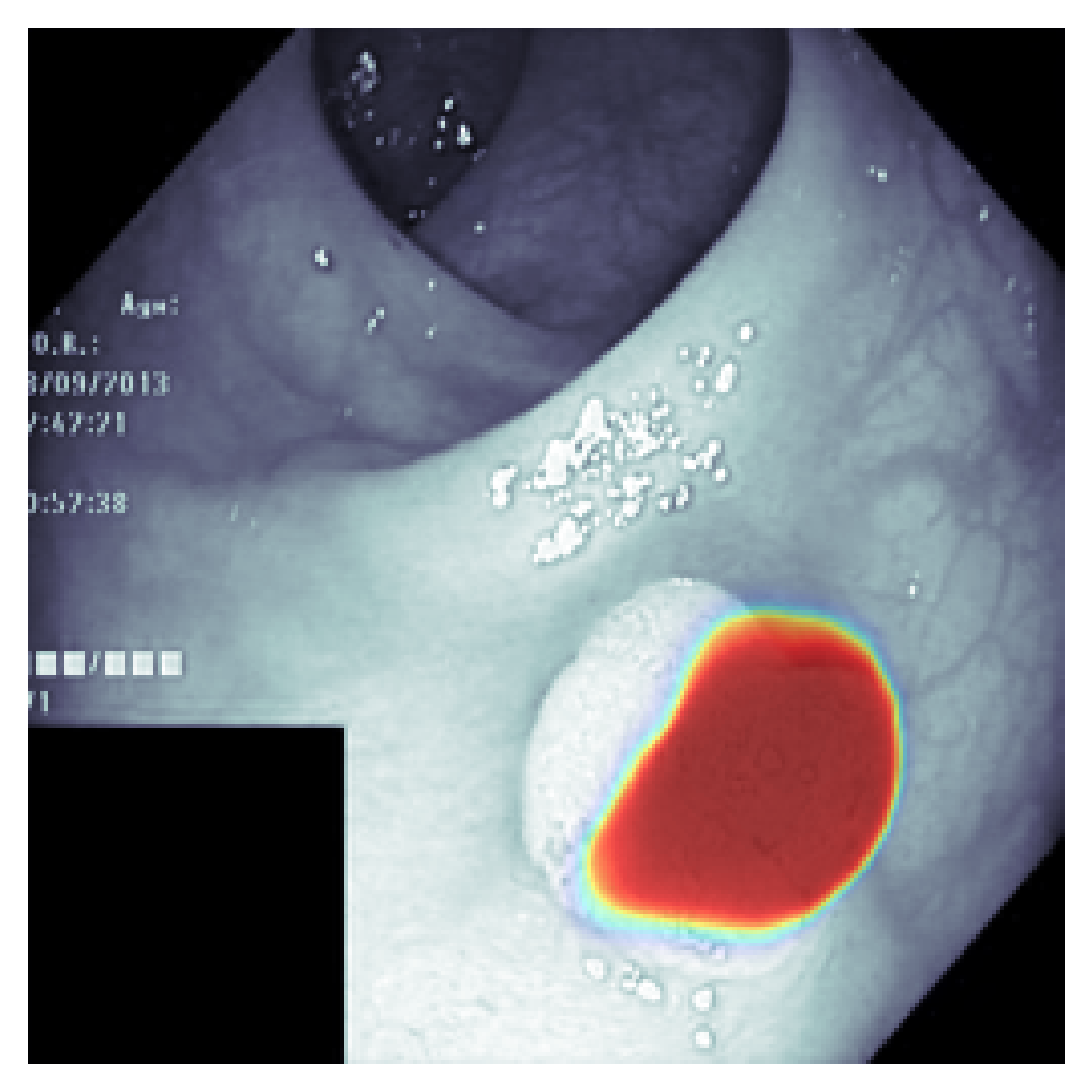} & \includegraphics[width=35mm] {predictions/Kvasir_gt.png} \\ (u) MSE+Softsign & (v) Ground truth
\end{tabular}}
\label{fig:allpred2}
\caption{All $21$ predictions for a single image of the Kvasir-SEG dataset, with the last image being the ground truth.}
\end{figure}

\subsection{Medical Segmentation Decathlon (MSD)}

\begin{figure}[H]
\centering
\scalebox{.77}{\begin{tabular}{cccc}
  \includegraphics[width=35mm]{predictions/MSD-BCELoss-arctan_activation-resnet34-Unet.png} &   \includegraphics[width=35mm]{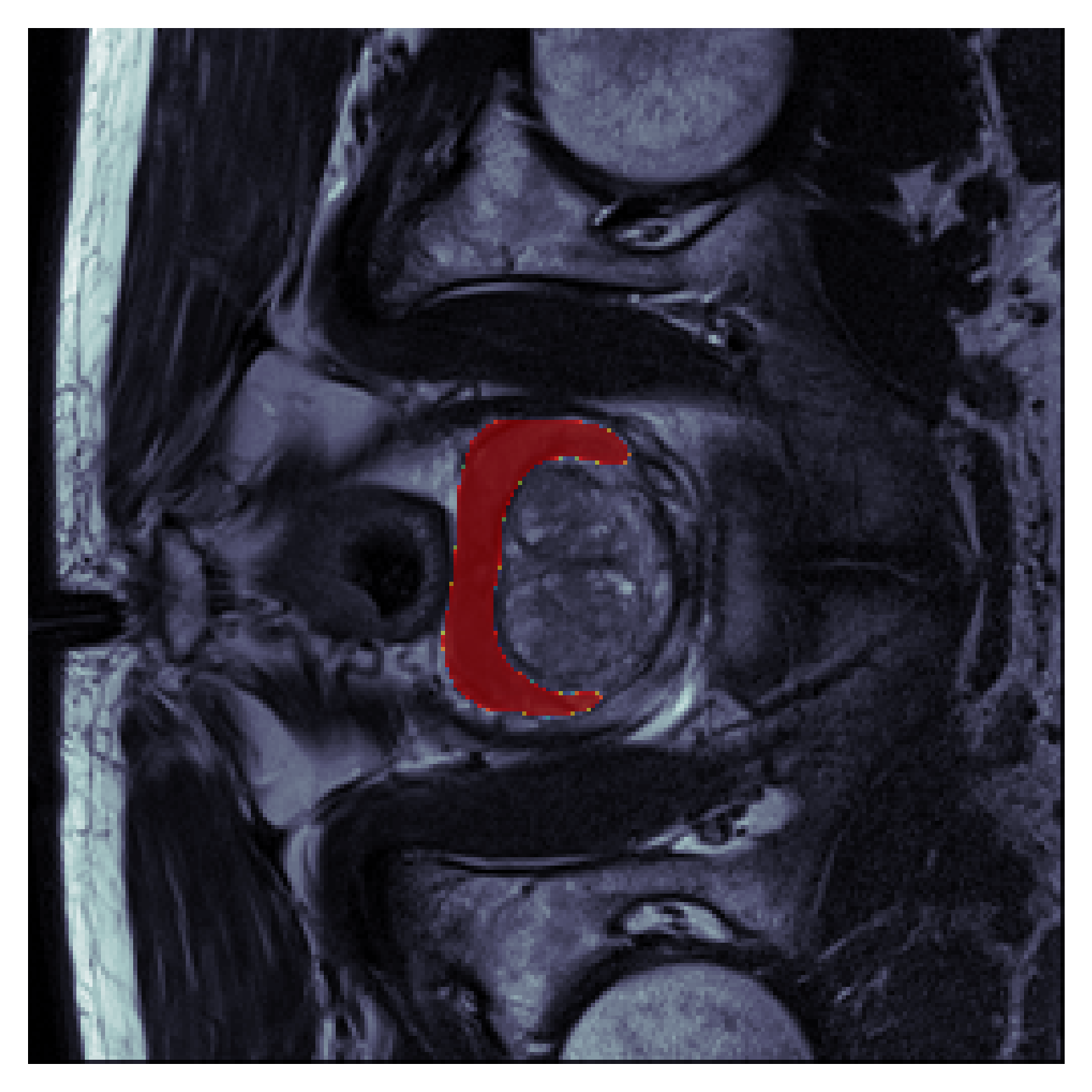} & \includegraphics[width=35mm]{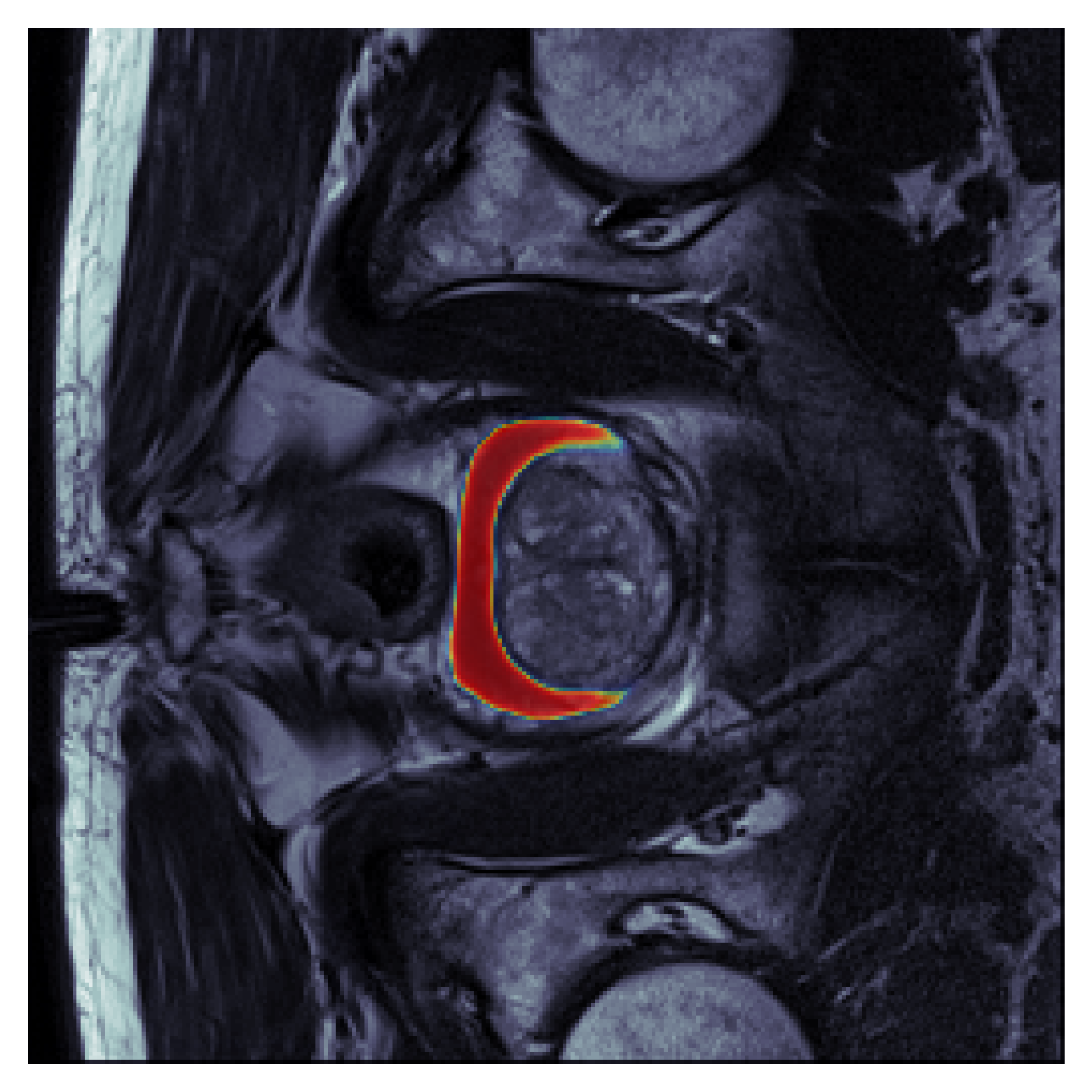} &   \includegraphics[width=35mm]{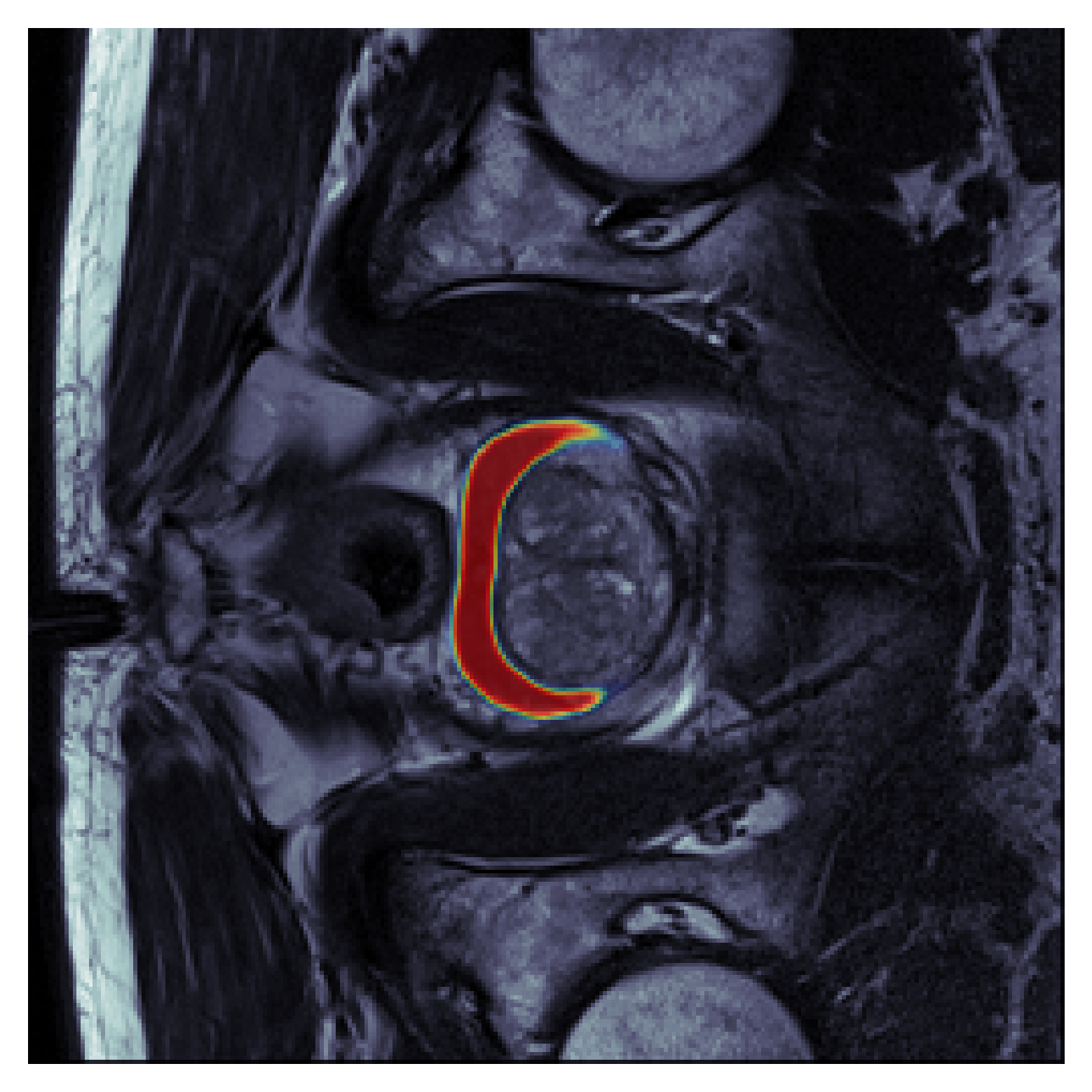} \\
  (a) BCE+Arctangent & (b) Dice+Arctangent & (c) MSE+Arctangent & (d) BCE+CDF \\[6pt]

  \includegraphics[width=35mm]{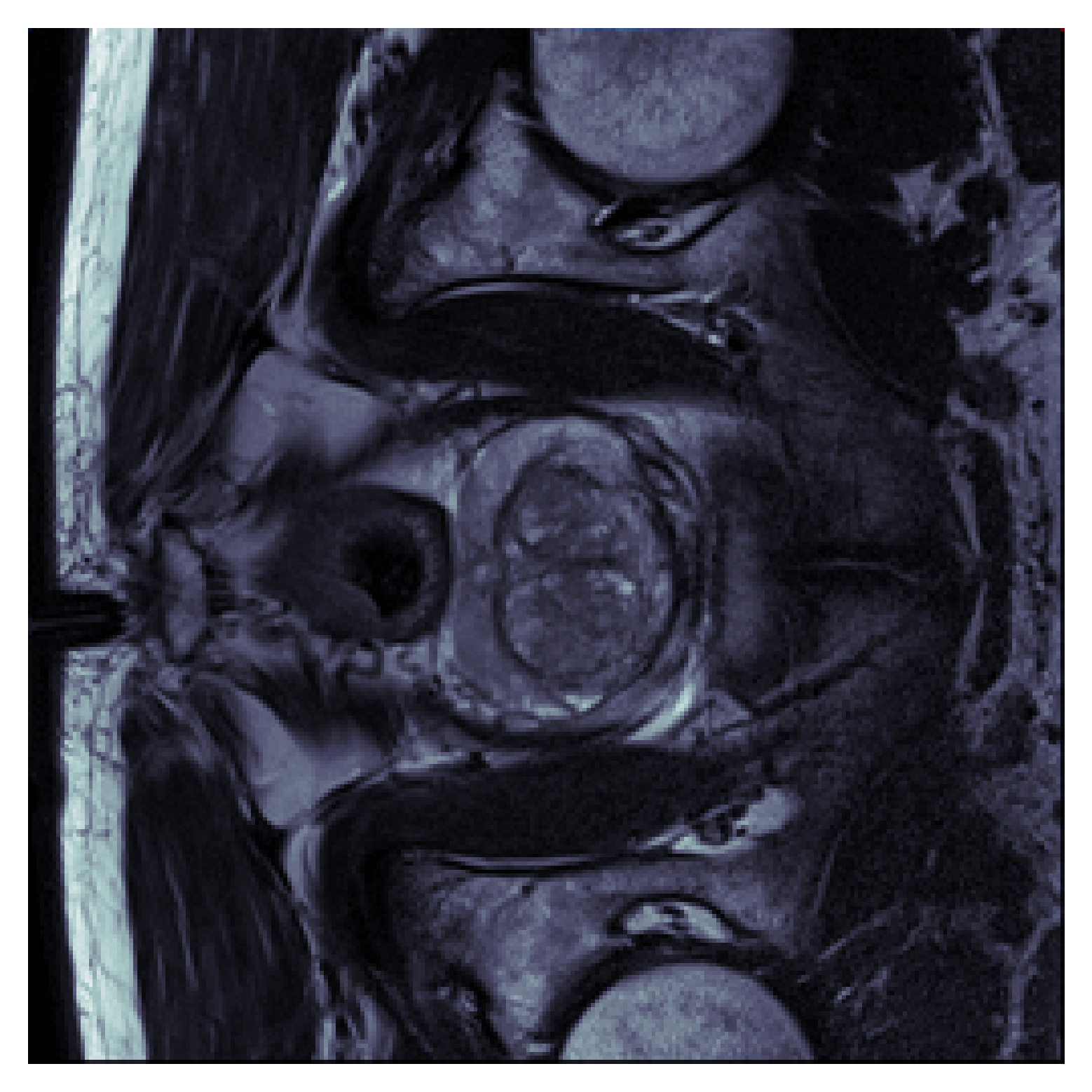} &   \includegraphics[width=35mm]{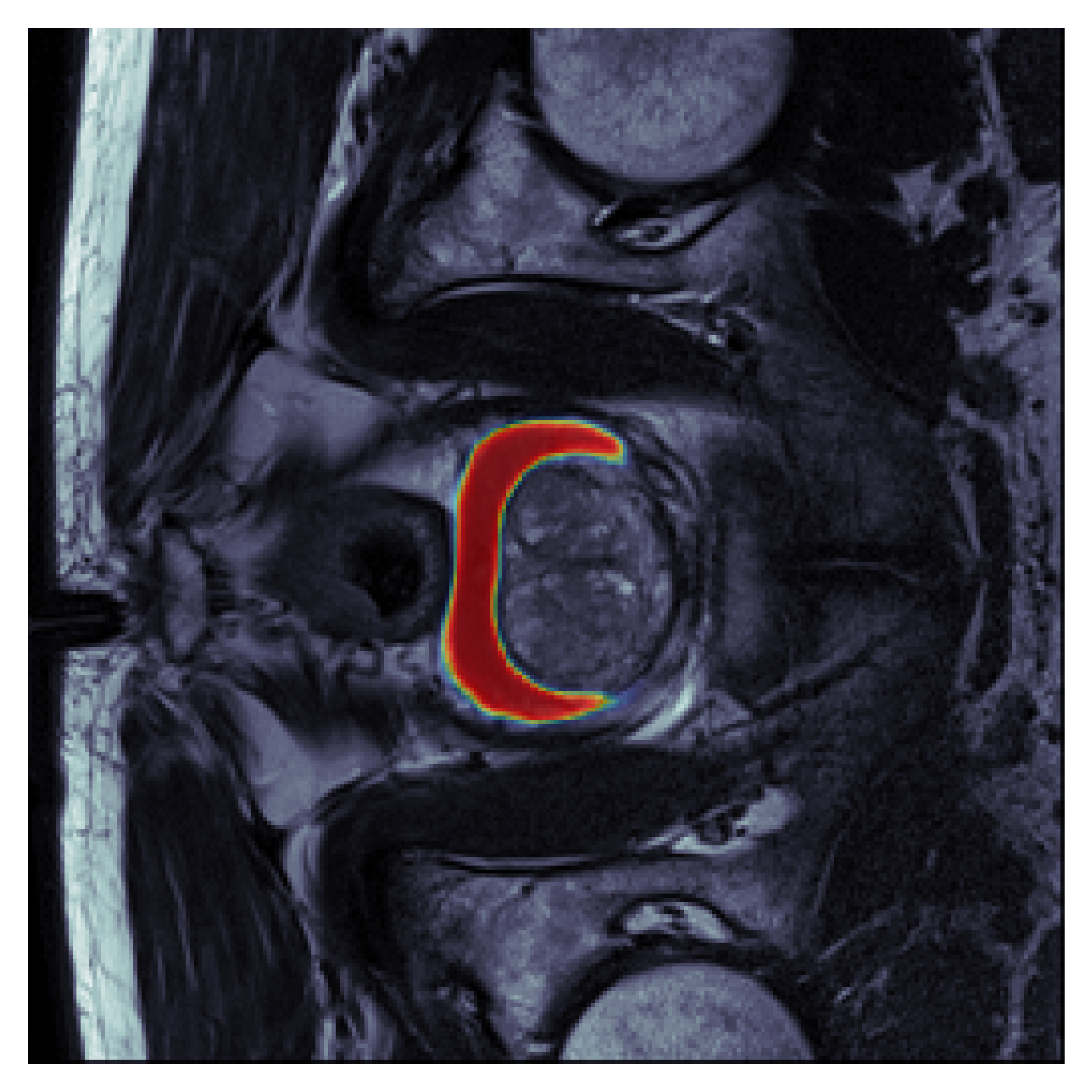} & \includegraphics[width=35mm]{predictions/MSD-BCELoss-hardtanh_activation-resnet34-Unet.png} &   \includegraphics[width=35mm]{predictions/MSD-DiceLoss-hardtanh_activation-resnet34-Unet.png} \\
  (e) Dice+CDF & (f) MSE+CDF & (g) BCE+HardTanh & (h) Dice+HardTanh \\[6pt]

  \includegraphics[width=35mm]{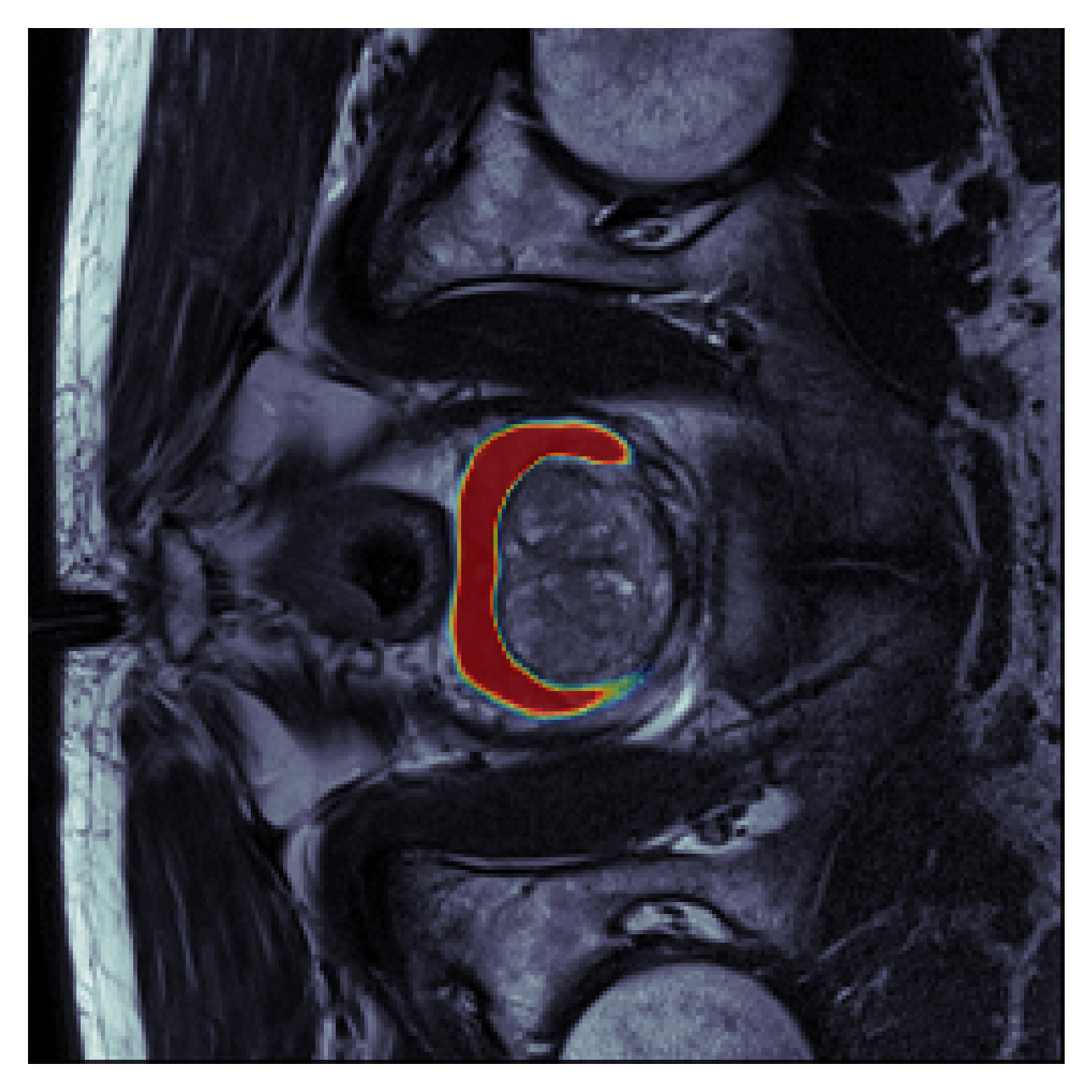} &   \includegraphics[width=35mm]{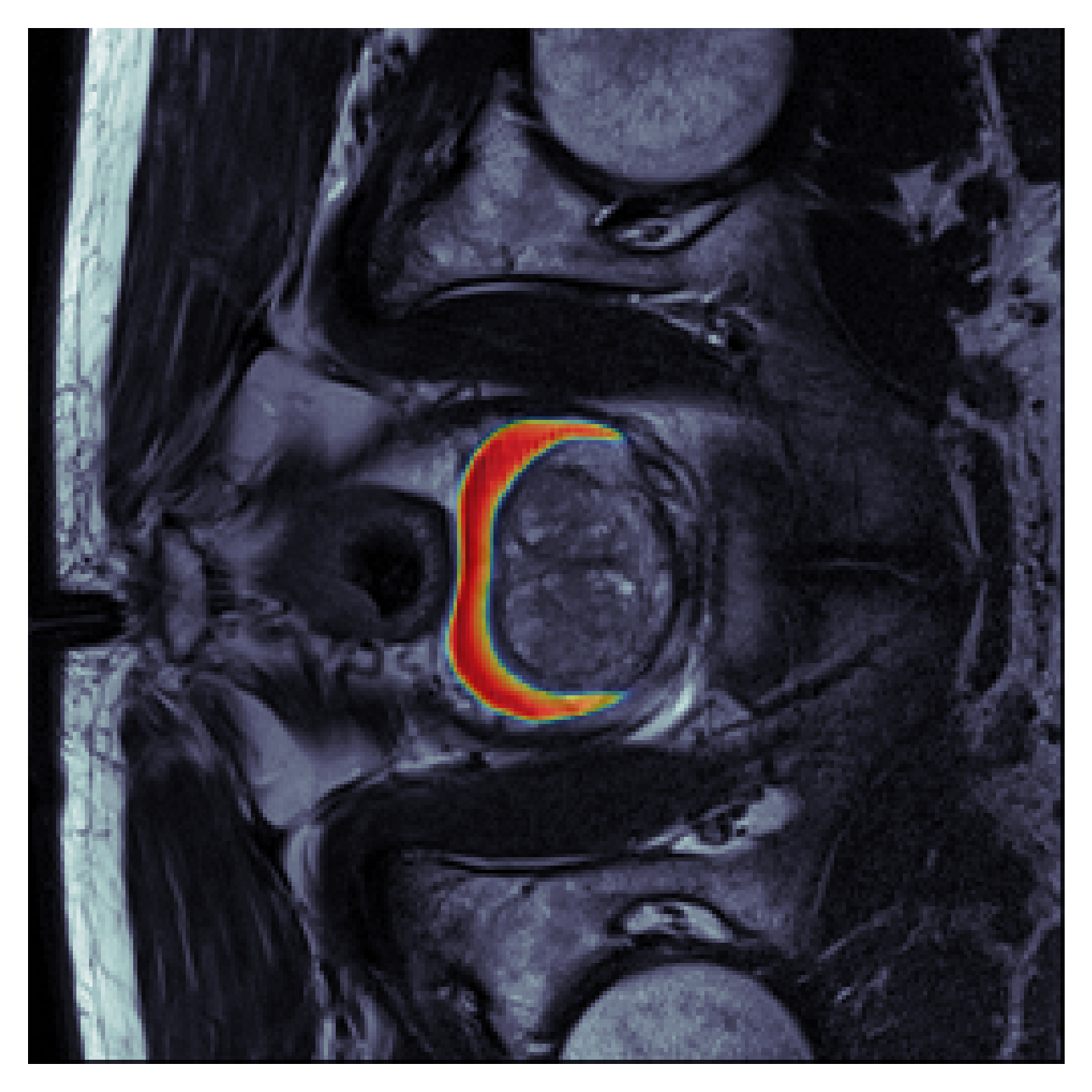} & \includegraphics[width=35mm]{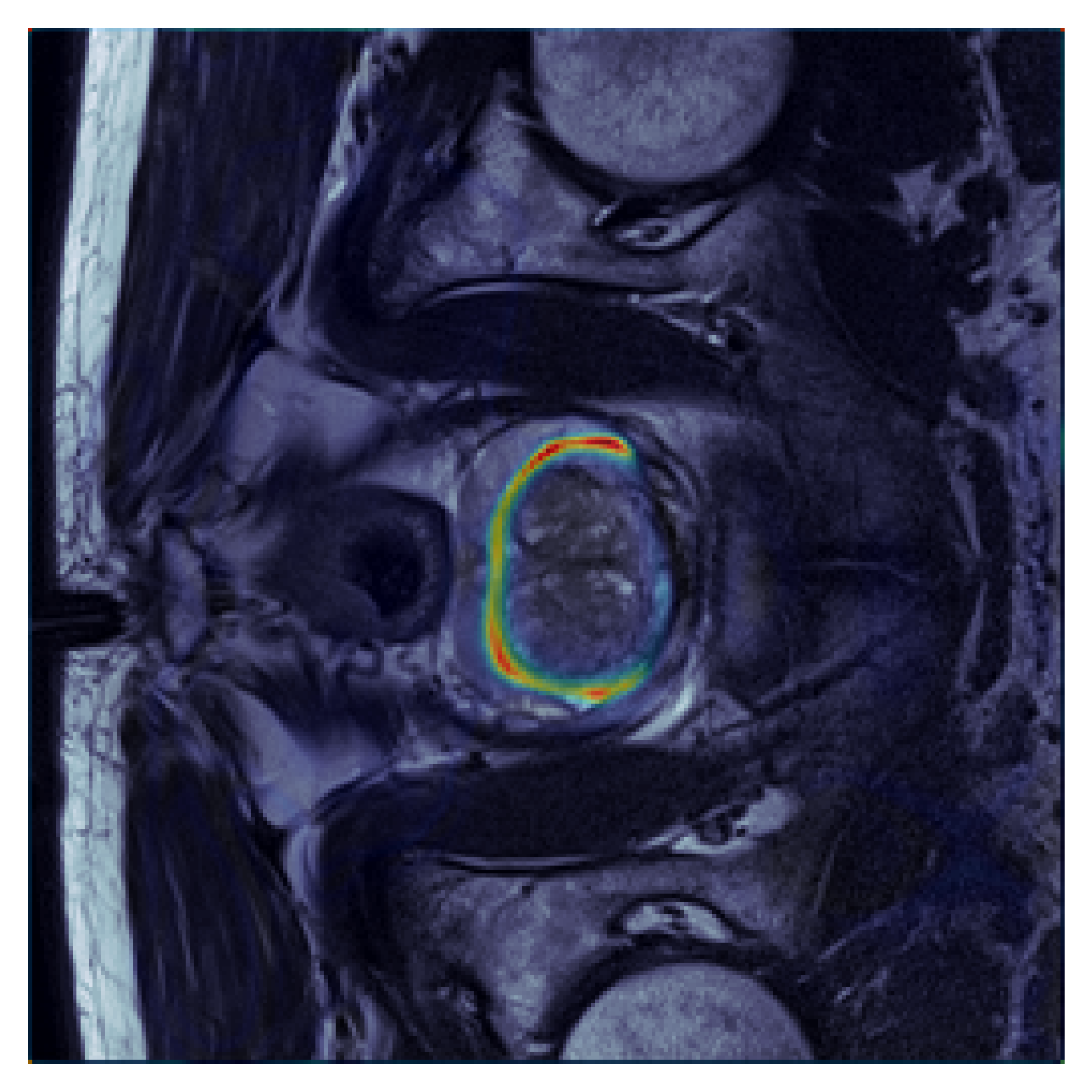} &   \includegraphics[width=35mm]{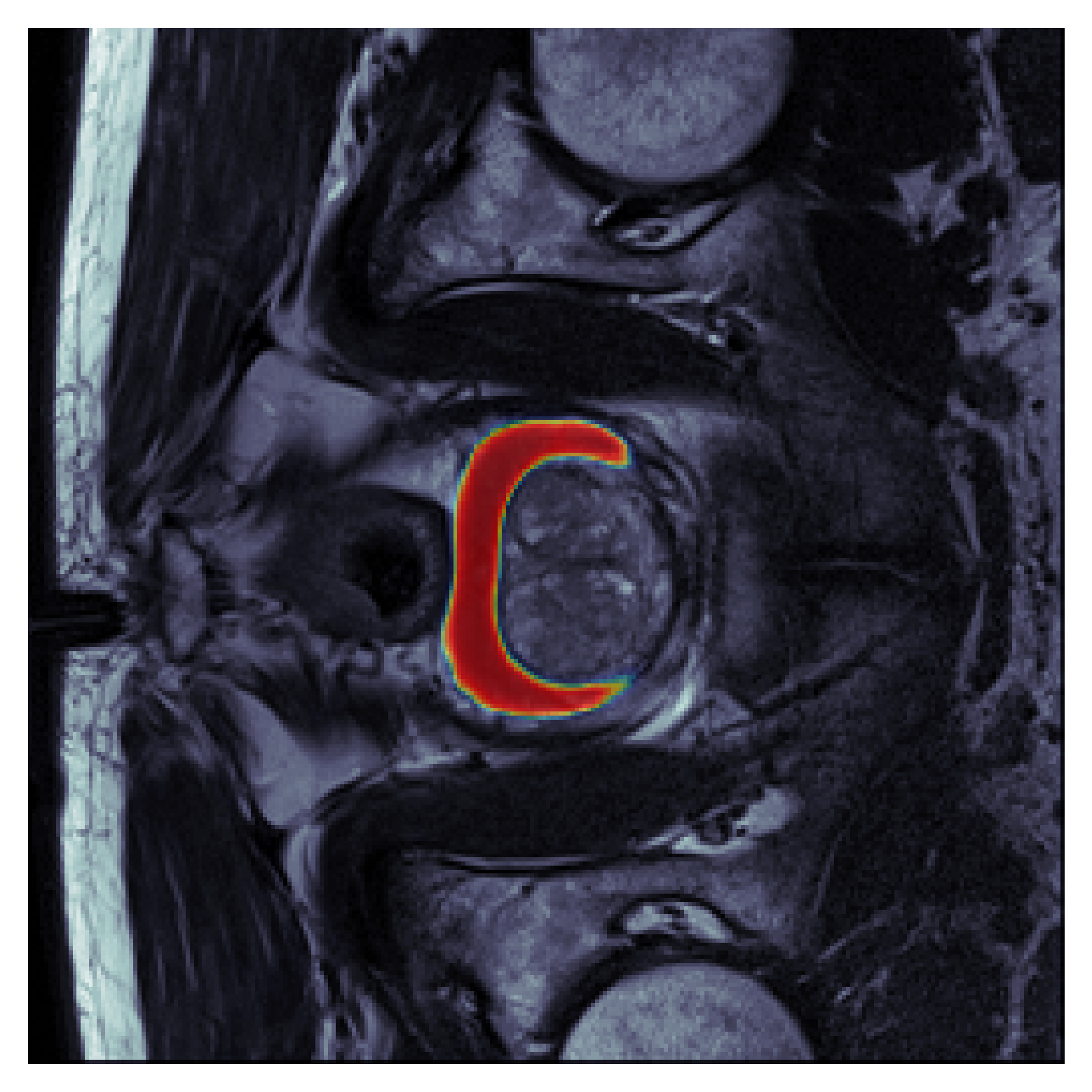} \\
  (i) MSE+HardTanh & (j) BCE+InvSquareRoot & (k) Dice+InvSquareRoot & (l) MSE+InvSquareRoot \\[6pt]

  \includegraphics[width=35mm]{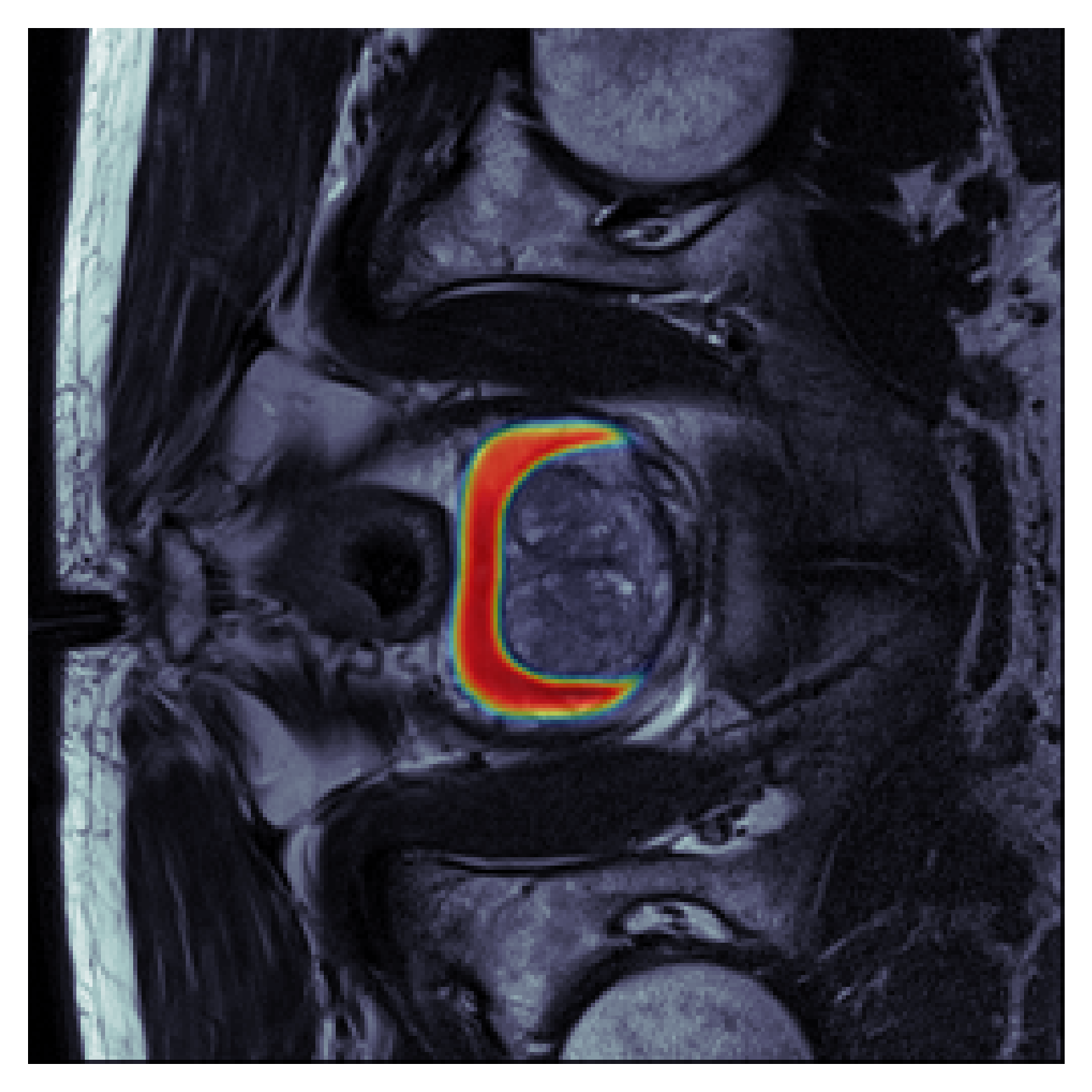} &   \includegraphics[width=35mm]{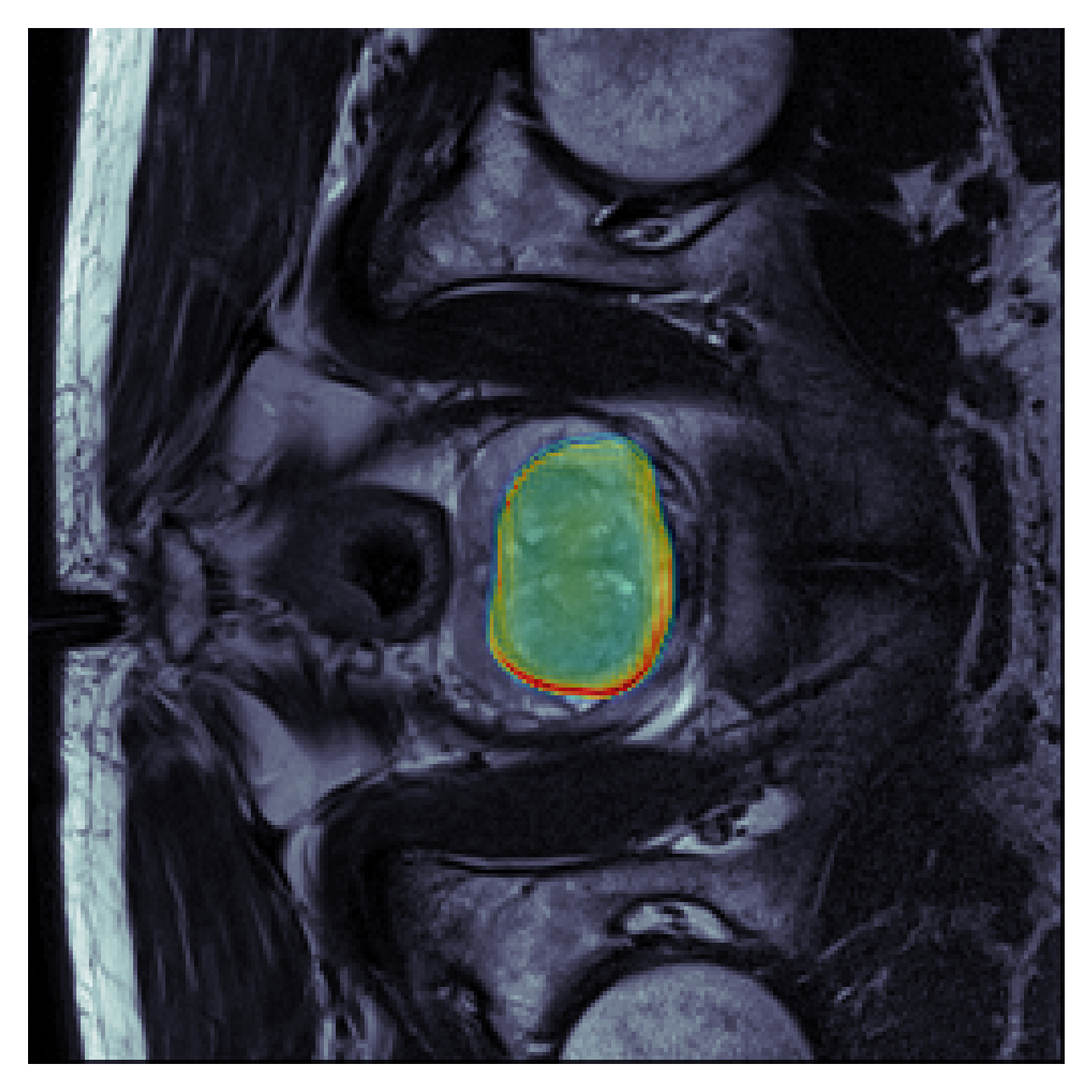} & \includegraphics[width=35mm]{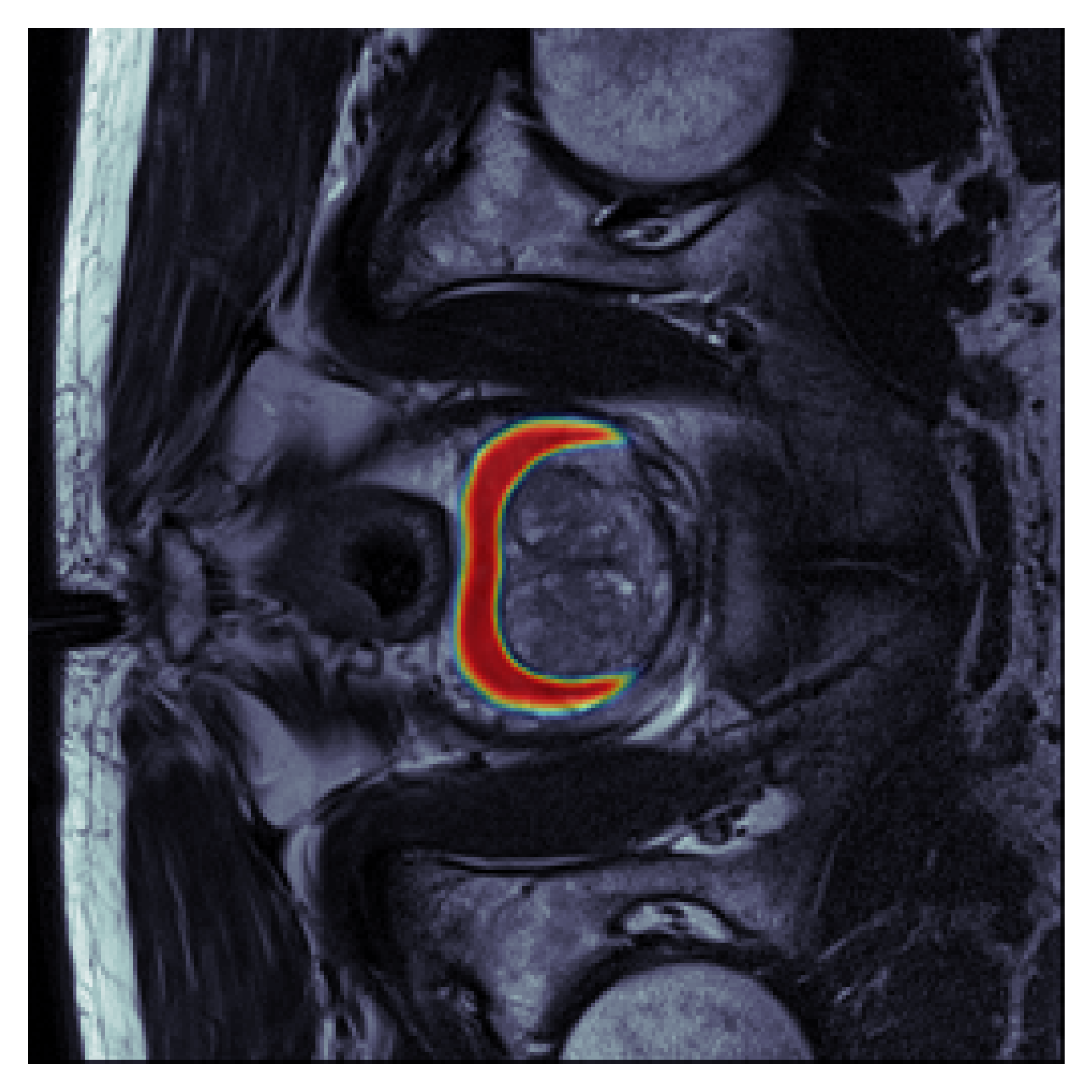} &   \includegraphics[width=35mm]{predictions/MSD-BCELoss-sigmoid_activation-resnet34-Unet.png} \\
  (m) BCE+Linear & (n) Dice+Linear & (o) MSE+Linear & (p) BCE+Sigmoid \\[6pt]

  \includegraphics[width=35mm]{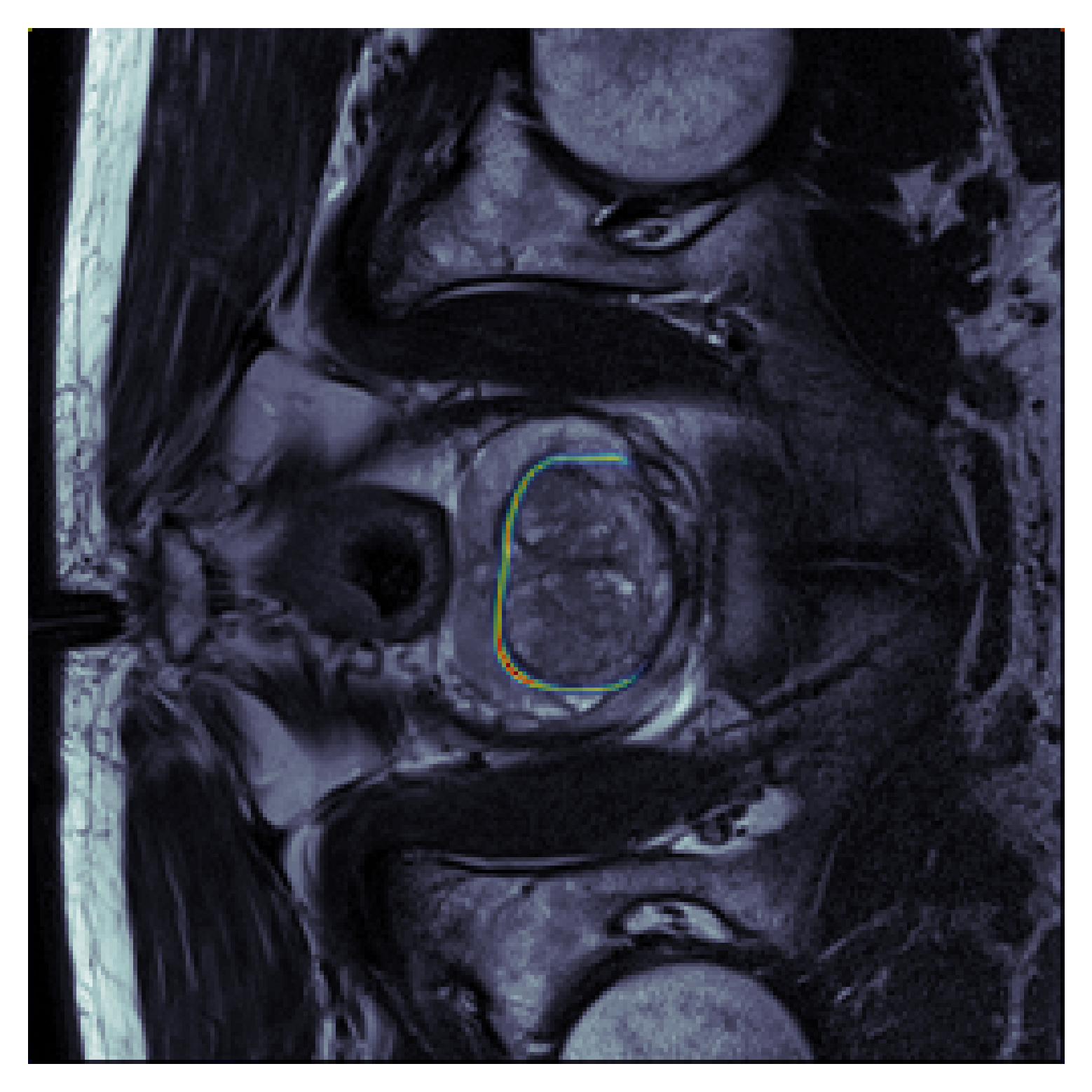} &   \includegraphics[width=35mm]{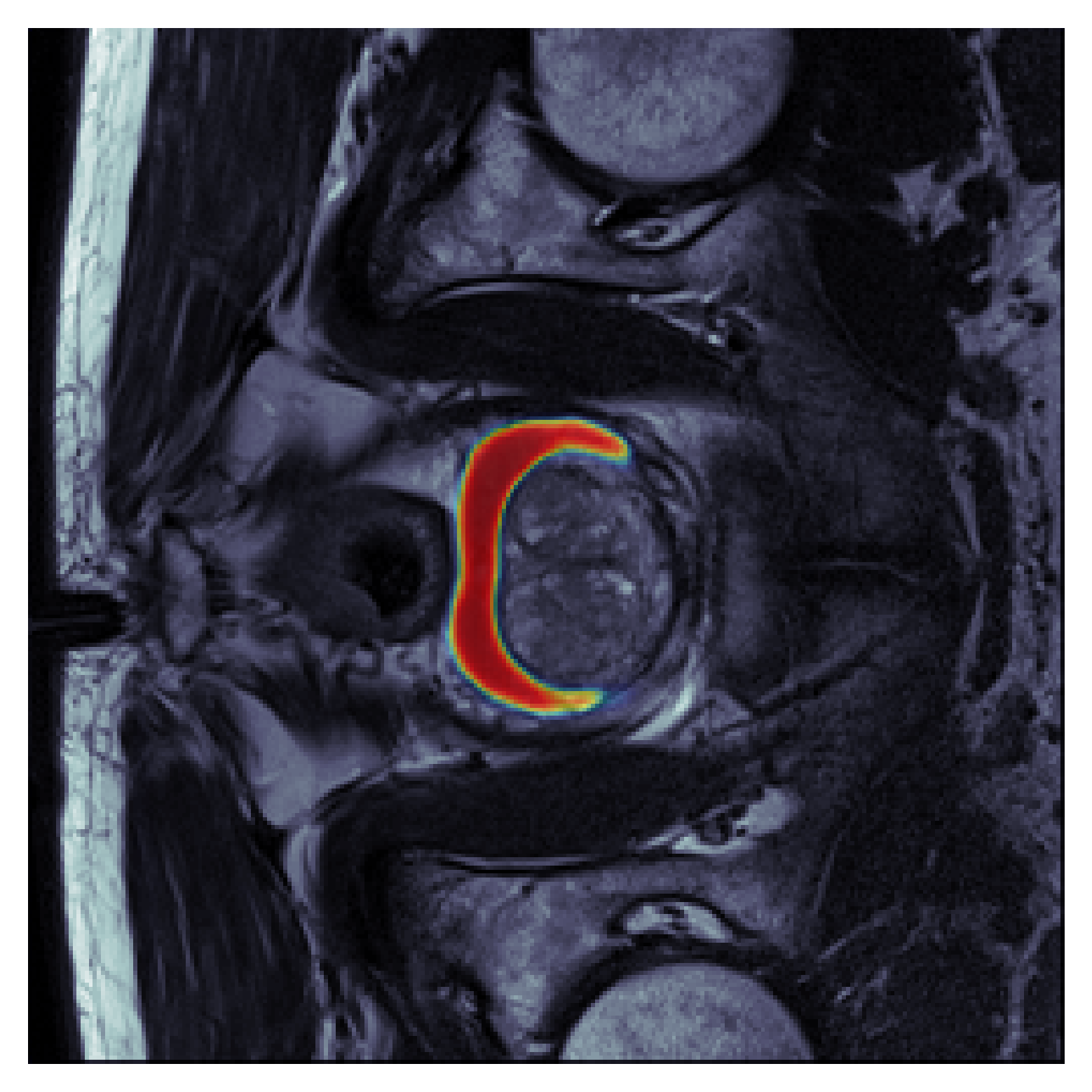} & \includegraphics[width=35mm]{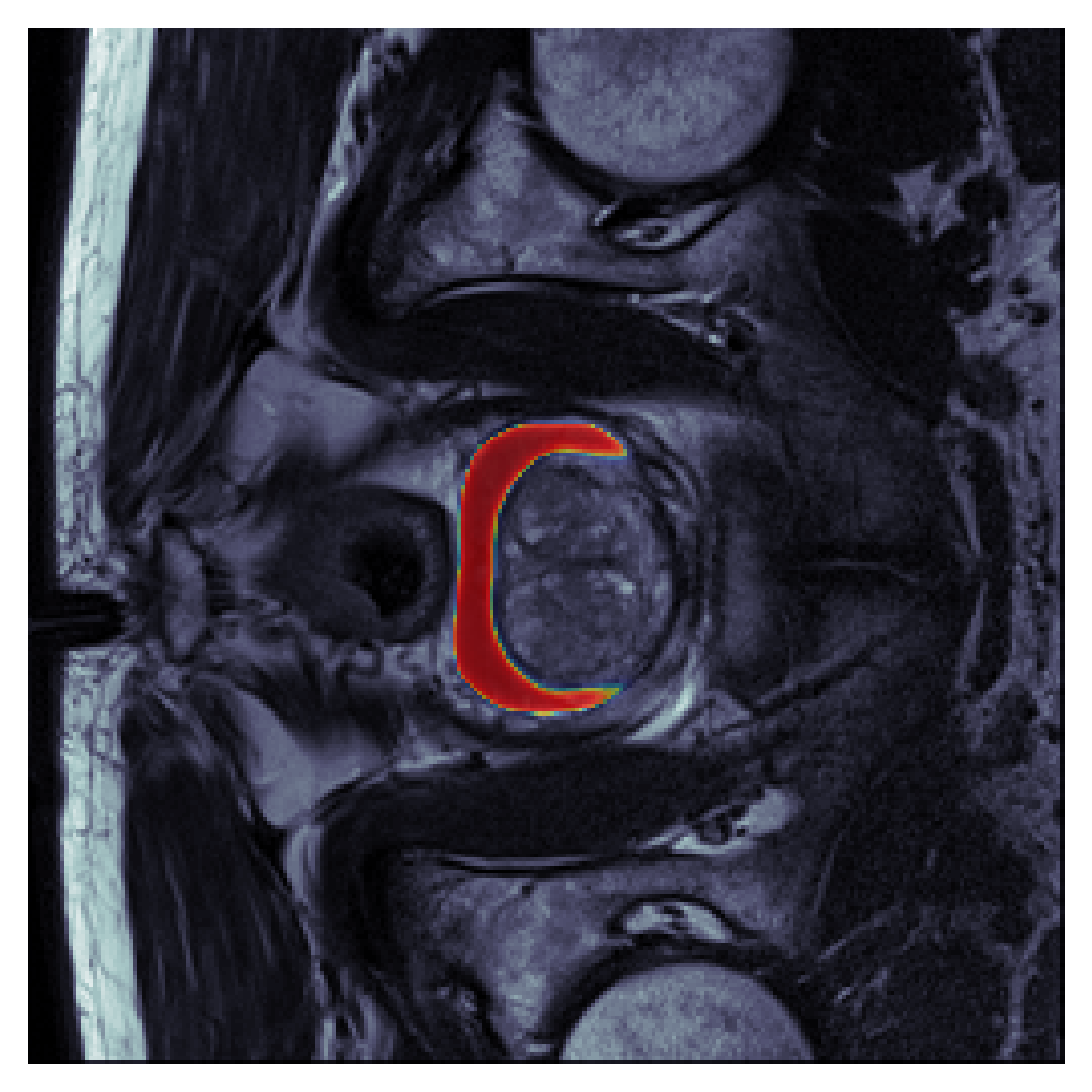} &   \includegraphics[width=35mm]{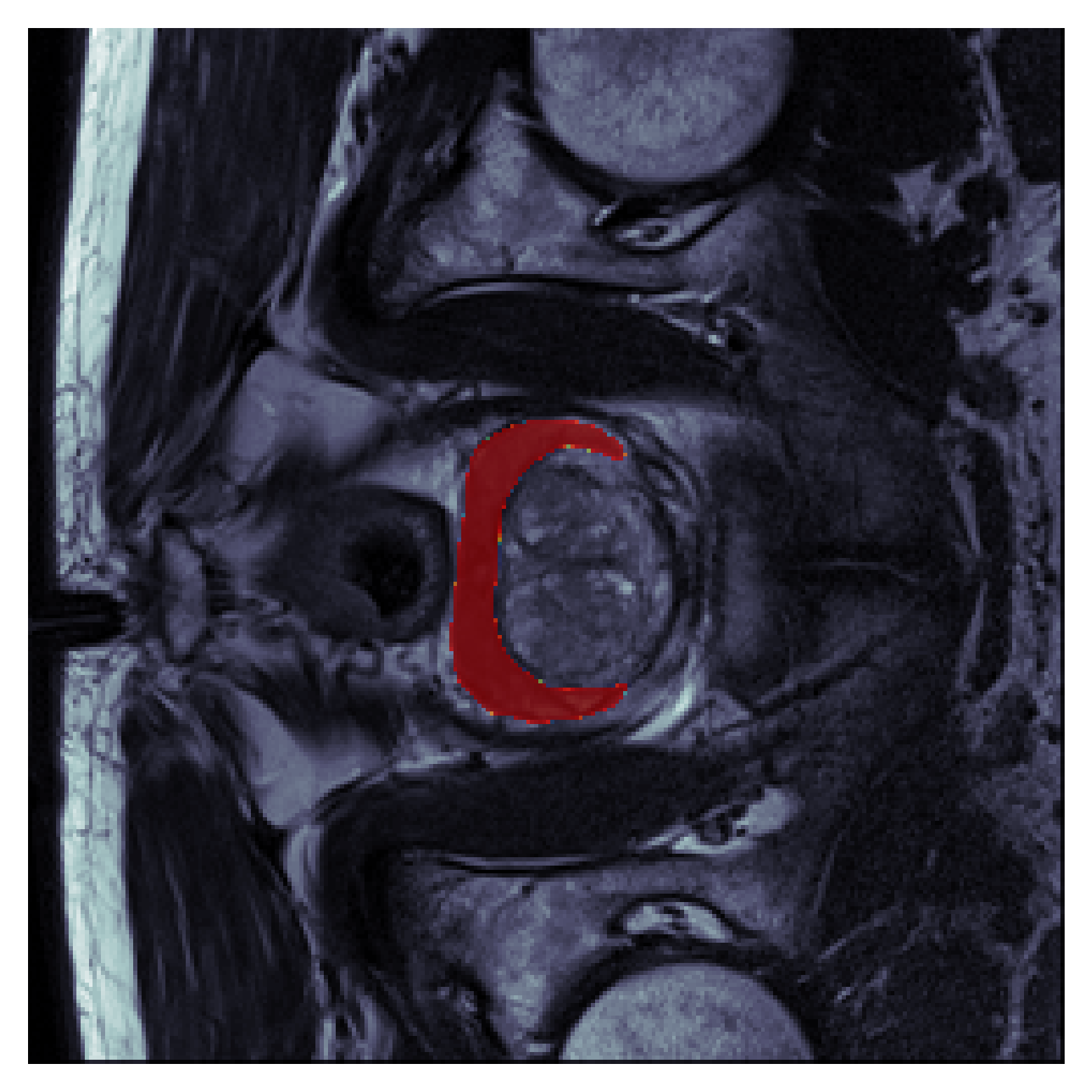} \\
  (q) Dice+Sigmoid & (r) MSE+Sigmoid & (s) BCE+Softsign & (t) Dice+Softsign \\[6pt]
  \includegraphics[width=35mm]{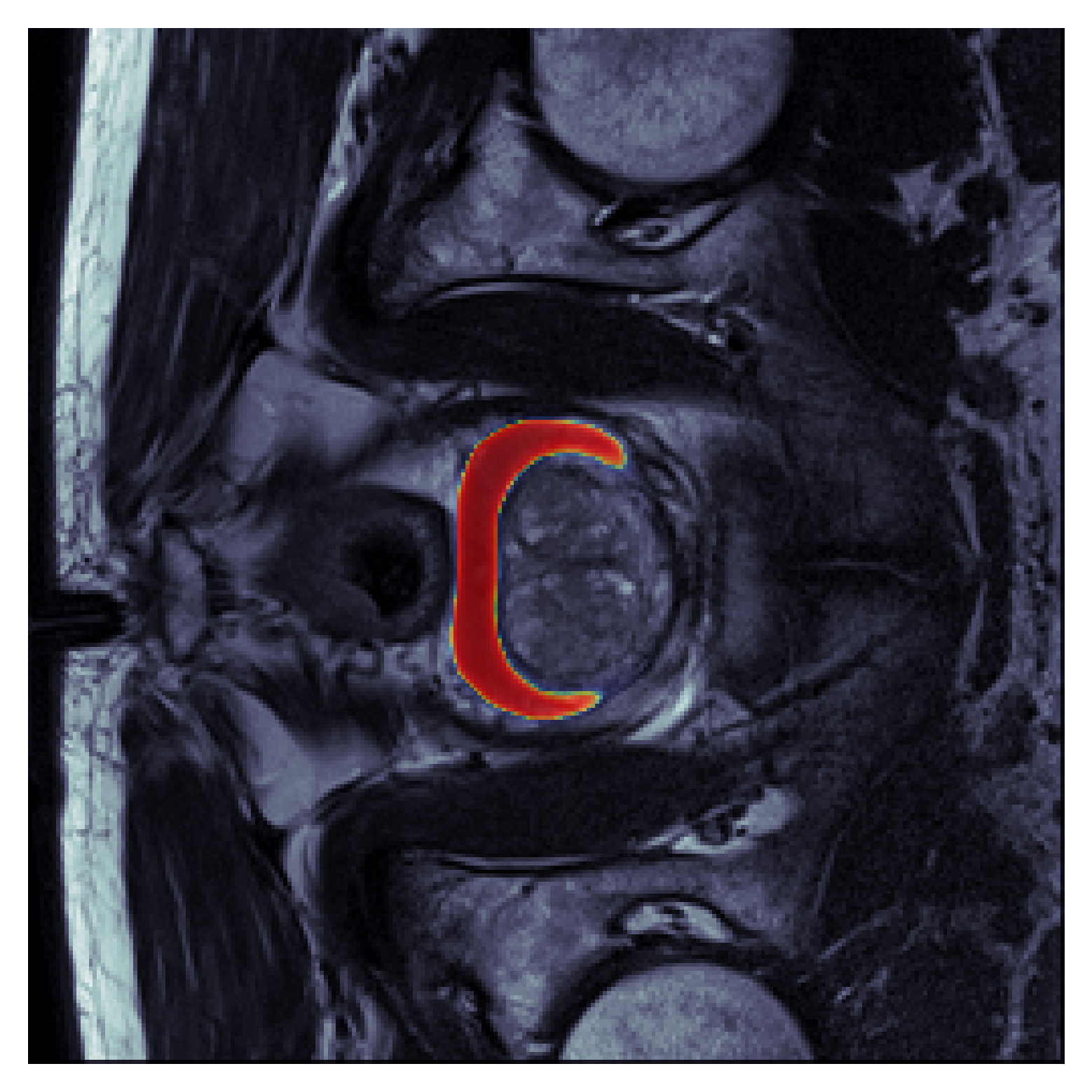} & \includegraphics[width=35mm] {predictions/MSD_gt.png} \\ (u) MSE+Softsign & (v) Ground truth
\end{tabular}}
\label{fig:allpred1}
\caption{All $21$ predictions for a single image of the MSD dataset, with the last image being the ground truth.}
\end{figure}

\bibliography{report} 
\bibliographystyle{spiebib} 

\end{document}